\documentclass{JHEP3} 
\usepackage{amsmath}
\usepackage{epsfig}
\usepackage{amssymb,amsfonts,nicefrac}
\newcommand{\oao}[2]{{#1\atopwithdelims[]#2}}

\def\zi{\mathbb{Z}}

\def\be{\begin{eqnarray}}
\def\ee{\end{eqnarray}}
\def\slr{$SL(2,\mathbb{R})\ $}
\def\slc{$SL(2,\mathbb{R})/U(1)\ $}
\def\sslash{\slash \hskip-1mm \slash}

\def\rrangle{\rangle \hskip-.5mm \rangle}
\def\nn{\nonumber}

\title{
\boldmath D-branes in Little String Theory\footnote{Research partially supported by the EEC 
under the contracts MRTN-CT-2004-512194 and  MRTN-CT-2004-005104.}
\unboldmath}
\author{Dan Isra\"el${}^{1,2}$, Ari Pakman${}^{2}$ and
Jan Troost${}^1$
\\  ${}^1$ Laboratoire de Physique Th\'eorique
de l'\'Ecole Normale Sup\'erieure\thanks{Unit{\'e} mixte  du
CNRS et de l'Ecole Normale Sup{\'e}rieure,
UMR 8549.}  \\ 24, Rue Lhomond  75231
Paris Cedex  05, France\\
 \\
 $ {}^2$ Racah Institute of Physics, The Hebrew University \\
Jerusalem 91904, Israel \\
\\
E-mail:  \email{israeld@phys.huji.ac.il}, \email{pakman@phys.huji.ac.il},
\email{troost@lpt.ens.fr}}

\abstract{We analyze in detail the D-branes in the near-horizon
limit of NS5-branes on a circle, the holographic dual of 
little string theory in a double scaling limit. 
We emphasize their geometry in the
background of the NS5-branes and show the relation with D-branes
in coset models. 
The exact one-point functions giving the coupling of the closed string
states with the D-branes and the spectrum of open strings are computed.
Using these results,
we analyze several aspects of Hanany-Witten setups, using exact CFT analysis.
In particular we identify the open string spectrum on
the D-branes stretched between
NS5-branes which confirms the low-energy analysis in brane constructions, and
that allows to go to higher energy scales. 
As an application we show
the emergence of the beta-function of the N=2 gauge theory on
D4-branes stretching between NS5-branes from the boundary states
describing the D4-branes. 
We also speculate on the possibility
of getting a matrix model description of little string theory 
from the effective theory on the D1-branes. 
By considering D3-branes
orthogonal to the NS5-branes we find a CFT incarnation of
the Hanany-Witten effect of anomalous creation of D-branes.
Finally we give an brief description of some non-BPS
D-branes.
}

\preprint{
LPTENS-05/08
\\hep-th/0502073}

\begin{document}
%%%%%%%%%%%%%%%%%%%%%%%%%%%%%%%%%%%%%%%%%%%%%%%%%%%%%%%
%%%%%%INTRO%%%%%%%%%%%%%%%%%%%%%%%%%%%%%%%%%%%%%%%%%%%%

\section{Introduction and summary}
The anti-deSitter/conformal field theory
correspondence is the best studied example of holography in string
theory \cite{Maldacena:1997re}.
The linear dilaton / little string theory (\textsc{lst})
map extends the holographic duality (which is often thought to be
generically valid in theories of quantum gravity
\cite{'tHooft:1993gx}\cite{Susskind:1994vu})
to further non-trivially curved
non-compact
backgrounds \cite{Aharony:1998ub}. It relates the theory
living on the NS5-branes to string theory on the near-horizon limit of
the background
created by such branes. The former is a rather mysterious non-gravitational
theory in 5+1 dimensions \cite{Seiberg:1997zk}, which in type IIB
superstring theory
has gauge group $U(k)$ (for $k$ NS5-branes) at low energies
and $\mathcal{N} = (1,1)$ supersymmetry,
and it shares many properties with string theory, like T-duality and
a Hagedorn transition at high temperature.
For coincident NS5-branes, the closed string background
is described \cite{Callan:1991dj,Kounnas:1990ud} by the exact superconformal field theory
$SU(2)_{k} \times \mathbb{R}_{Q^2 = 2/k}$. The background is a strongly
coupled string theory due to the linear dilaton.

It is possible
to obtain a perturbative description of the closed string dual to
Little String Theory by going to the Higgs phase
of Little String Theory\cite{Giveon:1999px},
where the gauge group is broken to $U(1)^k$ by
expectation values for the scalars on the worldvolume of the NS5-branes.
By scaling the string coupling to zero (decoupling gravity from the worldvolume)
while keeping the
W-boson mass (i.e. the mass of the D1-branes stretching between the
NS5-branes) fixed, one obtains a manageable theory, with a perturbative
closed string dual, called doubly scaled little string theory. 
In~\cite{Israel:2004ir}(using the results 
of~\cite{Sfetsos:1998xd}) the link to the
 neat geometrical description as the near-horizon limit of the supergravity
solution for a ring of $k$ five-branes was clarified.

We will study branes
in the closed string background corresponding to the doubly scaled Little
String Theory, both semi-classically and exactly. An important difference
with previous studies (see e.g. \cite{Elitzur:2000pq}\cite{Lerche:2000uy}\cite{Gava:2001gv})
is that we have recently gained more control over non-rational conformal
field theories, which allows for a precise analysis of the full string
theory background. Note that part of the exact boundary CFT analysis has been
done  in~\cite{Eguchi:2003ik,Eguchi:2004ik}, where the emphasis was put
 on the relation with singular CY compactifications.

In this paper we are mainly interested in the configurations of
D-branes and NS5-branes like those studied 
in~\cite{Hanany:1996ie,Witten:1997sc} to derive properties
of supersymmetric field theories. 
We will
study the non-trivial geometries of the D-branes in the curved background created by
the NS5-branes and relate them to the exact CFT analysis. We will find
a very good agreement between the qualitative picture that one can get by considering,
for instance, the S-dual configuration of D-branes at tree 
level (i.e. D-branes configurations
without taking into account the backreaction) and the exact analysis in 
the curved backgrounds
that can be taken to arbitrarily high curvatures (in the stringy perturbative 
regime).
In particular the quantization of various parameters of the D-branes 
have a natural geometrical
significance. We are also able to find a worldsheet realization of the
Hanany-Witten effect~\cite{Hanany:1996ie} of creation of D1-branes when a stack of D3-branes
crosses an NS5-brane.

Another important aspect is to study the field theory on the D-branes itself,
in the spirit of~\cite{Witten:1997sc}. For example from D4-branes suspended
between NS5-branes we obtain a four dimensional $\mathcal{N}=2$ SYM theory at
low
energy, and
we show how the boundary state encodes information about the beta-function
of this gauge theory. The appearance of new massless hypermultiplets when
two stacks of D-branes ending on both sides of an NS5-brane are aligned is
also proven. The finite D1-branes suspended between the NS5-branes in type IIB
are probably the most important objects to consider. Indeed they correspond
to the W bosons of the broken gauge symmetry $U(k) \to U(1)^k$ on the
five-branes. We shall argue that they may give a matrix model definition of Higgsed little string theory.

Our paper contains a review of the bulk theory in section \ref{bulk},
and additional remarks on instanton corrections are given
in appendix~\ref{localization-appendix}.
In section \ref{semiclass} we review and extend our understanding
of the semi-classical geometries of D-branes in coset theories, and use them
to construct non-trivial geometries for D-branes in the background of the ring of NS5-branes. 
In section \ref{exact} we construct the corresponding exact
boundary states for these branes, determine the precise open string
spectrum, and we show various applications as mentioned in the introduction.
We then conclude in section \ref{conclusions}. Additional data about characters and modular transformation
is gathered in appendices.

%%%%%%%%%%%%%%%%%%%%%%%%%%%%%%%%%%%%%%%%%%%%%%%%%%%%%%%%%%%%%%%%%%%%%%%
%%%%%%%%%%%%%%%%%%%%%%%%BULK%%%%%%%%%%%%%%%%%%%%%%%%%%%%%%%%%%%%%%%%%%%
%%%%%%%%%%%%%%%%%%%%%%%%%%%%%%%%%%%%%%%%%%%%%%%%%%%%%%%%%%%%%%%%%%%%%%%
\section{Bulk geometry}
\label{bulk}
We discuss in this section the bulk geometry of
string theory which
corresponds to the backreaction of the massless bulk fields
to the presence of NS5-branes
which are solitonic objects with mass proportional to $g_s^{-2}$ and
with magnetic charge under the NS-NS 3-form field strength $H$.
The generic NS-NS background corresponding to NS5-branes parallel
to the $x^{\mu=0,1,2,3,4,5}$-directions, in the string frame is:
\begin{eqnarray}
ds^2 &=& \eta_{\mu\nu} dx^{\mu} dx^{\nu} + H(x^i) dx^i dx_i \nonumber \\
e^{2 \Phi}  &=& g_s^2 \, H(x^i) \nonumber \\
H_{ijk} &=& - {\epsilon^l}_{ijk} \partial_l H(x^i)
\end{eqnarray}
where the harmonic function $H$ is given in terms of the positions of
the $k$ NS5-branes $x^{i=6,7,8,9}_a$ (indexed by the variable $a$) as:
\begin{eqnarray}
H(x^i) &=& 1 + \sum_{a=1}^k \frac{\alpha'}{|x^i-x^i_a|^2}.
\end{eqnarray}
In the following we will concentrate on $k$ NS5-branes spread evenly
on a topologically trivial circle of radius $\rho_0$ in the $(x^6,x^7)$
plane (see fig.~\ref{distribran}). We parameterize the transverse space as
\begin{eqnarray}
(x^6,x^7) &=& \rho \, (\cos \psi,\sin \psi)  \nonumber \\
(x^8,x^9) &=& R \, (\cos \phi,\sin \phi)
\end{eqnarray}
and for this distribution of NS5 branes, the harmonic function $H$ becomes~\cite{Sfetsos:1998xd}
\begin{eqnarray}
\label{formsofH}
H&=& 1+ \sum_{a=0}^{k-1} \frac{\alpha'}{R^2 +\rho^2+\rho_0^2-2
\rho_0 \rho \cos(\frac{2 \pi a}{k}-\psi)} \\
&=& 1 + \frac{\alpha'}{2 \rho_0 \rho} \sum_{a=0}^{k-1} \frac{1}{\cosh y - \cos(\frac{2 \pi a}{k}-\psi)} \nn \\
&=& 1 + \frac{\alpha' e^{-y}}{ \rho_0 \rho}   \sum_{m,n=0}^{\infty} \sum_{a=0}^{k-1} e^{-(m+n)y}
e^{i(m-n)(\frac{2\pi a}{k}-\psi)} \nn \\
&=& 1+ \frac{\alpha'\, k}{2 \rho \rho_0 \sinh y} \Lambda_k(y,\psi) \,, \nn
\end{eqnarray}
where
\begin{eqnarray}
\cosh y = \frac{R^2+\rho^2+\rho_0^2}{2 \rho \rho_0} \,,
\label{defy}
\end{eqnarray}
and the function $\Lambda_k(y,\psi)$, which keeps track of the location of the $k$ throats, is
\begin{eqnarray}
\label{lambda}
\Lambda_k(y,\psi) = 1 + \sum_{\pm} \sum_{m =1}^{\infty} e^{-m ky \pm imk \psi} =  \frac{\sinh ky}{\cosh ky-\cos k \psi} \,.
\end{eqnarray}
In most of this paper we will 
consider $\Lambda_k \rightarrow 1$ for the semi-classical analysis of
the D-branes, which is valid in the large $k$ limit.
The resulting function $H$ is still harmonic, and
corresponds to an homogeneous distribution
of k NS5 branes on the circle $\psi$, as follows from
\be
H &=& 1+ \frac{\alpha'\, k}{2 \rho \rho_0 \sinh y}
=1 + \frac{\alpha' k}{\sqrt{(R^2 + \rho^2 + \rho_0^2)^2 - 4 \rho^2 \rho_0^2}} \\
&=& 1 + \frac{\alpha' k}{2 \pi} \int_0^{2 \pi} \frac{d \phi}{R^2+ \rho^2 + \rho_0^2 -2 \rho \rho_0 \cos(\phi)}  \,. \label{rrr}
\ee
The infinite series discarded in (\ref{lambda}), which is responsible for the localization
of the NS5's along the circle, should appear as worldsheet instanton corrections to the 
$\mathcal{N}=(4,4)$ worldsheet non-linear sigma model, when  it is realized
as the low energy limit of a gauged linear sigma model. This instantonic localization phenomenon
has been proved explicitly in \cite{Tong:2002rq} for the case of an infinite array of NS5 branes along a line.
As we show in Appendix \ref{localization-appendix}, that setting
corresponds to a particular limit of our geometry, and the instanton
 corrections coincide in the limit.

To prepare for the double-scaling limit in which we keep the W-boson mass
fixed:
\begin{equation}
g_s, \rho_0 \to 0\ ; \ \ \  \alpha',
 \frac{\rho_0}{g_s } \ \ \mathrm{fixed},
\end{equation}
we parameterize the radial directions $(\rho, R)$
with new coordinates $(r,\theta)$, with $r \geq 0$ and $\theta \in [0, \pi/2]$:
\begin{eqnarray}
(x^6,x^7) &=& \rho_0 \cosh r \sin \theta \, (\cos \psi,\sin \psi)\,,  \nonumber \\
(x^8,x^9) &=& \rho_0 \sinh r \cos \theta \, (\cos \phi,\sin \phi)\,,
\label{cartcoords}
\end{eqnarray}
In these coordinates we have
\begin{eqnarray}
dx^i dx_i &=& \rho_0^2 (\cosh^2 r- \sin^2 \theta)
\left[ dr^2+d \theta^2 + \frac{ \tanh^2 r \  d \phi^2 +  \tan^2 \theta d
\psi^2 }{1+\tan^2 \theta \tanh^2 r}
\right] \nn \\
H &=& 1 + \frac{k \alpha'}{\rho^2_0 (\cosh^2 r- \sin^2 \theta)}
\end{eqnarray}
The double-scaling limit amounts to drop the constant
term  in the harmonic
function. From (\ref{rrr}) it is clear that this limit coincides with the geometry seen by
a near-horizon
observer,
i.e., $R \to 0, \, \rho \to \rho_0$. The resulting NSNS-background is
\begin{eqnarray}
ds^2 &=& dx^{\mu} dx_{\mu} + \alpha' k \,
\left[ dr^2+d \theta^2 + \frac{\tanh^2 r \ d \phi^2 + \tan^2 \theta \ d \psi^2}{1+\tan^2 \theta \tanh^2 r} \right] \, ,
\nonumber \\
e^{ 2 \Phi} &=& \frac{{g_{\textsc{eff}}}^2 }{\cosh^2 r - \sin^2 \theta} \, ,
\label{NS5geom} \\
B  &=& \frac{\alpha' k }{1+\tan^2 \theta \tanh^2 r} \ d \phi \wedge d \psi \,,
\nonumber
\end{eqnarray}
where the effective string coupling constant is
\begin{equation} g_\textsc{eff} = \frac{\sqrt{k \alpha'}g_s}{\rho_0} \end{equation}
and  we have chosen a gauge where all the other components of the $B$ field vanish.
Note that $(r,\theta,\psi,\phi)$ are dimensionless, and this is signaled by the factor $\alpha'$ in the metric.

The four-dimensional transverse space in (\ref{NS5geom}) is
an  exact coset CFT, corresponding to the null gauging 
$$\frac{SU(2)_k \times SL(2,\mathbb{R})_k}{U(1)_L \times U(1)_R}
$$ 
as  shown in \cite{Israel:2004ir}. Both
supersymmetric WZW models are at the same level $k$, corresponding
to the number of five-branes. 
\FIGURE{ \epsfig{figure=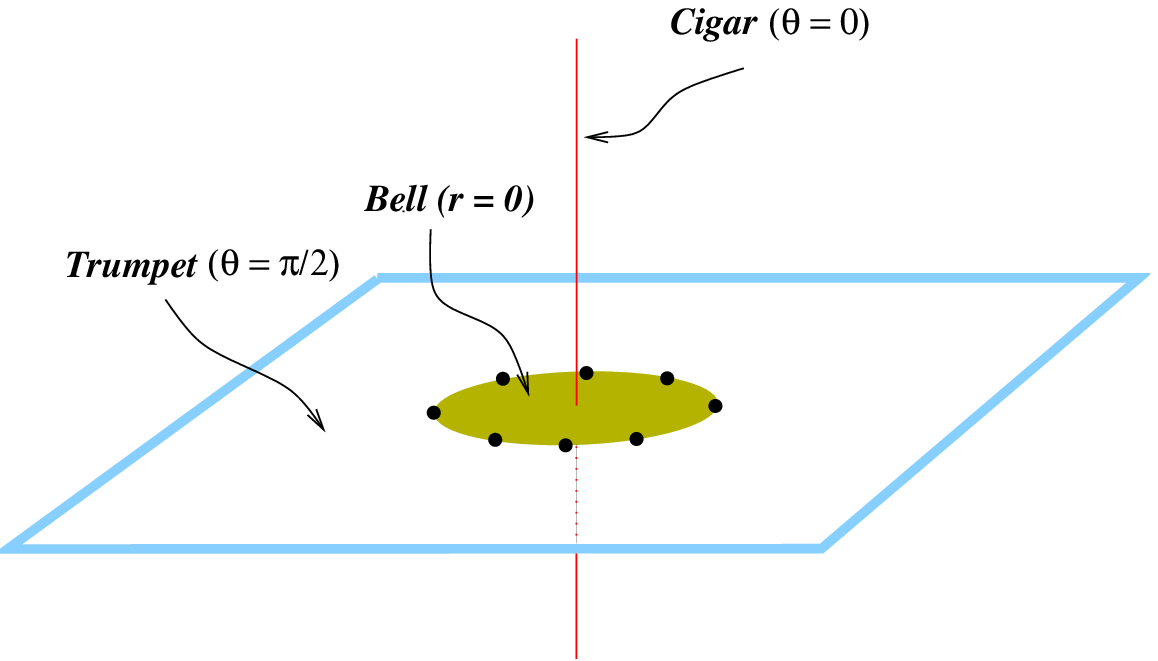, width=100mm}
\caption{Geometry of the ring of NS5-branes, and sections
corresponding to coset CFTs.} \label{cosetlimits}}
It is interesting to look at certain
two-dimensional sections, which have a geometry coinciding with
gauged $SU(2)/U(1)$ or $SL(2,\mathbb{R})/U(1)$ models (see 
figure~\ref{cosetlimits}):

\begin{itemize}
\item \underline{$x^8=x^9=0$ plane, inside of the NS5 circle: the bell}

This section is obtained by taking $r=0$ (see (\ref{cartcoords})),
and the two dimensional slice  is a  bell
\begin{eqnarray}
ds^2 = \alpha'k \, [ d\theta^2 + \tan^2 \theta d\psi^2 ] \,,
\end{eqnarray}
which corresponds to a $SU(2)/U(1)$ WZW model. The singularity at $\theta=\pi/2$ corresponds 
to the locus of the ring of five-branes. More generically,
each slice of the geometry for fixed $r$ can be viewed as a current-current 
deformation of $SU(2)$ of parameter $\tanh r$ (see
also~\cite{Kiritsis:2003cx,BFPR}).

\item \underline{$x^8=x^9=0$ plane, outside of the NS5 circle: the trumpet}

This section is obtained by taking $\theta=\pi/2$, and the
two dimensional slice  is  the trumpet
\begin{eqnarray}
ds^2 = \alpha'k \, [dr^2 + \frac{1}{\tanh^2 r} d \psi^2 ]\,,
\end{eqnarray}
which corresponds to a vector coset $SL(2,\mathbb{R})/U(1)$ model. 
Again, the singularity at $r=0$ corresponds 
to the locus of the ring of fivebranes.

\item \underline{$x^6=x^7=0$ plane: the cigar}

This section is obtained by taking $\theta=0$, and the
four-dimensional metric is reduced to the cigar
\begin{eqnarray}
ds^2 = \alpha'k \, [dr^2 + \tanh^2 r d \phi^2 ]\,,
\end{eqnarray}
which corresponds to an axial coset $SL(2,\mathbb{R})/U(1)$ model.
\end{itemize}

\noindent
Focusing in these two-dimensional sub-manifolds will be 
useful when we study the
shape of D-branes in the NS5 background.

\subsection{Two T-dual descriptions of the circle of NS5 branes}
Another connection with coset models appears when performing
T-duality along the isometries of the background (\ref{NS5geom}) in the directions of $\psi$ and $\phi$.

\subsection*{The T$_{\psi}$-dual geometry }

After a T-duality along the direction $\psi$
we obtain the dual torsionless solution (fig.~\ref{distribran}):
\be
ds^2 &=& \alpha' k \left[ d r^2 + \tanh^2 r \, \left( \frac{d\chi}{k}\right)^2
+ d\theta^2 + \mathrm{cotan}^2 \theta \, \left( \frac{d\chi}{k}-d\phi \right)^2 \right], \nn \\
e^{2 \Phi}&=& \frac{g^2_\textsc{eff}}{k} \ \frac{1}{\cosh^2 r \ \sin^2 \theta}
\label{cosetprod}
\ee
where $\chi$ is the coordinate T-dual to $\psi$.
The background is nothing but a vector $\zi_k$ orbifold of the coset
theories
$SU(2)_k /U(1) \, \times SL(2,\mathbb{R})_k /U(1)$, i.e. the cigar times
T-bell
 \footnote{
T-bell denotes the background T-dual to the standard parameterization of
the bell, with dilaton inversely proportional to $\sin \theta$ and where
the radial variable $\theta$ takes values in the interval
$\theta \in {[} 0 , \pi/2 ) $ --  in the geometrical picture for the
T-bell, the $\theta$ coordinate diminishes from center to border.
See subsection \ref{cosetgeometries}.}
background.
\FIGURE{
\epsfig{figure=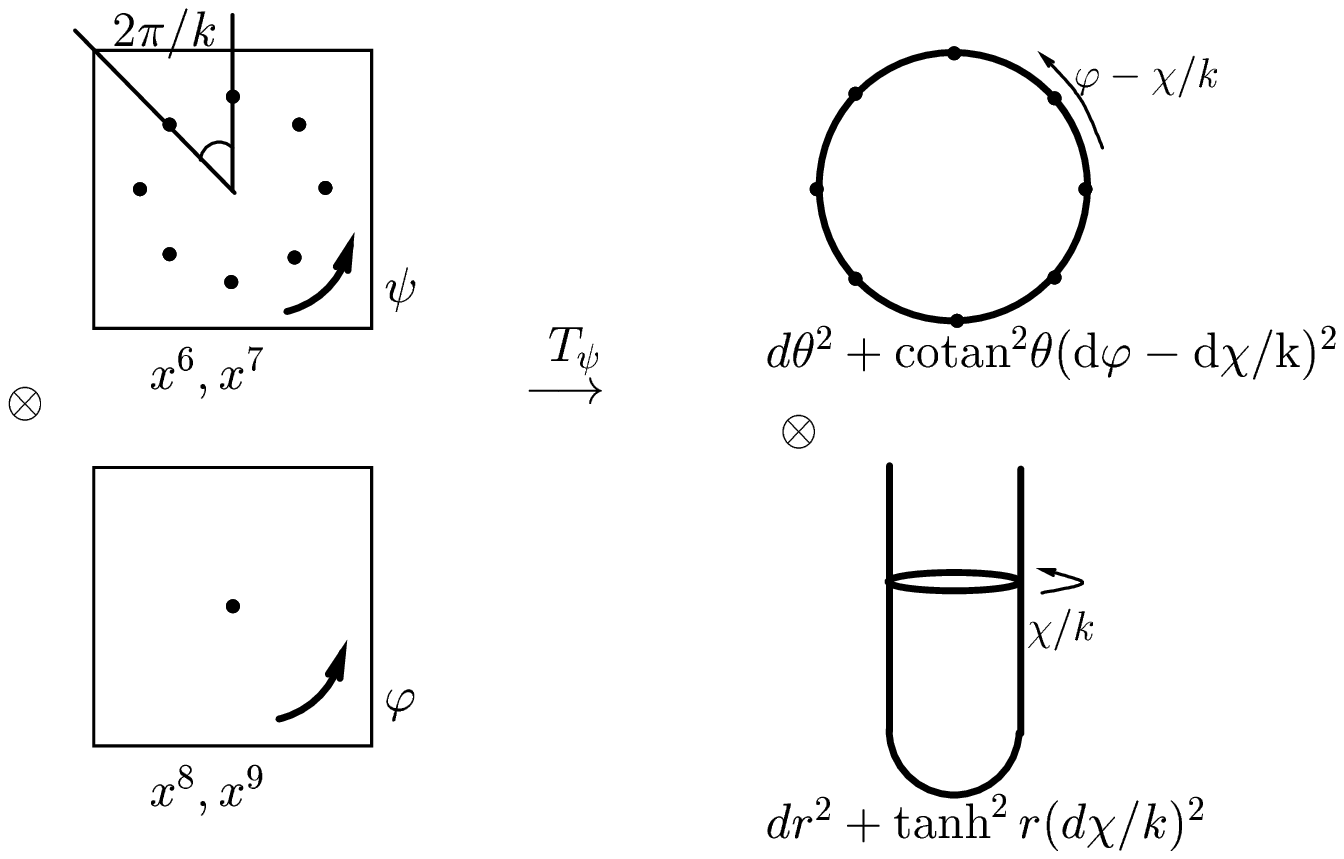, width=120mm}
\caption{Geometry of the brane setup and T$_{\psi}$-duality.}
\label{distribran}}

\subsection*{The T$_{\phi}$-dual geometry}
A T duality along $\phi$ in (\ref{NS5geom}) yields \cite{Ribault:2003hj, BFPR}:
\begin{eqnarray}
ds^2 &=& \alpha' k \left[ dr^2+d \theta^2+ \tan^2 \theta \left(\frac{d \omega}{k}\right)^2
+ \coth^2 r \left(\frac{d \omega}{k}+ d \psi\right)^2 \right] \nonumber \\
e^{2 \Phi}&=& \frac{g^2_\textsc{eff}}{k} \ \frac{1}{\sinh^2 r \ \cos^2 \theta}
\label{tphidual}
\end{eqnarray}
where $\omega$ is the coordinate dual to $\phi$ and we have written only the non-trivial
directions. This is a $\mathbb{Z}_k$ orbifold of the vector coset $SU(2)/U(1)$ (the bell) and the vector coset
\slc, the trumpet.

These T$_{\psi,\phi}$ duals will be the backgrounds in which we start out our
construction of the branes. In a second stage, we will
re-interpret them in the original  NS5 geometry.
We note at this point that when we consider
supersymmetric branes of even/odd space-dimension in the presence
of NS5-branes, we will work in type IIA/B string theory, and the
T-dual coset conformal field theory will be a background of type
IIB/A string theory.

To study these branes, it is useful to recall properties of the
D-branes in the coset backgrounds, which we will use as building
blocks for the branes in the background T-dual to the doubly
scaled little string theory.

\section{Semi-classical description of D-branes}
\label{semiclass}
We wish to obtain classical solutions of the Dirac-Born-Infeld action
for D-branes in NS5-brane backgrounds. We will concentrate on D-branes
that can be constructed out of D-branes of the coset theories
\slc and $SU(2)/U(1)$. Indeed the exact boundary states in these
gauged WZW theories are known and can be used to construct non-trivial
boundary states in the NS5-brane background. This will be the purpose of
the following sections. To that end, we will first review the D-branes in the coset
models and make some comments. Then we will move to the D-branes in the
background of five-branes obtained by T-duality.

\subsection{The geometry of D-branes in coset models: a review}
\label{cosetgeometries}
As a warm up exercise we review the D-brane taxonomy in the
axial and vector cosets $SL(2,\mathbb{R})_k /U(1)$ --~the cigar and
the trumpet~-- and the axial and vector cosets $SU(2)_k /U(1)$~-- the T-bell
and the bell (see e.g. \cite{Maldacena:2001ky}\cite{Fotopoulos:2003vc}).
We consider the following
parameterizations of the cigar, the trumpet, the bell and the T-bell
respectively:\footnote{In this section we set $\alpha' =1$ for convenience.}
\begin{eqnarray}
ds^2_c &=& k (dr^2 + \tanh^2 r \ d\psi_c^2) \qquad e^{\Phi} =
\frac{e^{\Phi_0}}{\cosh r} \nonumber \\
ds^2_t &=& k (dr^2 + \mbox{cotanh}^2 r \ d\psi_t^2) \qquad e^{\Phi} =
\frac{e^{\Phi_0}}{\sinh r} \nonumber \\
ds^2_b &=& k (d\theta^2 + \tan^2 \theta \ d\psi_b^2) \qquad e^{\Phi} =
\frac{e^{\Phi_0}}{\cos \theta} \nonumber \\
ds^2_T &=& k (d\theta^2 + \mbox{cotan}^2 \, \theta \ d\psi_T^2) \qquad e^{\Phi} =
\frac{e^{\Phi_0}}{\sin \theta}.
\end{eqnarray}
with $\psi_* \in [0,2\pi)$ and $\theta \in [0,\pi/2]$.
We note that the bell and the T-bell allow for an identical
geometrical interpretation, i.e. the $SU(2)/U(1)$ coset is geometrically
self-dual.
We wish to study the D-brane Born-Infeld action in these backgrounds:
\begin{eqnarray}
S_\textsc{dbi} &=& \tau_p \int d \xi^{\alpha} e^{- \Phi}
\sqrt{-det(g+B+2\pi  F)} + S_{WZ},
\end{eqnarray}
where $g$ is the induced word-volume metric, $B$ is the induced NS-NS
two-form,
and $F$ is the world-volume gauge field.

We will always study static branes in the following and we choose
static gauge for the time-coordinate: $\xi^0=x^0$ (i.e. worldsheet
time coincides, parametrically, with space-time time).
We suppose that the time-direction which is external
to the cosets is flat. To parameterize
the worldsheet actions, another coordinate system is sometimes convenient,
namely:
\begin{eqnarray}
u &=& \sinh r \ \text{(cigar)}, \qquad u=\cosh r \ \text{(trumpet)}, \qquad z = \sin \theta \ \text{(bell)},
\qquad z = \cos \theta \ \text{(T-bell)} \nonumber \\
ds^2_{c} &=& k \frac{du^2+u^2 d \psi^2}{1+u^2} \qquad e^{\Phi}
= e^{\Phi_0} (1+u^2)^{-1/2} \nonumber \\
ds^2_{t} &=& k \frac{du^2+u^2 d \psi^2}{u^2-1} \qquad e^{\Phi}
= e^{\Phi_0} (u^2-1)^{-1/2} \nonumber \\
ds^2_{b,T} &=& k \frac{dz^2+z^2 d \psi^2}{1-z^2} \qquad e^{\Phi}
= e^{\Phi_0} (1-z^2)^{-1/2}.
\end{eqnarray}
Thus we see that the cigar is conformal to the plane, the bell and the T-bell (which are the same since the
background is self-dual) are conformal to the unit disc and the trumpet is conformal to the complement of the
unit disc. 
\subsection*{Static D0-branes}
Static D0-branes in the cigar have an action proportional to:
\begin{eqnarray}
S_c & \propto & \int dt \cosh r
\end{eqnarray}
in other words, the D0-branes are only stable at the tip of the cigar.
Similarly it can be found that the
static D0-brane needs to live at the
$r \rightarrow 0$ singularity of  the trumpet, at the
boundary $\theta \to \pi/2$ of the bell, and also at the boundary $\theta \to 0$ of the T-bell.
\subsection*{Static D1-branes}
For static D1-branes we use the alternative coordinate system,
and we find the actions:
\begin{eqnarray}
S_{c,t} & \propto & \int  d \xi^1 \sqrt{u_{\xi^1}^2 + u^2
\psi^2_{\xi^1}} \nonumber \\
S_{b,T} & \propto &  \int d \xi^1 \sqrt{z_{\xi^1}^2 + z^2
\psi^2_{\xi^1}}
\label{DBId1}
\end{eqnarray}
for branes in the non-compact or compact coset respectively.
(We denoted with an lower index $\xi_1$ the
derivation with respect to $\xi_1$.)
Thus, the static D1-branes see a flat metric in these
coordinates.
They are straight lines in the  plane with polar coordinates $(u,\psi)$ or $(z,\psi)$.
In the original variables, these straight lines (or line
segments, or unions of half-lines) are parameterized as:
\begin{eqnarray}
\sinh r \sin(\psi_c-\psi_c^0) &=& c_c \nonumber \\
 \cosh r \sin(\psi_t-\psi_t^0) &=& c_t \nonumber \\
\sin \theta \sin(\psi_b-\psi_b^0) &=& c_b \nonumber \\
\cos \theta \sin(\psi_T-\psi_T^0) &=& c_T.
\label{d1profiles}
\end{eqnarray}
The geometrical interpretation of the D1-brane solutions in all instances
is clear. On the one hand, they try to follow a geodesic, to minimize
their tensional energy, but on the other hand, they prefer to pass
through a regime with strong coupling (i.e. large dilaton) to reduce
the tension itself. In the $u,z$ coordinates, the two tendencies
of the D1-branes are neatly encoded: the delicate balance is such
that the D1-branes follow straight lines in these auxiliary planes.
Classically,
we can fix their trajectories by specifying (e.g.) their direction at
infinity, and their point of nearest approach to the center of the
plane.

The seemingly flat behavior of these D1-branes has been exploited
in the Lorentzian context of black holes and cosmological singularities
in \cite{Yogendran:2004dm}\cite{Toumbas:2004fe}. Indeed, those
applications clearly illustrate that when the time-direction is
curved, we need to revisit our intuitive picture for D-brane
dynamics. We continue with a flat time direction in the
following.

\subsection*{Static D2-branes}
A usual route for the D2-branes would involve fixing the gauge
such that space-time and worldvolume coordinates entirely coincide.
We choose not to do so, because that gauge choice makes T-duality between
D1-branes and D2-branes less manifest, and we wish to prepare ourselves
for more involved T-dualities to be performed
later. To that end, we choose our gauge
as follows: $ \xi^0 = x^0$ and $\xi^2 = \psi_{c,t,b,T}$, while
we leave $\xi^1$ unfixed.
After a brief computation, we find the
actions:
\begin{eqnarray}
S_{c,t} & \propto & \int d\psi d \xi^1 \ \sqrt{u_{\xi^1}^2 + u^2
(\frac{2\pi}{k} F_{\psi \xi^1})^2} \nonumber \\
S_{b,T} & \propto &  \int d\psi d \xi^1 \ \sqrt{z_{\xi^1}^2 + z^2
(\frac{2\pi}{k} F_{\psi \xi^1})^2}.
\label{DBId2}
\end{eqnarray}
Since we wish to relate this D2 branes with D1 branes of the
T-dual geometries, we have used $u,z$ in their T-dual form, i.e.,
in the cigar $u=\cosh r_c$, in the bell $z=\cos \theta$, etc.
We thus see that the actions (\ref{DBId2}) go under T-duality along $\psi$ to the D1 actions (\ref{DBId1}) of the T-dual
geometries.\footnote{We recommend \cite{Simon:1998az} for a
clear and generic discussion of this worldvolume T-duality
 technique.}

\subsection*{The D2 on the cigar and its dual}
Let us discuss some of the solutions of the actions in a little more detail
in particular instances.
The solution to the equations of motion for the cigar
can be written as follows:
\begin{equation}
A = \frac{k}{2\pi} \ \left[ A_0 - \arcsin \left(\frac{c}{\cosh r} \right)\right] d\psi,
\label{potentialD2}
\end{equation}
which allows us to identify the gauge field $A_{\psi}$
 on the D2-brane in the
cigar with the $\psi_t$ coordinate of the D1-brane in the trumpet.
Both $c$ and $A_0$ are integration constants.
We will discuss first the solution with $c<1$.
The magnetic field strength on the D2-brane is:
\begin{eqnarray}
F &=& \frac{k}{2\pi} \left\{ [A_0 - \arcsin(c)] \delta (r)
 + \frac{c \tanh r}{\sqrt{\cosh^2 r - c^2}} \right\} \ dr \wedge d\psi.
\label{magD2}
\end{eqnarray}
The delta-function contribution can be identified by
using Stokes-theorem for a Wilson loop encircling the tip of the
cigar.\footnote{In \cite{Giveon:2004rw}, for example, the Wilson loop
is used
determine the electro-chemical potential
 of a charged two-dimensional black hole.}
Thus we see that we obtain a delta-function contribution to the origin
from the vortex-form for the gauge field, unless we put
\begin{eqnarray}
A_0 &=& \arcsin(c).
\end{eqnarray}
Note that near the tip of the cigar, the space-time approximates flat space.
In flat space, D0-brane flux spreads on a D2-brane worldvolume,
to form a bound state. Thus, the concentration of D0-brane charge near the
tip is energetically disfavored. To find the minimal energy solution
in a given super-selection sector (set by the total magnetic charge) we
thus fix
the Wilson line such that the magnetic field is non-singular at the
origin $A_0 =  \arcsin(c)$.
The integral of the magnetic field on the D2-brane
wrapping the cigar is then:
\begin{eqnarray}
\frac{1}{2 \pi} \int_{cigar} F &=&  \frac{k}{2\pi} \arcsin(c).
\end{eqnarray}
Because this flux is responsible for a D0-brane charge, it is likely
to be quantized. Indeed a quantization of the parameter
$\sigma = \arcsin (c)$ will be needed to interpret the result in the
context of the five-branes background.\footnote{This is a stronger
statement than the relative quantization discussed
in~\cite{Ribault:2003ss}.}
The D0-brane magnetic flux tends to spread
near the origin, as in flat space, but at larger distances the flux
(as an isolated D0-brane does) tends towards stronger coupling. This
leads to a local maximum for the magnetic field at a particular
radius ($\sinh^2 r = \sqrt{1-c^2}$).\footnote{This type of behavior
should have its analogue in the holographic set-up for
$N=1$ gauge theories in \cite{Klebanov:2004ya} where D3-branes in
six-dimensional non-critical superstring theories are accompanied
by D5-anti-D5 pairs.}
We can discuss the dynamics that leads to these particular static
solutions in some more detail.
Note that D2-branes covering the whole cigar (which have $|c|<1$)
are T-dual to D1-branes
in the trumpet background which bump into the disc cut out from the
$(u,\psi_t)$ plane (because of the identification $u=\cosh r$), in
other words, they are dual to the D1-branes that reach the open, strongly coupled
end of the trumpet.

Finally, note that the topologically trivial Wilson line in the
D2-brane reaching the tip of the cigar turns out not to be a zero-mode.
By T-duality, the angular position of the D1-brane
in the trumpet model is not a true zero-mode. This is familiar. Indeed,
we know that the naive isometry in the trumpet is broken in the quantum
theory, and that similarly, the winding number is not a fully conserved
quantum number in the cigar. Thus, indeed, the fact that the angular
variable of the D1-brane in the trumpet is not a true zero-mode is consistent
with the known symmetry-breaking patterns of the coset models. 

To further discuss the T-duality, let us start
with a D2-brane of the cigar with $c<1$, with a gauge field given by 
eq.~(\ref{potentialD2}). To avoid the vortex-like singularity at the origin, 
we choose $A_0 = \arcsin (c) = \sigma$. 
By T-duality
of the cigar, we obtain the following $\zi_k$ orbifold of the trumpet:
\begin{equation}
ds^2 = k \left[ dr^2 + \text{cotanh}^2 \, r \left( \frac{d\tilde{\psi}}{k}\right)^2 \right]
\end{equation}
with the identification $\tilde \psi \sim \tilde \psi + 2 \pi$. We have in this background a 
D1-brane of embedding equation $\sin \sigma = \cosh r \sin (\tilde{\psi}/k-\sigma)$. 
To recover the standard trumpet CFT --~i.e. the vector coset of the single cover of \slr-- one can 
go to the covering space of the orbifold by defining $\phi =\tilde{\psi}/k$ with the 
periodicity $\phi \sim \phi + 2 \pi$. This gives an extra freedom for the D1-brane, corresponding 
to the copy of the orbifold manifold we start with on the covering space.

To summarize we have argued
 that for the D1-branes of the trumpet with $c<1$
the angular position has to be quantized as $\psi^{0}_t =
 \arcsin(c)+\frac{2\pi \hat{p}}{k}$, with $\hat{p} \in \zi_k$, to 
be consistent with the T-dual picture. Since the D2-branes of the cigar with $c>1$ do not
reach the singularity, the parameter $\psi^{0}_t$ of the corresponding D1-branes in the trumpet
would not be
quantized. By the same reasoning we
find that the parameter $\psi^{0}_c$ for the D1-branes
of the cigar is not quantized, and that for the D1-branes of the bell the parameter
$\psi^{0}_b$ is quantized as $\psi^{0}_b=\theta_0 + \frac{2\pi \hat{p}}{k}$. The angles corresponding
to the two endpoints of the D1-brane
on the boundary of the disc are then
\begin{equation}
(\psi_1,\psi_2) =
\left( \frac{2\pi \hat{p}}{k}+2\theta_0,\frac{2\pi \hat{p}}{k}+\pi \right)
\label{posD1bell}
\end{equation}
We will see
later that these heuristic
rules will get a natural interpretation
 in the CFT of the ring of fivebranes.

Needless to say, the geometrical picture for the trumpet is known to
be corrected drastically
by worldsheet instantons and the physics is more
truthfully encoded in the sine-Liouville model -- we have just used the
trumpet to gain geometrical intuition on the D-brane solutions discussed
above. However by requiring the consistency of the D-branes profiles
with the T-duality we have gained some insight on the quantization
of the parameters of the branes, which would properly require to
be analyzed using the non-perturbative corrections to the sigma-models.
Now, we'll turn to applying these techniques in  more complicated examples.

\subsection{D-branes in the NS5-branes background}
Now we shall use the T-dual representations
of the background of NS5-branes
on a circle, i.e. the (orbifold of) the product
of coset models, eqs.~(\ref{cosetprod},\ref{tphidual}) in order to
construct a number of non-trivial D-branes in the original NS5-brane
background.

In this paper we are mainly interested by stable, BPS D-branes preserving
a fraction of supersymmetry. As will become clear in the CFT analysis,
we have then to choose, say in the $T_\psi$-dual background, the same A- or B-type boundary conditions for both axial
cosets, \slc (the cigar) and $SU(2)/U(1)$ (the T-bell).\footnote{The D-branes
constructed with different boundary conditions for the two cosets will
be symmetry-breaking branes with respect to
 the $\mathcal{N}=4$ superconformal
algebra on the worldsheet.} Under this condition the spectrum 
of open strings ending on one of these D-branes will be
supersymmetric, indicating their BPS nature. 
The supersymmetry of these D-branes
could also be checked at the level of the DBI action, using e.g. the 
techniques of~\cite{Ribault:2003sg}. 

\subsubsection*{Suspended D1-branes}
\label{d1}
These are D1-branes ending on both 
sides on a NS5-brane in type IIB superstrings.
They are of special interest, because they correspond
to the ``W-bosons'' of the Little String Theory, namely the D1-branes
stretched between the NS5-branes corresponding to the broken
gauge symmetry of the higgsed configuration, that remain massive in the double
scaling limit. The D1-branes
can be constructed from a D0-brane in the cigar and a D2-brane
in the T-bell:
\begin{equation}
r = 0 \ , \ \ F = \frac{k}{2\pi} \ \frac{\sin \theta_0 \, \text{cotan}\, \theta}{
\sqrt{\sin^2 \theta -
\sin^2 \theta_0}} \ d \theta \wedge d \chi.
\end{equation}
Since the D0-brane of the cigar sits at the tip (and the coordinate
$\chi$ plays the role of the angular coordinate on the T-bell at
$r=0$), by T-duality
we will simply get a D1-brane of the bell embedded in the
four-dimensional geometry of the transverse space:
\begin{equation}
\sin \theta \sin (\psi - \psi_0) = \sin \theta_0
\end{equation}
These D1-branes are straight lines in the $(x^6,x^7)$ plane, using the
coordinate transformations~(\ref{cartcoords}), ending on the ring of 
fivebranes. The parameter $\psi_0$ becomes quantized once we take into account
the T-duality considerations discussed in the previous section
or the exact CFT description. The parameter $\theta_0$ is also
quantized in the exact CFT, and it can be understood by advocating
the flux stabilization of the D2-branes in $SU(2)$~\cite{Bachas:2000ik}. With 
these quantization rules the D1-branes are then stretched 
between two NS5-branes out of the
$k$ NS5-branes that make up the background -- the configuration
is discussed in more detail below.

\subsubsection*{A first D3-brane from T$_{\psi}$ duality}
We are now ready to study a  D3-brane in the NS5-brane
background. We start in the T$_{\psi}$ dual background~(\ref{cosetprod}) and consider the product
of a D1-brane in the cigar times a D1-brane in the T-bell.
The equations for the branes are:
\begin{eqnarray}
\sinh r \sin \left( \frac{\chi}{k}-\psi_0\right) &=& \sinh r_0 \,,\nonumber \\
\cos \theta \sin \left( \frac{\chi}{k} - \phi+\phi_0 \right) &=& \cos \theta_0 \,.
\label{2d1}
\end{eqnarray}
To obtain a D3 brane in the NS5 background, we perform now a T duality along
$\chi$. We first solve for the coordinate $\chi$ which
by T-duality goes to a dual gauge field as $\chi \rightarrow 2\pi A_{\psi}$.
Secondly, we eliminate the coordinate $\chi$ on which we perform
T-duality from the equations~(\ref{2d1}) for the profile of the D3 brane. We obtain:\footnote{To be precise 
there are two branches of the solution.}
\begin{eqnarray}
\arcsin \left( \frac{\sinh r_0}{\sinh r} \right) &=& \arcsin
\left( \frac{\cos \theta_0}{\cos \theta} \right)
+ \phi-\phi_0 -\psi_0
\label{firstd3emb}\\
A  &=& \frac{k}{2\pi} \ \left[   \arcsin\left( \frac{\sinh r_0}{\sinh r}\right)+  \psi_0  \right] d\psi \,,
\label{firstd3a}\\
&=& \frac{k}{2\pi} \ \left[  \arcsin\left( \frac{\cos \theta_0}{\cos \theta}\right) + \phi -\phi_0 \right] d\psi\,.
\nonumber
\end{eqnarray}

The profile of the T-dual D3-brane is thus highly non-trivial and
non-factorized.
The two expressions (\ref{firstd3a}) for the gauge field are related by the embedding equation
(\ref{firstd3emb}).

The D3 brane worldvolume wraps the $\psi$ coordinate in (\ref{NS5geom}),
times a two-dimensional manifold defined by the constraint (\ref{firstd3emb}) in the
$(r,\theta,\phi)$ coordinates. This two-dimensional manifold
can be interpreted as straight lines in the
$(\sinh r,\phi)$ plane, which are shifted and tilted as a function
of $\theta$:
\begin{eqnarray}
\sinh r_0 \cos \theta &=& \sinh r \left[ \cos \theta_0 \cos (\phi
- \phi_0 -\psi_0) + \sqrt{\cos^2 \theta - \cos^2 \theta_0}
\sin (\phi - \phi_0 -\psi_0) \right],\nonumber \\
\label{firstd3profile}
\end{eqnarray}
and the gauge field gives the following magnetic field:
\begin{eqnarray}
F &=&  \frac{k}{2\pi} \, \frac{\text{cotanh}\, r}{\sqrt{\frac{\sinh^2 r}{\sinh^2 r_0} -1}} \
\ d \psi \wedge dr \,, \\
&=& \frac{k}{2\pi} \left[ d\phi \wedge d\psi + \frac{\tan \theta}{\sqrt{\frac{\cos^2 \theta}{
\cos^2 \theta_0}-1}} \ d\theta \wedge d\psi \right].
\label{D3firstf2}
\end{eqnarray}
Note that while the parameter $\phi_0$ is quantized (see above), the parameter $\psi_0$ is not;
so we can absorb the former by shifting the latter. 
The gauge for the B-field that is consistent with this expression for the magnetic field is implicitly 
fixed by the particular T-duality we have chosen to construct the D-branes.
The worldvolume of
this D-brane is restricted to:
\begin{equation}
r > r_0 \ , \ \ \theta < \theta_0.
\end{equation}
In view of the non-trivial nature of the solution, it is instructive to
directly check that it is a solution, using the D3-brane Dirac-Born-Infeld action in the
NS5-brane background. That is a non-trivial exercise which we performed
following the general ideas in \cite{Simon:1998az}.
In this way it is clearly possible to construct highly non-trivial
D3-brane configurations in the NS5-brane geometry.

\subsection*{Limit profiles}
To understand in more detail the profile of this D3-branes we can
go to some particular limits of the bulk coordinates or the D-brane
parameters:
\begin{itemize}
\item in the limit $\theta \to 0$ we focus on the two-plane
defined by $x^6=x^7=0$ (see e.g.~\ref{cartcoords}).
As we saw in section~\ref{bulk} the transverse space metric degenerates
to the cigar corresponding to coordinates $(r,\phi)$. Then the equation defining
the D3-brane becomes simply
$\sinh r_0 = \sinh r \cos (\phi - \phi_0 - \psi_0-\theta_0)$, i.e.
the D1-brane of the cigar. Going back to the Cartesian coordinates for
the transverse
space, eq~(\ref{cartcoords}), they are straight lines
defined by:\footnote{For simplicity we take $\phi_0+\psi_0+ \theta_0 = 0$.}
$x^6 = x^7 = 0$ and $x^8 =  \rho_0 \sinh r_0 $. Therefore
$r_0$ corresponds to the distance of closest approach of these D-branes with
respect to
the NS5-branes.
\item In the limit $r\to \infty$, the metric approaches the
solution for coincident NS5-branes, i.e. $\mathbb{R}_Q \times SU(2)_k$. 
It is an intermediate regime which is still near horizon but far enough 
to approximate the configuration with coincident fivebranes. 
The equation for the D-brane becomes $\cos \theta_0 = \pm \cos \theta
\sin (\phi - \phi_0-\psi_0)$, i.e. a symmetric $S^2$ D2-brane of $SU(2)$.
Strictly speaking we obtain two antipodal D2-branes corresponding to
the two branches of the solution.
\item for the D3-brane with parameter $r_0=0$, we can take the
$r\to 0$ limit to focus on the plane $x^8=x^9=0$ where the NS5-branes
live. The metric then degenerates to the bell $SU(2)/U(1)$, the
bell coordinates being $(\theta, \psi)$. The $k$ special points
on the boundary of the bell correspond to the position of the five-branes.
As the $\phi$ coordinate degenerates, the equation becomes simply
$\theta < \theta_0$, i.e. a D2-brane of the bell, carrying a gauge field
given by the second term of~(\ref{D3firstf2})
\end{itemize}
Note that the parameter $\phi_0$ is not quantized, which is consistent
with the geometrical fact that the $SO(2)$ rotational isometry
in the $(x^{8},x^{9})$ plane is not broken by the distribution of
fivebranes.
The D2-branes
of the coset $SU(2)/U(1)$ can be considered as bound states
of D0-branes sitting at the center of the bell~\cite{Maldacena:2001ky}.
Therefore
we can reinterpret this class of D3-branes as a bound state
of D1-branes transverse to the NS5-branes, defined
by the equations (for $\phi_0+\psi_0+ \theta_0 = 0$):
\begin{equation}
x^6=x^7=x^9=0 \ , \ \ x^8 = \rho_0 \sinh r_0.
\end{equation}

\subsubsection*{A second  class of D3-branes from T$_{\phi}$ duality}
\label{semiclassD3two}
A second, closely related D3-brane can be constructed by starting from D1 branes
 in the T$_{\phi}$ dual geometry (\ref{tphidual}). This exchanges
the roles of the \slr$/U(1)$ coset and the
$SU(2)/U(1)$ coset by their T-duals.

Explicitly we start from the following D1-branes
of the trumpet (the vector \slr$/U(1)$ coset) and the bell (i.e. the
vector $SU(2)/U(1)$ coset):\footnote{Note that this D3-brane can be obtained as well by starting with a
D2-brane of the cigar and a D2-brane of the T-bell in the first T-dual geometry of eq.~(\ref{cosetprod}).}
\begin{eqnarray}
\cosh r \, \sin \left(\frac{\omega}{k}+\psi-\psi_0 \right) &=&  c \nonumber\\
\sin \theta \sin \left(\frac{\omega}{k} -\phi_0 \right) &=& \sin \theta_0
\end{eqnarray}
It will be useful to distinguish once again the D1-branes
with $c>1$ which do not reach the $r=0$ end of
the trumpet -- we will name them "uncut" --, and the
D1-branes with $c<1$ which reach the end of
the trumpet -- "cut". For each case, we will use the parameterization:
\begin{equation}
c=\cosh r_0 \ , \ \ c\geqslant 1 \quad \text{and} \quad c=\sin \sigma \ , \ \ c <1
\end{equation}
We revisit this distinguishing characteristic in the exact construction of the boundary states.

Going through the same steps
of the T-duality (this time along $\phi$),
we find the following non-trivial D3-branes:\footnote{We have again two branches for the solution.}
\begin{eqnarray}
c \, \sin \theta &=& \cosh r \left[ \sin \theta_0 \cos (\psi
+\phi_0- \psi_0 ) + \sqrt{\sin^2 \theta - \sin^2 \theta_0}
\sin (\psi +\phi_0- \psi_0 ) \right],\nonumber\\
F &=& \frac{k}{2\pi}\ \frac{\text{cotan}\, \theta}{
\sqrt{\frac{\sin^2 \theta}{\sin^2 \theta_0} -1}} \
d \phi \wedge d \theta \nonumber \\ &=&  \frac{k}{2\pi} \left[ d \phi \wedge d\psi  + \frac{\tanh r}{\sqrt{\frac{\cosh^2 r}{
c^2}-1}} \ d \phi \wedge dr \right]
\label{profilehwD3}
\end{eqnarray}
We have a non-trivial magnetic field on the D3-brane. The worldvolume of
this D-brane is restricted to:
\begin{eqnarray*}
\text{Uncut:}&&\quad
r > r_0 \ , \ \ \theta > \theta_0 \\
\label{class2w}
\text{Cut :}&&\quad
\theta > \theta_0
\label{class3w}
\end{eqnarray*}
We can as in the previous case consider various limits of the solution.
\paragraph{The uncut branes}
\begin{itemize}
\item in the limit $\theta \to \pi/2$ we focus on the two-plane
defined by $x^8=x^9=0$, and in the region outside the ring
of five-branes, i.e. $x^6,x^7 \geq \rho_0$. Then (see sect.~\ref{bulk}) the metric
degenerates to the trumpet, with coordinates $(r,\psi)$.
We then get from this class of D3-branes "uncut" D1-branes  of
the trumpet of equation $\cosh r_0 = \cosh r \cos (\psi+\phi_0-\psi_0-\theta_0)$.
\item In the $r\to \infty$ limit we obtain a D2-brane of $SU(2)$ as for the other kind of D3-branes,
but with a different position.
\end{itemize}
\paragraph{The cut branes}
In this case first, 
as we argued in our discussion of the
D-branes in the coset models, 
the parameter $\psi_0$ will be quantized as $\psi_0 = \sigma+2\pi \hat{p}/k$.\footnote{This will be
confirmed by the CFT analysis.}
Then the picture is modified as follows.
\begin{itemize}
\item
In the $\theta_0=0$ case, in $\theta \to 0$ limit,
we get the smooth D2 of the cigar with gauge field strength
$2\pi F_{r\phi} =  k \sin \sigma \tanh r / \sqrt{\cosh^2 r - \sin^2 \sigma}$
getting its maximal value at some  finite distance from the tip.
\item In the $\theta \to \pi/2$ limit we get an cut D1-brane of
the trumpet, of embedding equation:
$\sin \sigma = \cosh r \sin (\psi+\phi_0-\psi_0+\theta_0)$, with all the parameters
quantized. To be precise we get a second copy of this D1-brane rotated by $\pi-2\theta_0$.
\item in the $r\to \infty$ limit, we get again a D2-brane of $SU(2)$.
\item For all these cut branes we can take the $r \to 0$ limit;
then we go to the plane defined by $x^8=x^9=0$, where the metric of the five-branes degenerates to the bell
(of coordinates $ds^2 = d\theta^2 + \tan^2 \theta d\psi^2$).
and we get a D1-brane of the bell (more precisely, again two D1-branes), of embedding equation:
$\sin  \theta_0 = \sin \theta  \sin (\psi_0+\sigma - \phi_0 - \psi ) $.
This suggests strongly that the parameter $\sigma$ has to be quantized as well.
\end{itemize}
\subsubsection*{Semi-infinite D1-branes}
The previous class of D3-branes contains as a special case "D-rays",
which are semi-infinite D1-branes coming in from infinity.
They correspond to D3-branes with parameter $\theta_0=\pi/2$. Their worldvolume is
thus restricted to $\theta=\pi/2$, corresponding to the plane $x^{8}=x^{9}=0$,
but {\it outside} the ring of NS5-branes; then the metric degenerates to the trumpet.
These D1-branes are given by the embedding equation:
\begin{equation}
 \cosh r   \cos (\psi+\phi_0 - \psi_0 ) = c
\end{equation}
They consist in straight lines in the plane $(x^6,x^7)$,
in the domain $\sqrt{(x^{6})^2 + (x^{7})^2} > \rho_0$ outside the ring. If for example
we choose $\phi_0-\psi_0 = 0$, the equation for the D1-brane is $x^6 = c/ \rho_0$,
with the condition $\sqrt{c^2 + (\nicefrac{x^{7}}{\rho_0})^2} > 1$.

Let us consider first the case $c<1$, i.e. the "cut" D-branes. These D-branes are made
of two semi-infinite D1-branes ending on the ring. 
The parameter $\psi_0$ is quantized for the "cut" branes as 
$\psi_0 = \sigma+2\pi \hat{p}/k$. Then, if we insist that these D1-branes end on NS5-branes, we have to choose only
D-branes with $\sigma \in \pi \zi /k$. This will be confirmed by the CFT analysis.
We have obtained a semi-infinite D1-brane that ends on the ring precisely on a specific
NS5-brane. The physics of D4-branes T-dual of these D1-branes along 
$x^{1,2,3}$ has been discussed in~\cite{Elitzur:2000pq}. 

On the contrary the "uncut" D-branes (i.e. with $c>1$) will give in the NS5-brane background
infinite D1-branes avoiding the ring of five-branes entirely.
For these infinite D1-branes avoiding the ring of NS5-branes there is no reason for a quantization
of $\psi_0$.  Note finally that this picture will be a little bit modified in the exact CFT analysis,
since strictly speaking there is no exact D-brane with $\theta_0 = \pi/2$. Thus our picture is valid
in the semi-classical large $k$ limit, but at finite $k$ the picture is fuzzy, 
since
these D1-branes have some extension in their transverse directions.

\FIGURE{\centering
\epsfig{figure=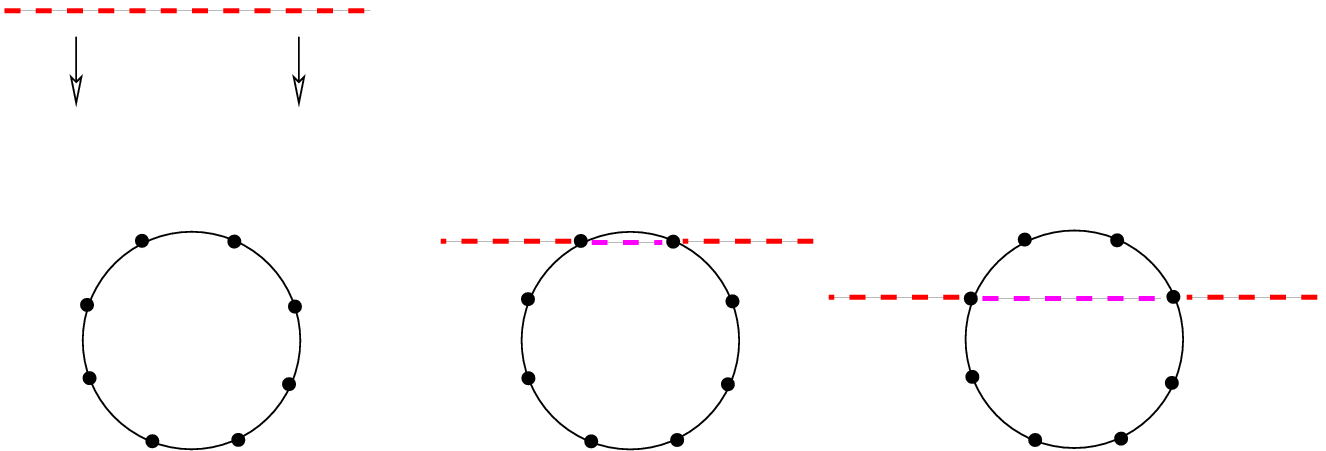, width=140mm}
\caption{Geometry of a D1-brane for different values of $c$. For $c>1$ (left) the infinite
D1-brane avoids the ring of fivebranes and corresponds
to an "uncut" D1 of the trumpet. For $c<1$ (right) the D1-brane intersects the ring; its two
semi-infinite halves correspond to a "cut" D1 of the trumpet and the finite part inside the
ring to a D1-brane of the bell.
}
\label{d1break}}

In the course of our analysis we observe
a remarkable phenomenon, see fig.~\ref{d1break}. The interior and
the exterior of the ring of fivebranes (both in the plane $x^8=x^9=0$) are both described by exact coset CFTs,
respectively the bell $SU(2)/U(1)$ and the trumpet (the vector coset \slc). We can bring
D1-branes from infinity (i.e. for $c>1$) to the ring of fivebranes;
when the D1-branes intersects the ring of fivebranes they break into
two parts: a D1-brane of the trumpet corresponding to the two semi-infinite halves
of the D-string outside the ring --~and ending on NS5branes~--
 and a D1-brane of the bell corresponding to
the finite D1-branes suspended between the NS5-branes. Thus we have an exact
CFT realization of the configurations discussed in~\cite{Hanany:1996ie},
in which a D-brane can split into two halves when it crosses a NS5-brane.

In this section, we used the geometrical picture for branes
in coset models to construct non-trivial branes in the NS5-brane background. The T-duality
involved in developing the correct geometrical picture was often
intricate. We have considered all the combinations of coset branes in the T-dual picture
that would lead to BPS D-branes in the background of fivebranes.
It would be worthwhile to examine the set of
semi-classical brane profiles that cannot be obtained using the techniques explained
in this section, i.e. that cannot be expressed in the T-dual geometry as product
of D-branes of \slc and $SU(2)/U(1)$.
We now turn to constructing the exact boundary
states corresponding to the branes we discussed. 

%%%%%%%%%%%%%%%%%%%%%%%%%%%%%%%%%%%%%%%%%%%%%%%%%%%%%%%%%%%%%%%%%%%%%%%%%%%%%%%%%%%%%%
%%%%%%%%%%%%%%%%%%%%%%%%%%%%%%CFT
%%%%%%%%%%%%%%%%%%%%%%%%%%%%%%%%%%%%%%%%%%%%%
%%%%%%%%%%%%%%%%%%%%%%%%%%%%%%%%%%%%%%%%%%%%%%%%%%%%%%%%%%%%%%%%%%%%%%%%%%%%%%%%%%%%%

\section{Exact construction of branes}
\label{exact}
To construct branes consistent with the bulk theory, it is useful to recall
the spectrum of closed strings in the bulk  -- the cylinder amplitude in the
doubly scaled little string theory background will need to be consistent
with the bulk spectrum obtained from the torus partition function. For simplicity,
we mostly work with branes in type IIB theory. D-branes of type IIA are obtained 
by an odd number of T-dualities along the flat directions $x^{1,2,3,4,5}$.
\subsection*{Bulk spectrum}
The most non-trivial feature in the
spacetime-supersymmetric spectrum of the type II doubly scaled little string
theory is the spectrum of the $SL(2,\mathbb{R})/U(1)$ superconformal field theory
 \cite{Hanany:2002ev}\cite{Eguchi:2004yi}\cite{Israel:2004ir}.
In the full NS5-brane background it leads to the identification of the \emph{discrete spectrum}, built with the
discrete representations of $SL(2,\mathbb{R})_k$ of spin $j$,
together with representations of $SU(2)_k$ of spin $j'$:
\begin{eqnarray}
Z_{d} = \frac{1}{2} \sum_{a,b \in \zi_2} (-)^{a+b} \
\frac{1}{2} \sum_{\bar a,\bar b \in \zi_2} (-)^{\bar a+\bar
b+\varepsilon \bar{a} \bar{b}}
\frac{\vartheta^2 \oao{a}{b} \bar{\vartheta}^2 \oao{\bar a}{\bar b}}
{(8\pi^2 \tau_2)^2 \eta^4 \bar{\eta}^4}
\ \sum_{2j'=0}^{k-2} \sum_{2j=1}^{k+1} \Upsilon (2j,1)
\sum_{m,\bar m \in \zi_{2k}} \sum_{w_L , w_R \in \zi} \nonumber\\
\mathcal{C}^{j'}_{m}\oao{a}{b} \bar{\mathcal{C}}^{j'}_{-\bar m}
\oao{\bar a}{\bar b}
ch_d (j,\frac{m}{2}+kw_L -j-\frac{a}{2}) \oao{a}{b}
ch_d (j,-\frac{\bar{m}}{2}+kw_R -j-\frac{\bar a}{2}) \oao{\bar a}{\bar b} \nonumber \\
\end{eqnarray}
where $\varepsilon = 0$ in type IIB superstrings and $\varepsilon=1$ in
type IIA. We have also defined $\Upsilon (2j,1)=\nicefrac{1}{2}$ if
$2j=1 \mod k$, $\Upsilon (2j,1)=1$ otherwise.

This expression involves the characters $\mathcal{C}^{j}_{m}
\oao{a}{b}$ of the N=2 minimal models
and the discrete characters $ch_d (j,r) \oao{a}{b}$ of $SL(2,\mathbb{R})/U(1)$
(see appendix~\ref{modtrans}), associated to the non-flat coset
factors $SU(2)/U(1)$ and $SL(2,\mathbb{R})/U(1)$ respectively. The sum
over left and right winding separately is necessary to allow for
the twisted sectors of the orbifolded cosets. Note that this partition
function is obtained by performing first a $\zi_k$ diagonal
orbifold of the product of the two cosets $SU(2)_k/U(1)$ and $SL(2,\mathbb{R})_k/U(1)$,
and then a specific $\zi_2$ projection to obtain odd-integral
$N=2$ charges, see~\cite{Eguchi:2004yi}~\cite{Israel:2004ir} for
details.\footnote{This second
step is responsible for the appearance of only even
left and right spectral flows $w_{L,R}$.}

Then we have a \emph{continuous spectrum} constructed with the continuous
representations of $SL(2,\mathbb{R})$, of spin $j=1/2 + i P$:
\begin{eqnarray}
Z_{c} = \frac{1}{2} \sum_{a,b \in \zi_2} (-)^{a+b}
\frac{1}{2} \sum_{\bar a,\bar b \in \zi_2} (-)^{\bar a+\bar
b+\varepsilon \bar{a} \bar{b}}
\frac{\vartheta^2 \oao{a}{b} \bar{\vartheta}^2 \oao{\bar a}{\bar b}}{
(8\pi^2 \tau_2)^2 \eta^4 \bar{\eta}^4}
\ \sum_{2j'=0}^{k-2} \ \sum_{m,\bar m \in \zi_{2k}}
\mathcal{C}^{j'}_{m}\oao{a}{b} \bar{\mathcal{C}}^{j'}_{- \bar m} \oao{\bar a}{\bar b} \nonumber \\
\int_{0}^{\infty} dP \sum_{w_L , w_R \in \zi}
\rho(P,m+2kw_L,-\bar{m}+2kw_R;a,\bar{a}) \phantom{aaaaaaaaaaaaaaaaaaaa}\nonumber \\
ch_c (P,\frac{m}{2}+kw_L)\oao{a}{b}
ch_c (P,-\frac{\bar{m}}{2}+kw_R) \oao{\bar a}{\bar b} \nonumber\\
\label{contspecbulk}
\end{eqnarray}
The $SL(2,\mathbb{R})/U(1)$ part, written with
continuous characters, involves a non-trivial density of states
$\rho$ which consists of a term proportional to the infinite volume and a
regulator dependent sub-leading term which is the derivative
of the phase-shift of the regularizing potential
\cite{Eguchi:2004yi}\cite{Israel:2004ir}.
In both cases we denote the primary operators of the doubly scaled LST
background as, for states in the NS vacuum:
\begin{equation}
V_{j'm\bar{m}\ jw_L w_R,{\bf p}}  = e^{-\phi - \tilde{\phi}} \ e^{i {\bf
p \cdot X}}
\Phi^{SU(2)/U(1)}_{j',\ m,\ -\bar{m}} \widetilde{\Phi}^{SL(2,R)/U(1)}_{j,\
  \frac{m}{2}+k w_L , \ -\frac{\bar{m}}{2} + k w_R}~,
\end{equation}
where the first factor corresponds to the bosonized NS-NS ground state of the
super-reparameterization ghosts. In general the operators carry
fermionic labels. For convenience, we will use in the following
the formalism in which the fermions are bosonized to a compact $U(1)_2$
boson, at level $2$ which carries $\zi_4$ valued chiral momenta,
 as explained in
appendix~\ref{modtrans}.

To orient the reader and fix our conventions, we recall that the
 non-trivial part of the two-point function for two NS-NS primaries is
given by:
\begin{eqnarray*}
\langle V_{j'm\bar{m}\ jw_L w_R,{\bf p}}
V_{\tilde{j}'\tilde{m}\tilde{\bar{m}}\
  \tilde{j}\tilde{w}_L \tilde{w}_R {\bf p'}}  \rangle \sim
\delta^{(6)} ({\bf p} - {\bf p'})
\, \delta_{j',\tilde{j}'} \, \delta_{m,\tilde{m}}
 \, \delta_{\bar{m},\tilde{\bar{m}}}  \, \delta_{w_L,\tilde{w}_L}
\delta_{w_R
  ,\tilde{w}_R} \nonumber \\
\times \ \left[ \delta ( j + \tilde{j}+1 ) +
  R(j,\frac{m}{2}+k w_L,-\frac{\bar{m}}{2}+k w_R ) \delta (j - \tilde{j} )
  \right]\, ,
\end{eqnarray*}
in terms of the reflection coefficient:
\begin{eqnarray}
R \left( j,\frac{m}{2}+k w_L,-\frac{\bar{m}}{2}+k w_R \right) =
\phantom{aaaaaaaaaaaaaaaaaaaaaaaaaaaaaaaaaaaaaaaaaa}
\nonumber \\ \nu_{k}^{\frac{1}{2} -j}
\frac{\Gamma (1-2j)}{\Gamma(2j-1)} \frac{\Gamma (j + \frac{m}{2}+kw_L )
\Gamma  (j+\frac{\bar{m}}{2} -kw_R)}{\Gamma (1-j + \frac{m}{2} + kw_L )
\Gamma  (1-j+\frac{\bar{m}}{2}-kw_R)} \frac{\Gamma (1 + \frac{1-2j}{k}
)}{\Gamma (1  - \frac{1-2j}{k} )} \qquad
\end{eqnarray}
with $\nu_k = \Gamma (1-\nicefrac{1}{k})/\Gamma (1+\nicefrac{1}{k})$.

We recalled a few aspects of the bulk theory, and we now wish to
analyze the introduction of boundaries in the worldsheet conformal
field theory, i.e. we will construct examples of D-branes, and their
corresponding one-point function. In this exact analysis we roughly follow
the scheme we set out in the previous section, where the semi-classical
behavior of these D-branes was analyzed. It will be useful to keep the
semi-classical pictures in mind while analyzing the exact boundary states.

%%%%%%%%%%%%%%%%%%%%%%%%%%%%%%%%%%%%%%%%%%%%%%%%%%%%%%%%%%%%%%%%%%%%%%%%%%%%%%%%
%%%%%%%%%%%%%%%%%%%%%%%%%%%%%%%D1-branes%%%%%%%%%%%%%%%%%%%%%%%%%%%%%%%%%%%%%%%%
%%%%%%%%%%%%%%%%%%%%%%%%%%%%%%%%%%%%%%%%%%%%%%%%%%%%%%%%%%%%%%%%%%%%%%%%%%%%%%%%

\subsection{Boundary states for suspended D1-branes}
\label{exactd1}
Let us now construct the exact D1-branes consistent with the closed
string theory
reviewed above. We are considering first the D1 branes, or W-bosons,
stretched between the NS5 branes of which we analyzed the semi-classical
behavior in subsection \ref{d1}. We take the NS5-branes and the D1-branes
to span the following directions (schematically):\footnote{More precisely 
these D1-branes have one Neumann and one Dirichlet directions in the $(x^{6},x^{7})$ plane 
but for generic parameters they do not coincide with the $x^6$ and $x^7$ directions. 
This will hold for the other D-branes.}
$$
\begin{array}{c|cccccccccc}
& 0 & 1 & 2 & 3 & 4 & 5 & 6 & 7 & 8 & 9 \\
\hline
\text{NS5} & \sslash & \sslash & \sslash & \sslash & \sslash & \sslash &
\perp & \perp & \perp & \perp \\
\text{D1} & \sslash & \perp & \perp & \perp & \perp & \perp & \sslash
& \perp & \perp & \perp \\
\end{array}\\
$$
In that case, as explained is section~\ref{semiclass}, the D1-brane geometries are
factorisable in terms of coset branes and the T-duality acts straightforwardly on them.
The D1-branes consist of a product 
of D1-branes of the N=2 minimal model (i.e. the coset $SU(2)/ U(1)$) and the D0-brane of the cigar. 
We refer to fig.~\ref{Wbosfig}) for a drawing of the geometry.

\FIGURE{
\epsfig{figure=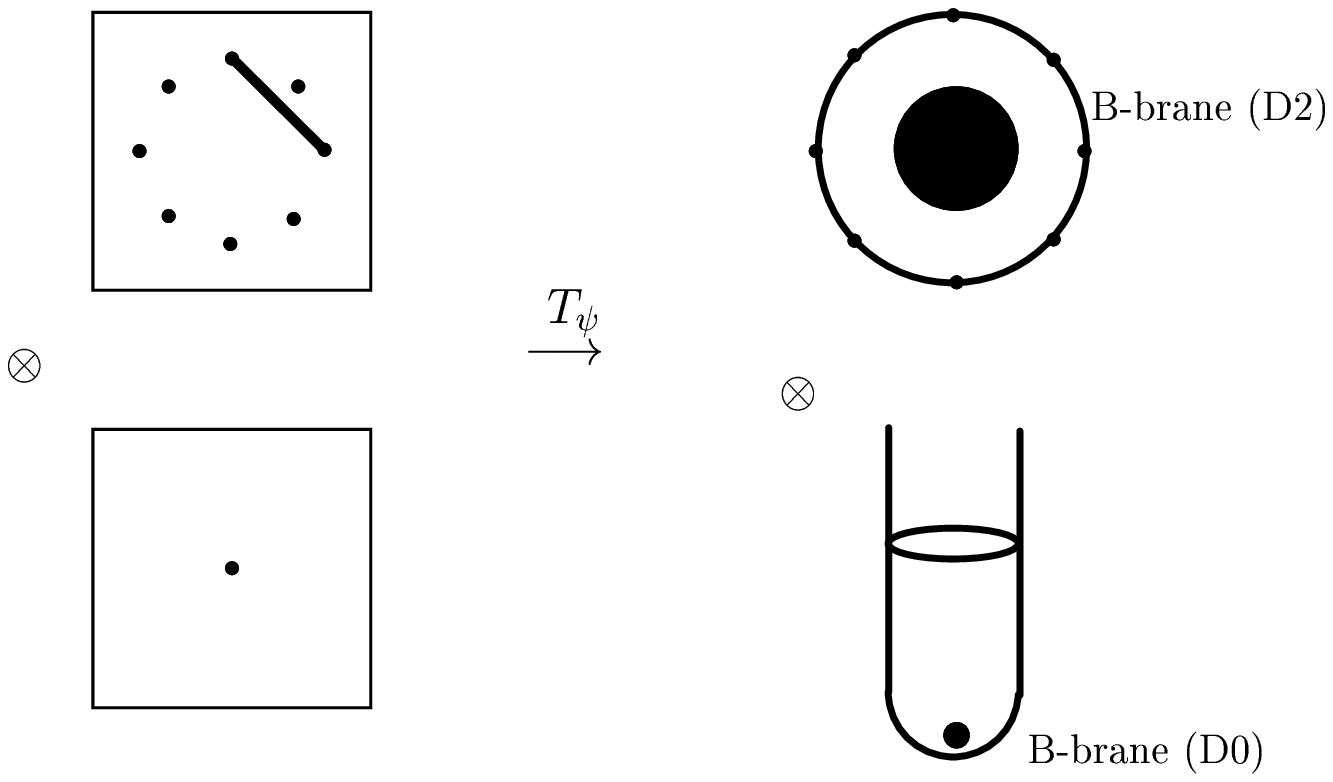, width=120mm}
\caption{Geometry of the D1-branes and their T-duals.}
\label{Wbosfig}}

The D1-branes in the N=2 minimal model  are straight lines connecting the $k$ special points on the boundary
corresponding to the localization of the NS5-branes. The positions will be
labeled by $\hat{n}$  which indicates the middle point
of the brane on the boundary of the disc,
and the Cardy label $\hat{\jmath}$ giving the angle spanned by the whole
D1-brane. Indeed the D1-branes is extended between the angles(see~\cite{Maldacena:2001ky}):
\begin{equation}
\psi_1 = \frac{\pi (\hat{n}-2\hat{\jmath}-1)}{k} \ \ \text{and} \ \
\psi_2 = \frac{\pi (\hat{n}+2\hat{\jmath}+1)}{k}
\end{equation}
Thus (up to a rotation of $\pi$) these labels are related to the labels discussed in sect.~\ref{cosetgeometries} as
follows, see eq.~(\ref{posD1bell}):
\begin{equation}
\frac{\pi}{2}-\theta_0 = \frac{\pi(2\hat{\jmath}+1)}{k} \ \ 
\text{and} \ \ \hat{p} = \frac{\hat{n}+2\hat{\jmath}+\hat{s}}{2}
\label{corrparamD1}
\end{equation}
with $2\hat{\jmath} = 0,1,\ldots,k-2$ and $\hat{n} \in \zi_{2k}$. 
Note that $\hat{p}$ is indeed integer, thanks to the selection rule~(\ref{selruleMM}).
There is an overall rotation of the picture by $\pi /k$
depending on the parity of $\hat{s}$. However, since
D-branes with values of $\hat{s}$ of different parities cannot appear simultaneously
(see appendix~\ref{boundfree}), it doesn't alter the geometrical interpretation of the branes.
These fermionic Cardy labels $\hat{s}_i$ for $i=1,2,3,4$ indicate
the particular boundary conditions or the projections performed on the
worldsheet fermions.

The D0-branes of the cigar (i.e. $SL(2,\mathbb{R})/U(1)$)
are point-like objects sitting at the tip, with only a
fermionic label, which we will take to correspond to $\hat{s}_4$.
They have been obtained in~\cite{Eguchi:2003ik,Ahn:2003tt,Israel:2004jt}.\footnote{
Actually there is a whole class of D0-branes in the cigar with an integer label $u$
giving in the open string channel finite representations of \slr
of spin $j=-(u-1)/2$, see~\cite{Ribault:2003ss,Israel:2004jt}.
However only the trivial representation (i.e. $u=1$) is unitary.
Thus we use only the corresponding D0-brane to build our D1-branes in the NS5 background.}
In the NS5-branes geometry, the fact that the D0-brane of the cigar
lives at its tip is interpreted as the fact
 that the D1-branes stretched between the NS5-branes are
localized on the two-planes in which the fivebranes live.

\subsection*{The one-point function}
Now that we have a reasonable idea of what kind of labels to expect
for our boundary states, we
turn to the exact expression for the one-point function.
We will work in the following in the light-cone gauge.
As explained in~\cite{Green:1996um,Gaberdiel:2000jr}
we put Dirichlet boundary conditions on the lightcone directions.
Moreover, in the directions transverse to the NS5-brane the D-branes
are of type A in the minimal model directions and of type B in the
remaining two directions (where we use the nomenclature of type A and
B as applying to factor conformal field theories, and not to the whole
of the D-brane). An analysis of the relation between type A/B branes
and Neumann and Dirichlet boundary conditions (as the one summarized
in appendix~\ref{boundfree} for branes in flat type II superstrings) then teaches
us that in the remaining four directions of the worldvolume of the
NS5-branes,
the boundary conditions can be chosen to be of
type A (label $\hat{s}_1$) for the
coordinates $(x^2 , x^3)$ and type B (label $\hat{s}_2$) for the directions
$(x^4,x^5)$. After Wick rotation back to ordinary Minkowski space, this allows
for the $1+1$ dimensional worldvolume of the D-string.

By combining the known one-point function for the branes for a free scalar,
the $N=2$ minimal model, and the $N=2$ non-compact conformal field theories,
and implementing the orbifold procedure on these one-point functions,
we obtain the following
one point-function for the W-bosons of Little String Theory (i.e.
the suspended D1-branes):\footnote{Here and in the following we have
  suppressed the $(z,\bar{z})$ dependence. One should read 
$\langle V_{\alpha,\bar{\alpha}} (z,\bar{z}) \rangle_{\hat{\beta}} =
  |z-\bar{z}|^{-\Delta_{\alpha}-\Delta_{\bar{\alpha}}} \Psi_{\hat{\beta}} (\alpha,\bar{\alpha}) 
$ and the coefficient 
$\Psi_{\hat{\beta}} (\alpha,\bar{\alpha})$, including the selection rules, is given in the text.} 
\begin{eqnarray}
\langle \ V_{j'm\bar{m}j w_L w_R,{\bf p}}^{(s_i)\
 (\bar s_i)}\
 \rangle^\textsc{w}_{\hat{\jmath},\hat{n},\hat{s}_i,{\bf \hat{y}}}
=   \frac{\nu_{k}^{\frac{1}{2}-j}\phantom{\!\!\!\!\!}}{k} \ \
\delta_{m,\bar{m}}
\delta_{m+ 2kw_L,\bar{m}-2k w_R} \
\delta_{s_1,\bar{s_1}} \delta_{s_2, -\bar{s_2}} \delta_{s_3,\bar{s_3}}
\delta_{s_4,-\bar{s_4}} \, \delta (p^5 )
\nonumber\\ e^{i \sum_{i=0}^{4} p^i \hat{y}^i}
e^{i\frac{\pi}{2} \sum_{i} s_i \hat{s}_i }\
e^{-i\pi \frac{m \hat{n}}{k}}  \
\frac{\sin \pi \frac{(1+2j')(1+2\hat{\jmath})}{k}}{
\sqrt{\sin \pi \frac{1+2j'}{k}}} \ \frac{\Gamma \left( j + \frac{m-s_4}{2}+kw
\right) \Gamma \left( j - \frac{m-s_4}{2}-kw \right)}{\Gamma (2j-1)
\Gamma ( 1 - \frac{1-2j}{k} )}.\nonumber\\
\label{oneptW}
\end{eqnarray}
On the left hand side, we have labels for the primary field of the minimal
model, with given spin and left and right momentum, and likewise for
the primary of the non-compact $N=2$ model. Moreover the labels $s_i$
and $\bar{s}_i$ indicate the chiral momenta of the bosonized pairs of
fermions. The lower indices are the (generalized) Cardy labels of the
boundary state. In the right hand side, we have substituted $w=w_L=-w_R$
(which follows from the constraints on the quantum numbers
to have a non-trivial one-point function). We have moreover
explicitly denoted the flat space one-point functions, which are well-known.
This expression is valid both for continuous representations,
$j= 1/2 + i P$, and for the discrete ones with $j$ real. Indeed,
the one-point function has poles for the discrete representations,
which correspond to couplings with states localized
on the NS5-branes plane (no poles arise due to the infinite volume of
the non-compact $N=2$ theory, since the D0-brane is a localized object). 
The couplings to these discrete representations are then given by the residues at the poles. 
We perform a Cardy-type consistency check on this one-point function next.
\subsection*{The open string partition function}
We start with the following integrand for the annulus amplitude
 for open strings
stretched between two D1-branes of the doubly scaled LST:
\begin{eqnarray}
Z^\textsc{w}_{\text{open}} = \sqrt{-i\tau}\ \frac{q^{\frac{1}{2} \left( \frac{\bf \hat{y} -
\hat{y}'}{2\pi}
\right)^2}}{\eta (\tau)^4}
\sum_{ \{ \upsilon_i \}  \in (\zi_2 )^4}
\frac{1}{2} \sum_{a,b=0}^{1} (-)^b (-)^{a (1+ \sum_{i} \upsilon_i)}
\chi^{(b+2\upsilon_1 + \hat{s}_{1}' - \hat{s}_{1}
)} \chi^{(b+2\upsilon_2 + \hat{s}_{2}' - \hat{s}_{2})} \nonumber\\
\sum_{2j=0}^{k-2} \sum_{n \in \zi_k}
N^{j}_{\hat{\jmath} \ \hat{\jmath\,}'} \
\mathcal{C}^{j \ (b+2\upsilon_3+
\hat{s}_{3}' - \hat{s}_{3})}_{2n+b+\hat{n} - \hat{n}'}  (\tau)\
Ch_\mathbb{I}^{(b+2\upsilon_4 + \hat{s}_{3}' - \hat{s}_{3})} (n;\tau ),\nonumber\\
\label{partopenD1}
\end{eqnarray}
written in terms of the $SU(2)_{k-2}$ fusion rules (see appendix~\ref{modtrans}), given through the
Verlinde formula in term of the modular S-matrix:
$ N^{j}_{\hat{\jmath} \ \hat{\jmath\,}'} = \sum_{j'}
S^{\hat{\jmath}}_{\ j'} S^{\hat{\jmath\,}'}_{\ j'} S^{j}_{\ j'}
/S^{0}_{\    j'}$. 
The $SL(2,\mathbb{R})/U(1)$ factor corresponds to the character 
of the identity representation, see appendix~\ref{modtrans}.
Note that the discrete momentum takes even values in
the NS sector and odd values in the R sector. 
An similar open string partition function has been found
in~\cite{Eguchi:2003ik}, using modular bootstrap techniques.
We discuss it in slightly more detail here in view of its importance
for Little String Theory.

Our goal is to show that this (physically sensible)
open string spectrum is consistent with the one-point function recorded
above. As
in flat space, the labels $\hat{s}_i$ of the branes has to obey two
kind of constraints. First, for all the fermions to be
either \textsc{r} or \textsc{ns}, one has to impose $\hat{s}_{i}' -
\hat{s}_i
= 0$ mod 2, $\forall i$. Second, the open string spectrum is supersymmetric
provided that:
\begin{equation}
\sum_{i} (\hat{s}_{i}' - \hat{s}_i ) - \frac{2(\hat{n}'-\hat{n})}{k}
= 0 \ \text{mod} \ 4.
\end{equation}
This last condition can be geometrically interpreted as the fact that
the two D1-branes between which we compute the open string spectrum
should be parallel and of the same orientation to obtain a supersymmetric answer.
Then we modular transform the open-string amplitude to go to the closed
string
channel, using the formulas for the modular transforms of the individual
characters recalled in appendix~\ref{modtrans}. Firstly we obtain a
continuous part ($\tilde{\tau} = -1/\tau$):
\begin{eqnarray}
Z_{closed}^{cont} (\tilde{\tau} ) = \sqrt{\frac{2}{k}} \
\int d^5 p \ 
\frac{\tilde{q}^{\frac{1}{2} {\bf p}^2}}{\eta (\tilde{\tau})^4}
\ e^{i {\bf p \cdot (\hat y - \hat{y}')}} \nonumber\\
\frac{1}{2} \sum_{a,b=0}^{1}
\sum_{ \{ \nu_i \}  \in (\zi_2 )^4} (-)^a (-)^{b (1+ \sum_{i} \nu_i)}
e^{i\frac{\pi}{2} \sum (a+2\nu_i) (\hat{s}_{i}' - \hat{s}_i )}
\chi^{(a+2\nu_1)} \chi^{(a+2\nu_2)} \nonumber\\
\sum_{m} \sum_{j'=0}^{k-2}
\frac{S_{\hat{\jmath}}^{\ j'} S_{\hat{\jmath\,}'}^{\ j'}}{S_{0}^{\   j'}}
e^{i\pi \frac{m(\hat{n} - \hat{n}')}{k}} \mathcal{C}_{m}^{j'\ (a+2\nu_3)}
(\tilde{\tau})\nonumber\\
\int_{0}^{\infty} dP\
\frac{2\sinh 2\pi P \, \sinh 2\pi P/k}{\cosh 2\pi P + \cos
 \pi (m-a)} Ch_{c}^{(a+2\nu_4)} (P, \frac{m}{2} ; \tilde{\tau}).
\label{cardyWcont}
\end{eqnarray}
Secondly we obtain a discrete part:
\begin{eqnarray*}
Z_{closed}^{discr} (\tilde{\tau} ) = \sqrt{\frac{2}{k}} \
\int d^5 p \
\frac{\tilde{q}^{\frac{1}{2} {\bf p}^2}}{\eta (\tilde{\tau})^4}
\ e^{i {\bf p \cdot  (\hat y - \hat{y}')}} \nonumber\\
\frac{1}{2} \sum_{a,b=0}^{1}
\sum_{ \{ \nu_i \}  \in (\zi_2 )^4} (-)^a (-)^{b (1+ \sum_{i} \nu_i)}
e^{i\frac{\pi}{2} \sum (a+2\nu_i) (\hat{s}_{i}' - \hat{s}_i )}
\chi^{(a+2\nu_1)} \chi^{(a+2\nu_2)} \nonumber\\
\sum_m \sum_{j'=0}^{k-2}
\frac{S_{\hat{\jmath}}^{\ j'} S_{\hat{\jmath\,}'}^{\ j'}}{S_{0}^{\   j'}}
e^{i\pi \frac{m(\hat{n} - \hat{n}')}{k}} \mathcal{C}_{m}^{j'\ (a+2\nu_3)}
(\tilde{\tau})\nonumber\\
\sum_{2j=2}^{k}\ \sum_{w\in \zi} \  \sum_{r \in \zi_k+\frac{a}{2}} \ \delta_{2j+2r+a+2\nu_4,m}
\ \sin \left(\pi \frac{1-2j}{k}\right)\,   ch_{d}^{(a+2\nu_4)}
(j,r+kw;\tilde{\tau})
\end{eqnarray*}
It can be straightforwardly shown using techniques which are minor
generalizations of those
used in \cite{Ribault:2003ss}\cite{Israel:2004jt}
that these two expressions correspond precisely to the overlap of two boundary
states, defined by eq.~(\ref{oneptW}).
To obtain the contribution of a discrete
representation,  we have to take the residue of the one-point function
at the corresponding pole.
In the above we have made manifest the coupling of the discrete
representations to the D1-branes stretching between the NS5-branes.

\subsection{The effective action on the D1-branes}
We constructed the exact boundary states for the
D1-branes stretching between NS5-branes. This determines
exactly the spectrum of open strings on the D1-branes. A
goal is to obtain an alternative description of physics of
little string
theory in the Higgs phase
in terms of the degrees of freedom living on these
D-branes.
\paragraph{Relation to the M(atrix)-theory}
This model can be related to the standard M-theory matrix model as follows.
Under T-duality, as discussed in~\cite{Giveon:1999px}, the ring of five-branes is mapped to a
resolved $A_{k-1}$ singularity. Then the relevant branes
capturing the degrees of freedom in the decoupling limit
are the D2-branes of type IIA string theory (lifted to membranes of M-theory)
wrapping the two-cycles of the resolved singularity (this is closely related to the discussion 
of~\cite{Douglas:1996xg,Diaconescu:1997br}); 
in the orbifold limit of the singularity they correspond to fractional D0-branes~\cite{Douglas:1996sw}. 
This type IIA string theory contains also D0-branes localized "far away" from the
singularity --~whose bound states correspond to Kaluza-Klein modes of the eleventh-dimensional 
graviton~-- but the decoupling limit (more precisely the
double scaling limit) captures only the degrees of freedom living
near the resolved singularity.
This gives at low energies a matrix quantum mechanics, which is specified by the partition
of the D2-branes onto the various two-cycles they can wrap.

When we take an infinite number of such wrapped D2-branes ($P \to \infty$) 
one describes the
full dynamics of M-theory on resolved $A_{k-1}$ in the decoupling limit when one goes to the
infinite momentum frame, using the same arguments as
for the standard  M(atrix)-theory in flat space~\cite{Banks:1996vh}.
A main difference is that in the present example the matrix model
describes a non-gravitational theory. The matrix model obtained is specified by a set of parameters
$\{ \rho_a = p_a /P \}$; they give the densities of D2-branes
wrapped on each two-cycle (in the perturbative regime,
these are in one-to-one correspondence with the
parameters $\{ \hat{\jmath}_a \}$ of the D1-branes).
Note that we discuss here a matrix
model description of higgsed $\mathcal{N}=(1,1)$ LST (i.e. arising from type IIB
fivebranes); a matrix model description for $\mathcal{N}=(2,0)$ LST
is proposed in~\cite{Aharony:1997th}.

\subsubsection*{Spectrum of light states}
Let us now determine the spectrum of lightest bosonic modes on these D-branes,
in order to derive the effective action. We will take
any supersymmetric collection of such D1-branes, i.e. such
that (for instance) $\hat{n}=\hat{s}=0$ for all the stacks of D1-branes.
Then the configuration is given simply in terms of the
occupation numbers $\{ n_{\hat{\jmath}} \}$ ( with either
$\hat{\jmath}=0,1,\ldots$ or
$\hat{\jmath}=\nicefrac{1}{2},\nicefrac{3}{2},\ldots$)
giving the numbers of D1-branes at each allowed position.\footnote{for the first series
the D1-branes are extended between NS5-branes separated by an even number of
NS5-branes, and for the second series  they are separated by an odd number. 
A supersymmetric configuration is built with all the D-branes of one kind.} 
We start from the open string spectrum of eq.~(\ref{partopenD1}) and 
we will make use of the properties of the characters recalled in
appendix~\ref{modtrans}.

First we consider the sectors of open strings with both ends
on the same D1-brane, with multiplicity $n_{\hat{\jmath}}$.
The first type of states that survive the GSO projection have
$\nu_1 = 1$ or $\nu_2=1$, the other $\nu_i$ being zero. Then the
\slc sector imposes $n=0$, and we obtain states of mass
$m^2 = \nicefrac{j(j+1)}{\alpha' k}$. They correspond to the
usual action of flat space oscillators $\psi^{\mu}_{-1/2} |0\rangle_\textsc{ns}
\otimes |j,0\rangle \otimes |n=0\rangle$ with $\mu=0,1,\ldots,5$. For
$j=0$ we obtain  (using the classification of D=4, $\mathcal{N}=2$ supersymmetry)
a $U(n_{\hat{\jmath}})$ gauge multiplet reduced to 0+1 dimensions.
The vevs of the five scalars correspond to the position of the D1-branes
along the worldvolume directions of the NS5-brane.
The second kind of states have $\nu_3=1$ and correspond to excitations along the
$SU(2)/U(1)$ factor. For $j \geqslant 1$ we obtain also states of mass
$m^2 = \nicefrac{j(j+1)}{\alpha' k}$ (again the \slc factor enforces $n=0$), and for $j=0$ we get a
 very massive state of mass
$m^2= 1/\alpha'$. For excitations along the \slc factor, we have necessarily
$n\neq 0$, and we get states of mass
$m^2 = \nicefrac{j(j+1)}{\alpha' k}+ \nicefrac{|n|-1}{\alpha'}$ for $0<|n|\leqslant 2j$
(states with $2j<n$ are hypermassive). Thus we get two scalars of mass
$m^2 = \nicefrac{j(j+1)}{\alpha' k}$ for nonzero values of $j$.

To summarize, the spectrum of light
states (i.e. those surviving the low energy limit
$\alpha' \to 0$, $\alpha' k$ fixed) for the self-overlaps
are a gauge multiplet of D=4, $\mathcal{N}=2$ reduced to 0+1;
for the massive states starting at $m^2 = \nicefrac{2}{\alpha' k}$ we obtain 
massive multiplets of $\mathcal{N}=2$, by adding the degrees of freedom of 
adjoint hypermultiplets corresponding to excitations along the directions transverse
to the NS5-branes. The maximal mass for these multiplets corresponds to 
$j_\text{max}=\text{min} (2\hat{\jmath},k-2\hat{\jmath})$.

We concentrate now on sectors of open strings stretched between different
D1-branes, of parameters $\hat{\jmath}$ and $\hat{\jmath\,}'$,
with $\hat{\jmath\,}' - \hat{\jmath} \in \zi$. The main difference
with the self-overlaps is that now $j\geqslant j_\text{min}=|\hat{\jmath\,} ' - \hat{\jmath}\, |$.
The rest of the analysis is quite similar, and we obtain the
massive multiplets described above, starting at  $m^2 = \nicefrac{j_\text{min}(j_\text{min}+1)}{\alpha' k}$.
They transform in the bifundamental representation of $U(n_{\hat{\jmath}})\times U(n_{\hat{\jmath\,}'})$.

\subsubsection*{Low energy effective action}
According to the previous analysis the low energy effective action is
described simply by the dimensional reduction of $d=6$, $\mathcal{N}=1$
SYM to 0+1 dimensions. The effective interactions are presumably trivial in the transverse
directions (since only the "identity" is involved in the relevant boundary
three-point functions, both for the $SU(2)$ and the \slr part).
The above reasoning is a microscopic version of the reasoning
of~\cite{Hanany:1996ie} that determined the low-energy theory for D-branes
stretching between NS5-branes. For finite $k$ (i.e. a fixed number
of NS5-branes) only these fields survive the low energy limit.
It is interesting to note that
this matrix quantum mechanics comes from dimensional reduction of six-dimensional gauge theory,
while the type IIB LST itself flows at low energies to a six-dimensional
$\mathcal{N}=(1,1)$ gauge theory.

Now, the exact boundary state description determine the full open string
spectrum for the D-branes stretching between the NS5-branes, including
all massive modes and all descendants. It would be very interesting to study
the associated open string field theory, which encodes physics of two
separated NS5-branes.\footnote{A simpler analogue for this
type of change of perspective on worldvolume
physics would be the Nahm formulation of the monopole
solution, which can be interpreted as giving the profile of D1-branes when they
open up into a D3-brane \cite{Diaconescu:1996rk}, in an $SU(2)$ gauge theory
Higgsed to $U(1)$.} But here, we do not zoom in on the low
energy physics only.

We can in principle go beyond this approximation (which was low energies compared
to the inverse string length, and the inverse string length divided by 
the square root of the number
of NS5-branes), and concentrate on energies below the inverse of the
string length, but higher than the inverse of the string length divided
by the square root of the number of NS5-branes (in the limit of a large number of NS5's).
This brings us in the regime where the boundary three-point function in the
compact minimal model becomes relevant while the non-compact coset still
yields trivial interactions because it involves only the trivial representation
of the \slr algebra. There is at least a specific regime where this effective
action can be computed, as we shall see below.

\paragraph{Large N dynamics of LST}
We are interested in the $k\to \infty$ limit of the theory living on the
D1-branes in the DSLST background. We hope it will give some insights
about the large N limit (in our notations,
large $k$) of the little string theory.
The 't Hooft coupling of the little string theory is, for $k$ fivebranes $\lambda = \alpha' k$.
We study the LST in a point in the moduli space where the
gauge group $U(k)$ is broken to $U(1)^k$
at a scale $\rho_0$, corresponding to the radius
of the circle of fivebranes. In the double scaling limit:
\begin{equation}
g_s \to 0 \ , \ \
U_0 = \frac{\rho_0}{g_s \sqrt{\alpha'}}\ \ \text{fixed}
\end{equation}
we scale the Higgs vev to zero in string units.
In the holographic description in the bulk,
the effective string coupling is finite and given by
$g_{s}^\textsc{eff} = \sqrt{k}/U_0$.
Thus to stay in the perturbative description in the bulk for large $k$ we should consider the regime:
\begin{equation}
k \ \ \text{large} \ , \ \ \frac{k}{U_{0}^2}=
\left( \frac{g_s}{\rho_0} \right) \lambda \ \ \text{fixed and small}.
\label{effcouplconstr}
\end{equation}
Let us consider a configuration of $P$ such D1-branes of parameters
$(\hat{\jmath}_a,\hat{n}_a,\hat{s}_a)$ stretched between the
NS5-branes; the D1-branes stacks have to be parallel to preserve
some supersymmetry. As we said previously we can choose for instance
$\hat{n}_a=\hat{s}_a=0 \ \forall a$. We add Chan-Paton factors
to the D1-branes, then each stack of D1-brane contains a $U(p_a)$
gauge group, with the constraint $\sum p_a  = P$.

The mass of a D1-brane of label $\hat \jmath$ is:
\begin{equation}
M_\textsc{w}^a
\ \stackrel{k \gg 1}{\simeq}\
\frac{1}{g_{s}^\textsc{eff}} \sqrt{\alpha ' k} \ \Delta \Psi
\ = \ \sqrt{\frac{\alpha '}{k}} \,U_0
\ \frac{2\pi(1+2\hat \jmath )}{\sqrt{k}}
= \frac{\rho_0}{g_s} \ \frac{2\pi(1+2\hat \jmath )}{k}
\end{equation}
Note that the length of the D1-brane $L=  \sqrt{\alpha ' k} \Delta \Psi $
does not depend on the radius of the circle $U_0$ since this parameter
drops from the five-brane metric in the double scaling limit.
\FIGURE{\epsfig{figure=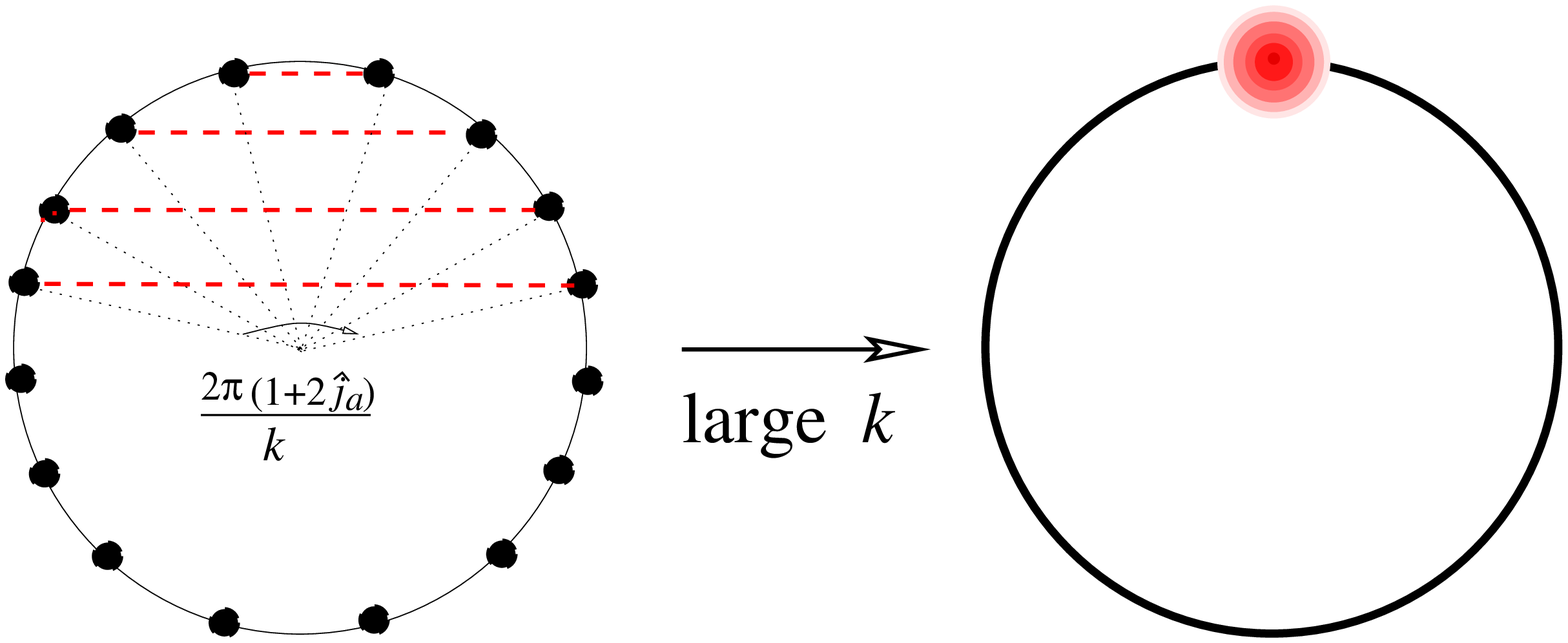, width=100mm}
\label{largek}
\caption{Large k limit of a BPS D1-brane configuration. All the
D1-branes collapse to a point.} }
To probe the low energy behavior at large N we can go to the {\it Alekseev-Recknagel-Schomerus limit}
(\textsc{ars} limit)~\cite{Alekseev:2000fd} where we consider the theory at energies $E$ such that:
\begin{equation}
\alpha ' E^2 \to 0 \ , \ \ \lambda \ E^2 =  \alpha' k \ E^2 \to \infty \ , \ \
\hat {\jmath}_a \ \text{fixed}.
\end{equation}
In this limit all the D1-branes collapse to a point
on the boundary of the disc (fig.~\ref{largek}).
{From} the LST
point of view we are in the regime of strong 't Hooft coupling.
Thus we expect to get in this limit some matrix quantum mechanics
containing degrees of freedom of strongly coupled Little String Theory at
large $k$. We have chosen the limit such that all the states with
masses below the inverse of the string length contribute.

We want now to compute the effective action in the ARS limit.
All the massive states that we described at the beginning of these section
(those with small mass compared to the inverse of the string length)
will then contribute because their mass goes to zero.
The non-trivial interactions come from
the boundary three-point function. Since the \slc part of the
CFT is always in the trivial representation it presumably
won't give non-trivial
contribution. On the other hand the $SU(2)$ part is non-trivial
but reduces in this limit to the matrix
multiplication~\cite{Alekseev:2000fd} of a fuzzy gauge theory.
However this beautiful construction
does not directly apply to our case. Indeed according to our
analysis of the spectrum of light states on 
the D1-branes, for any $j$ only states with $n=0$
of the representation survives the low energy limit.\footnote{
The results are rather different from the results for the coset $SU(2)/U(1)$;
this originates in the
$\zi_k$ orbifold with the trivial representation of \slc.} Therefore there is
no 
standard enhancement of (fuzzy)
gauge symmetry on the W-bosons of LST.

However in the present case we still have a
 ring of massless states labeled by the spin
$j$ of the $SU(2)$ representations coming from the various massive multiplets discussed above.
Their fusion rule is given by:
\begin{equation}
V_j (x_1) V_{j'} (x_2) = \sum_{j''} \left[ \begin{array}{ccc}j & j' & j'' \\ 0&0&0 \end{array} \right]
\left\{
\begin{array}{ccc}j & j' & j'' \\ \hat{\jmath}&\hat{\jmath}& \hat{\jmath}\end{array}
 \right\} V_{j''} (x_2 )
\end{equation}
in terms of the Clebsch-Gordan coefficients $[ \cdots ]$ and the 6$j$-symbols $\{ \cdots \}$.
Using this data it should be possible 
to write the matrix quantum mechanics corresponding
to the effective action, much as in~\cite{Alekseev:2000fd}. As we argued it may contain the
non-trivial dynamics of higgsed $\mathcal{N}=(1,1)$ Little String Theory at large N.

%%%%%%%%%%%%%%%%%%%%%%D4-branes%%%%%%%%%%%%%%%%%%%%%%%%%%%%%%%%%%%%%%%%%%%%%
%%%%%%%%%%%%%%%%%%%%%%%%%%%%%%%%%%%%%%%%%%%%%%%%%%%%%%%%%%%%%%%%%%%%%%%%%

\subsection{D4-branes and the beta function of D=4 SYM}
Now that we have the exact description of D-branes between
NS5-branes, we can ask whether we can recuperate familiar properties
from these exact descriptions. Let's consider the following
gauge theory physics \cite{Witten:1997sc}.
If we consider D4-branes stretching between NS5-branes in type IIA
string theory, then the D4-branes pull the NS5-branes and cause a logarithmic
bending of the NS5-branes. Since the gauge coupling for the four-dimensional
gauge theory on the D4-branes is inversely proportional
to the distance between the NS5-branes, we thus recuperate the logarithmic
running of the four-dimensional gauge coupling.

Here, we wish to show
how the boundary one-point function of the D4-branes stretched between the
NS5-branes encodes the beta-function of the gauge theory living on the
D4-branes. To realize this, we need to add one step to the above reasoning:
the one-point function of the D4-branes encodes the logarithmic bending
of the NS5-branes. It is this extra step that we want to demonstrate in this
subsection.
Let's consider then our familiar background of NS5-branes, in the near-horizon
limit, and the  D4-branes
stret\-ched between them following~\cite{Witten:1997sc}.

Schematically,
the branes fill out the following directions:
$$
\begin{array}{c|cccccccccc}
& 0 & 1 & 2 & 3 & 4 & 5 & 6 & 7 & 8 & 9 \\
\hline
NS5 & \sslash & \sslash & \sslash & \sslash & \sslash & \sslash &
\perp & \perp & \perp & \perp \\
D4 & \sslash & \sslash & \sslash & \sslash & \perp & \perp & \sslash
& \perp & \perp & \perp \\
\end{array}\\
$$
As we recalled above, the presence of the D4-branes will bend the NS5-branes on
which
they are attached. More precisely we expect that their position in the
$(x^6,x^7)$ plane will become a function of the coordinates
$(x^4,x^5)$. To measure the bending, we will use a holographic
reasoning\footnote{ We thank David Kutasov for an interesting discussion on
this point.}.
We know that the position of the fivebranes is
encoded in the expectation value of the scalars $X^{6,7,8,9}$, which live
on the NS5-brane worldvolume (at low energy, in type IIB). Traceless
symmetric combinations of the scalars are
holographically dual to particular BPS closed string vertex operators.
Using the holographic dictionary of \cite{Giveon:1999px}\cite{Aharony:2004xn}
we can thus translate the computation of an expectation value of the transverse scalars on the
NS5-brane worldvolume (i.e. the computation of the profile of the NS5-branes)
into a computation in the bulk worldsheet conformal field theory. 
However this dictionary is known only in type IIB (i.e. for $\mathcal{N}=(1,1)$ little 
string theory) thus we have to start from this T-dual type IIB setup.
The relevant operators are:
\begin{equation}
\tilde{\text{Tr}} (B^{2j+2} ) \sim
V_{j} (p^\mu ) = e^{-\varphi - \tilde{\varphi}}
\Phi^{SU(2)/U(1)\ (2,2)}_{j,j+1,j+1} \
\Phi^{SL(2)/U(1)\ (0,0)}_{\tilde{\jmath},j+1,j+1} \
e^{i p_\mu X^\mu}
\label{dicthol}
\end{equation}
in terms of the complexified scalar $B=X^6+iX^7$. The 
value of $\tilde{\jmath}$ is fixed by the on-shell condition of the 
string theory. From the holographic point of view, these 
non-normalizable vertex operators correspond to off-shell operators in the dual little string theory, 
see~\cite{Aharony:2004xn} for details. To find the (change of the) expectation values of these observables 
of little string theory, we compute the coupling of the boundary states for the D4-branes
to the localized states corresponding to~(\ref{dicthol}) in the bulk theory.
Thus, we calculate the following quantity:
\begin{equation}
\delta \langle \tilde{\text{Tr}} (B^{2j+2} ) \rangle (p^4 , p^5 )  \sim \langle
V_{j} (p^\mu ) | D | \hat{\jmath},\hat{n},\hat{s}_i,p^4,p^5 \rangle\, ,
\end{equation}
where $D$ is the closed strings propagator in the NS sector:
\begin{equation}
D = \int_{|z|\leqslant 1} \frac{dz d\bar{z}}{z\bar{z}}
z^{L_0 -1/2} \bar{z}^{\bar{L}_0 - 1/2}.
\end{equation}
This corresponds to the one point function for the closed string state
$|V_{j}(p^\mu )\rangle $ in the presence of the boundary state $|
B_{\hat{\jmath},\hat{n},\hat{s}_i} \rangle$. In the following,
we will pull the closed
string vertex
to infinity, i.e. concentrate on the long distance effect
of the boundary state, which signals the presence
of a  closed string tadpole.
We will use the change of variables $z=e^{-\pi t+ i\theta}$, and using
the formula for the
one-point function, eq~(\ref{oneptW}), we find the expression:
\begin{eqnarray}
\delta \langle \tilde{\text{Tr}} (B^{2j+2} ) \rangle (p^4 , p^5 )
\sim \pi \int_{0}^{\infty} dt \quad e^{-2\pi t \frac{j(j+1)-(j+1)^2}{k} }
\langle \Phi^{SU(2)/U(1) \ (2,2)}_{j,j+1,j+1} \Phi^{SL(2)/U(1)\ (0,0)}_{\tilde{\jmath},j+1,j+1}
| \hat{\jmath}, \hat{n} , \hat{s} \rangle \nonumber\\
e^{-\pi t p_\mu  p^\mu } \langle p_4, p_5 | \hat{x}^4, \hat{x}^5
\rangle_{D} \ \langle p_0 , \cdots , p_3 | 0 \rrangle_{N}\nonumber\\
\nonumber\\
= \pi \int_{0}^{\infty} dt \quad e^{-2\pi t
  \frac{j(j+1)-\tilde{\jmath}(\tilde{\jmath}-1)}{k} + \frac{1}{2}(p_{4}^2
+ p_{5}^2 )} \hskip5cm \nonumber\\
(-)^{\hat{s}}\,
\frac{\nu_{k}^{\frac{1}{2}-\tilde{\jmath}}\phantom{\!\!\!\!\!}}{k} \
e^{i (p_4 \hat{x}^4 + p_5 \hat{x}^5 )}
e^{-2i\pi \frac{(j+1) \hat{n}}{k}}  \
\frac{\sin \pi \frac{(1+2j)(1+2\hat{\jmath})}{k}}{
\sqrt{\sin \pi \frac{1+2j}{k}}} \ \frac{\Gamma \left( \tilde{\jmath} + j+1
\right) \Gamma \left(  \tilde{\jmath} - (j+1) \right)}{
\Gamma (2\tilde{\jmath}-1) \Gamma ( 1 + \frac{2\tilde{\jmath}-1}{k}
)}\nonumber\\
\end{eqnarray}
In the case $\tilde{\jmath} = j+1$, we have a pole of this expression 
(of the LSZ type in the classification of~\cite{Aharony:2004xn})
corresponding to a massless localized mode, and we take the residue to 
find the coupling. Then we can integrate over $t$ and Fourier transform on $p_4$ and $p_5$.
To compute the expectation value, we take the residue at the discrete
pole in
the second gamma function in the numerator. It leads to the expression:
\begin{eqnarray}
\langle \tilde{\text{Tr}} (B^{2j+2} )\rangle  (x^4 , x^5 ) =
- \frac{\log v}{2\pi}
(-)^{\hat{s}}\,
\frac{\nu_{k}^{-\frac{1}{2}-j}\phantom{\!\!\!\!\!}}{k} \
e^{-2i\pi \frac{(j+1) \hat{n}}{k}}  \
\frac{\sin \pi \frac{(1+2j)(1+2\hat{\jmath})}{k}}{
\sqrt{\sin \pi \frac{1+2j}{k}}} \ \frac{\Gamma \left( 2j+2
\right)}{
\Gamma (2j+1) \Gamma ( 1 + \frac{2j+1}{k}
)}\, , \nonumber\\
\text{with} \ v^2 = (x_4 - \hat{x}_4)^2 + (x_5 - \hat{x}_5 )^2.\nonumber\\
\label{hologbend}
\end{eqnarray}
\FIGURE{
\phantom{fffffffffffffffffffffffff} \epsfig{figure=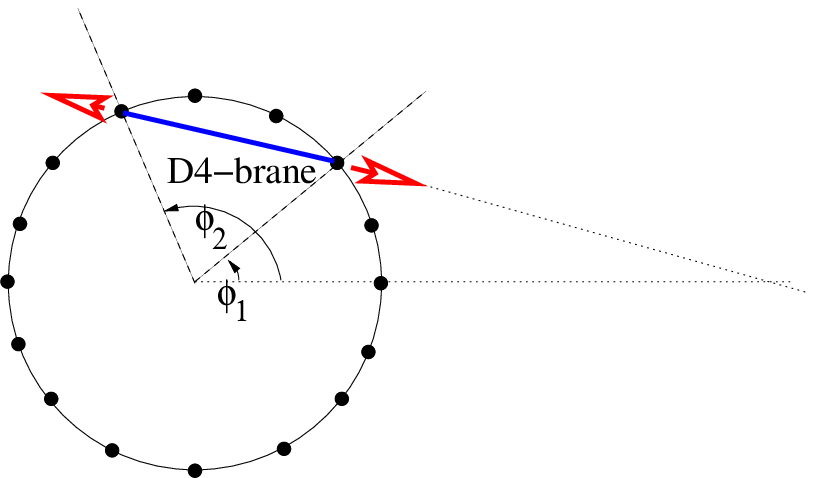, width=120mm}
\caption{Deformation of the ring of NS5-branes by a D4-brane.}
\label{pullfig}}
Let us now interpret this result geometrically.
We started out with a configuration of
five-branes
on a circle, corresponding to the following expectation value for
the complex scalar field $B$:\\
\begin{equation}
\langle B \rangle = \rho_0 \ \text{diag} (1,e^{2i\pi/k}, \cdots, e^{2i\pi
(k-1)/k} ),\end{equation} before we put any D4-branes into the system.
Then we stretched
a D4-brane of parameters $\hat{\jmath},\hat{n}$ between
two NS5-branes of the ring,
located at $\phi_1 = \frac{\pi(\hat{n}-2\hat{\jmath}-1)}{k}$
and $\phi_2 = \frac{\pi(\hat{n}+2\hat{\jmath}+1)}{k}$ (see
fig.~\ref{pullfig}).
Then, following the geometrical picture,
we expect that the expectation value of the operator $B$
will be modified as follows:
\begin{equation}
\langle B \rangle \sim \text{diag}\left( 1, \cdots, e^{i
\pi(\hat{n}-2\hat{\jmath}-1)/k}
- i \lambda e^{i\pi \hat{n}/k} \log v   , \cdots ,
e^{i \pi(\hat{n}+2\hat{\jmath}+1)/k }
+ i \lambda e^{i\pi \hat{n}/k} \log v   , \cdots \right)\, ,
\end{equation}
where $\lambda$ is an expansion parameter.
The variation in the trace of $B^{2j+2}$ is:
\begin{equation}
\delta \langle \text{Tr} (B^{2j+2}) \rangle  \sim 2 i \lambda (2j+2)
e^{2i\pi \frac{\hat{n} (j+1)}{k}} \sin \frac{\pi}{k} (2j+1)
(2\hat{\jmath}+1) \log v.
\end{equation}
This has to be compared with the holographic computation,
eq.~(\ref{hologbend}). In particular, the functional dependence
in the geometrical parameters $(\hat{\jmath},\hat{n})$ matches
precisely. This convincingly demonstrates that the one-point function
on the D4-branes allows for the computation of the logarithmic bending
of the NS5-branes, which was the extra step we wished to demonstrate.
As a side remark, note that
we can also check that there is no tadpole for the scalar
$A = X^8 + i X^9$. In fact using the dictionary,
they correspond to the operators:
\begin{equation}
\tilde{\text{Tr}} (A^{2j+2} ) \sim
V_{j} (p^\mu ) = e^{-\varphi - \tilde{\varphi}}
\Phi^{SU(2)/U(1)\ (2,2)}_{j,j+1,-(j+1)} \
\Phi^{SL(2)/U(1)\ (0,0) }_{\tilde{\jmath},j+1,-(j+1)} \
e^{i p_\mu X^\mu}
\end{equation}
which do not couple to the boundary state defined by the one-point
function~(\ref{oneptW}). This is expected since the NS5-branes are 
not pulled in the $x^{8},x^{9}$ directions by the D4-branes.

Note that an alternative interpretation of the computation would be
as follows. In~\cite{DiVecchia:1997pr} the
change of the background fields around flat space by a D-brane was computed
with roughly the same methods. In fact, the computation can be interpreted
as providing the massless closed string tadpoles in the string effective
action, which can then be used to determine a new on-shell closed string
background. (Reasoning in
this way does not require the holographic dictionary.) Both types
of reasoning lead to the same conclusion.

In the configuration described here, there is an infrared divergence because
the bending of the NS5-branes has a logarithmic profile. As in~\cite{Witten:1997sc}
we could avoid this problem by considering a "balanced" configuration,
i.e. such that the same number of D4-brane end on each side of any NS5-brane.
Then the bending of the NS5-branes falls of asymptotically (i.e. for large $v$).
However in our case --~because the NS5-branes are distributed on a circle and not on
a line~-- it is not possible to obtain such a supersymmetric configuration
with only the suspended D4-branes discussed here. We would need to attach to
the other side of the NS5-branes semi-infinite D4-branes that will be discussed
below.\footnote{They are  obtained from the radial D1-branes by T-dualities
along the flat directions of the worldvolume of the NS5-branes.}

\subsection*{Further remarks on the beta-function}
For completeness, we briefly recall from~\cite{Witten:1997sc}
the precise relation with the computation of the beta-function.
For $n$ coincident D4-branes suspended between NS5-branes, we obtain a
four-dimensional gauge theory with $\mathcal{N}=2$ supersymmetry.
It contains an $SU(n)$ gauge multiplet\footnote{As explained in~\cite{Witten:1997sc}
a $U(1)$ multiplet is  frozen} and no hypermultiplets. It is also possible
to add fundamental matter to the gauge theory by adding D6-branes to the
setup --~which are obtained from the D3-branes discussed below by T-duality
along $x^{1,2,3}$.

Calling $\ell (v)$ the distance between a point on the NS5-branes
and the point where the
D4-branes are attached, the running of the coupling constant is given by:
\begin{equation}
\frac{1}{g_{\textsc{ym}}^2 (v)} = \frac{\ell (v)}{g_{string}}.
\end{equation}
To fix the precise relative coefficient we
first complexify
the gauge coupling:
\begin{eqnarray}
\tau &=& \frac{\theta}{2 \pi} + \frac{4 \pi i }{g^2}.
\end{eqnarray}
We then observe that the monodromy that the gauge coupling picks
up as we change branch in the log-function should correspond to a
trivial operation in the gauge theory. The elementary monodromy
naturally corresponds to the smallest trivial shift of the theta-angle:
$\theta \rightarrow + 2 \pi$. So, for one D4-brane, we put:
\begin{eqnarray}
-i \tau(v) &=& -2 \log (v).
\end{eqnarray}
(To fix the smallest multiple possible,
we have made use of the fact that we can add other D4-branes
to this picture that will contribute flavors to the ${\cal N}=2$
 low-energy gauge
theory \cite{Witten:1997sc}. That leads to the prefactor of 2
in the above formula.)  The sign of the beta-function is
determined by considering the direction of the bending of the NS5-branes
due to the attached D4-branes.
When we generalize this picture to $N$ D4-branes, we obtain:
\begin{eqnarray}
-i \tau(v) &=& -2N \log (v).
\end{eqnarray}
which coincides with the beta-function of $SU(N)$ $N=2$ SYM.
The above can be interpreted as a way of fixing the overall coefficients
in our computation. It should be clear that if we perform our
computation very carefully, we would be able to obtain the precise prefactor,
and the beta-function of $N=2$ SYM.
(See e.g.\cite{DiVecchia:1997pr} for the relevant techniques and a flat
space example.) It would be interesting to nail down the overall coefficient,
but we haven't attempted to do so.

In this context, we would moreover like to recall yet another way to compute
the beta-function of $N=2$ SYM, using a different
holographic set-up in which fractional
branes carry the gauge theory that is holographically dual to their
corresponding
supergravity solution. The supergravity solution
is then shown to encode the beta-function \cite{Bertolini:2000dk}.

Finally, we would like to
mention the intriguing
possibility of determining the $N=1$ SYM beta-function from
the backreaction of a D3-brane (which can be thought of as the tensor
product BCFT of a D3-brane in four-dimensional flat space and the
D0-brane of the cigar) in non-critical six-dimensional string theory
(see e.g.\cite{Kuperstein:2004yk}\cite{Klebanov:2004ya}\cite{Alishahiha:2004yv})\footnote{We thank
Emiliano Imeroni and especially
Sameer Murthy for discussions on this point}. It
will be interesting to apply the technique we used here for determining
the beta-function in this less supersymmetric setting.
The general idea is that the one-point function of the brane encodes
sufficient information to determine the running of the coupling that
lives on the brane (by open-closed string duality).

%%%%%%%%%%%%%%%%%%%%%%%%%%%%%%%%%%%%%%%%%%%%%%%%%%%%%%%%%%%%%%%%%%%%%%%%%%%%%%%%%
%%%%%%%%%%%%%%%%%%%%%%%%%%%%%%D3-branes%%%%%%%%%%%%%%%%%%%%%%%%%%%%%%%%%%%%%%%%%%
%%%%%%%%%%%%%%%%%%%%%%%%%%%%%%%%%%%%%%%%%%%%%%%%%%%%%%%%%%%%%%%%%%%%%%%%%%%%%%%%%

\subsection{Boundary states for the cylindrical D3-branes}
\label{boundcyld3}
In this section we construct  exact descriptions of the first class of D3-branes orthogonal to 
the NS5-branes in type IIB string theory, 
of which we discussed the semi-classical description in section~\ref{semiclass}.
We concentrate on the D3-branes which we constructed in the T$_\psi$-dual picture described by
the orbifold of the product of cosets, see eq.~(\ref{firstd3profile}).
In this T-dual type IIA string theory, we combine the two D1-branes in the
respective cosets, and look for the regular brane in the orbifold,
i.e. we construct the brane by summing
a particular brane and all its images
under the orbifold operation. It corresponds to the regular
representation of the orbifold group.
\FIGURE{
\epsfig{figure=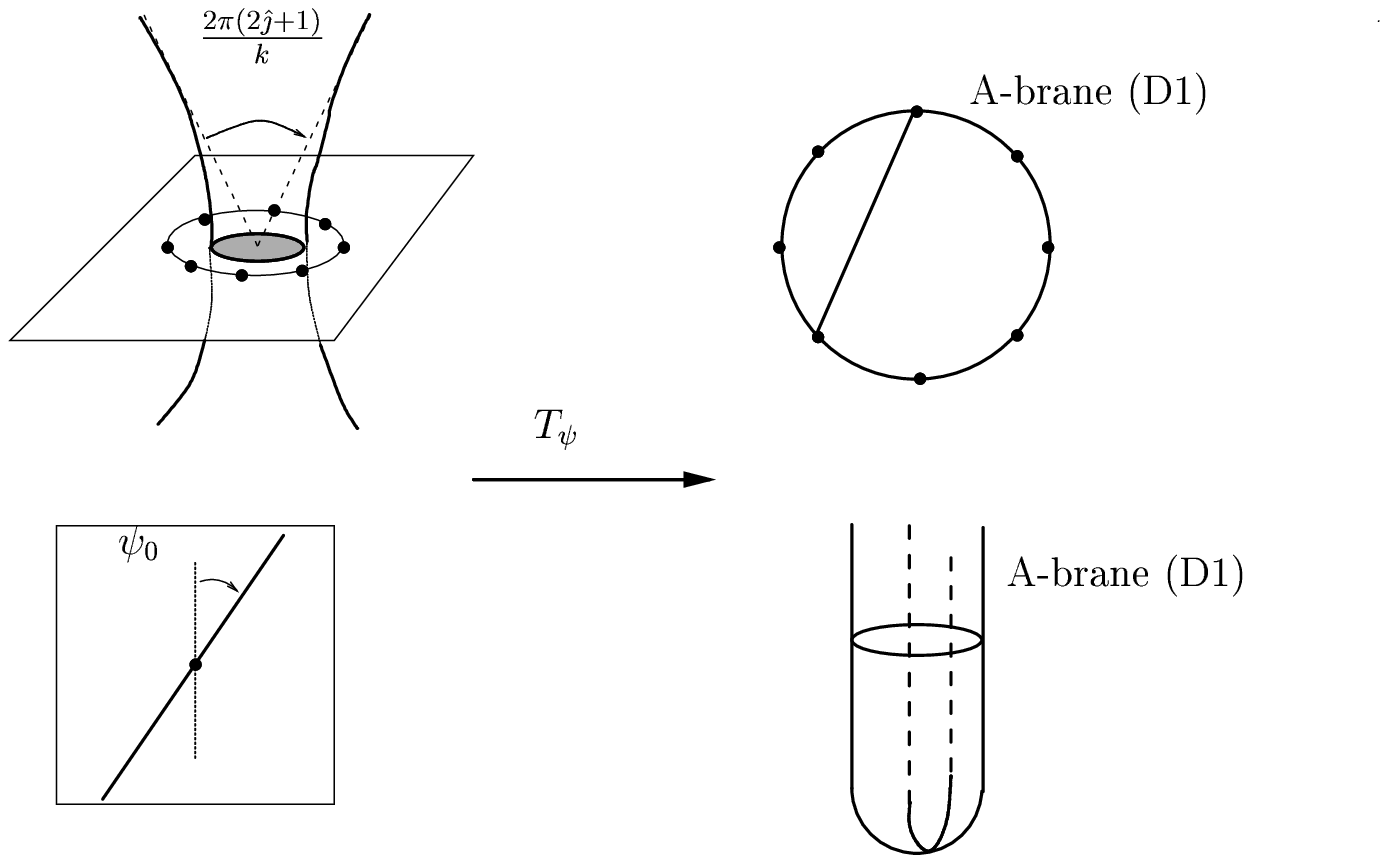, width=110mm}
\caption{Geometry of the D3-branes and
its T-dual, in the case $r_0=0$. For large $r$ the
D3-brane asymptotes a  cone, corresponding to a symmetric D2-brane of $SU(2)$ and
a Neumann D-brane (FZZ) of the linear dilaton direction. The intersection
of the D3-brane with the plane $x^8=x^9=0$ is a disc, i.e. a D2-brane of the
coset $SU(2)/U(1)$.}
\label{D3fig}}
 The semi-classical picture of the brane
in the NS5-brane background is:
$$
\begin{array}{c|cccccccccc}
& 0 & 1 & 2 & 3 & 4 & 5 & 6 & 7 & 8 & 9 \\
\hline
NS5 & \sslash & \sslash & \sslash & \sslash & \sslash & \sslash &
\perp & \perp & \perp & \perp \\
D3 & \sslash & \perp & \perp & \perp & \perp & \perp & \sslash
& \sslash & \sslash
 & \perp \\
\end{array}\\
$$
We combine the profiles of the coset D-branes as indicated in the semi-classical analysis
(see fig~\ref{D3fig}). What we obtain from eq.~(\ref{firstd3profile})  
is a non-trivial D3-brane with, for $r_0=0$,
a "cylinder" section near the plane $x^8=x^9=0$ going through the ring of five-branes,
and connecting smoothly with a  cone of opening 
angle $2\pi (2\hat{\jmath}+1)/k$ on both sides.
For $r_0 \neq 0$, the geometry is more involved and in particular the D3-brane
avoids the plane $x^8=x^9=0$.

To construct this D3-brane, we have first to consider a D1-brane in the cigar
with the profile (with parameters $\psi_0$ and $r_0$):
\begin{eqnarray}
\sinh r \sin(\chi-\psi_0) &=& \sinh r_0
\end{eqnarray}
and then a D1-brane in the T-bell with geometry
 (with parameters $\phi_0$ and $\theta_0$):
\begin{eqnarray}
\cos \theta \sin( \chi - \phi+\phi_0) &=& \cos \theta_0.
\end{eqnarray}
Their exact one-point functions (or Cardy states) are
described in terms of the one-point functions. For the $SU(2)/U(1)$ part, we have
(e.g. in the NS sector)
\begin{eqnarray}
\langle V^{j'}_{n'} \rangle_{\hat{\jmath} \hat{n}} &=&
\frac{1}{k} \frac{
  \sin \left( \frac{\pi(2\hat{\jmath}+1)(2j'+1)}{k}\right)}{
\sqrt{\sin(\frac{\pi(2j'+1)}{k})}}
e^{\pi i \hat{n} n'/k}
\label{D1Tbell}
\end{eqnarray}
with $\theta_0$ and $\phi_0$ given by~(\ref{corrparamD1}). 
For the \slc part, the one-point function reads:
\begin{eqnarray}
\langle \Phi^j_{nw} \rangle_{r_0,\psi_0}
&=& \delta_{w,0} \nu^{1/2-j} \frac{1}{\sqrt{2}}
e^{i n \psi_0} \nonumber \\
&& \left[ e^{-r_0(-2j+1)}+(-1)^n e^{r_0(-2j+1)} \right]
\frac{\Gamma(1-2j)\Gamma(1+\frac{1-2j}{k})}{\Gamma(1-j+\frac{n}{2})\Gamma(1-j-\frac{n}{2})}
\label{D1cigar}
\end{eqnarray}
where the chiral momenta are $m=(n+kw)/2$ and $\bar{m}=-(n-kw)/2$.\footnote{For states
which are not NS primaries it will be modified  (in the $\zi_4$ formalism) as
$m+s/2=(n+kw)/2$ and $\bar{m}-\bar{s}/2=-(n-kw)/2$.}

We can now sum over $\zi_k$ copies of these branes to obtain boundary
states of regular branes
described by the product of one-point functions of the cosets. The one-point
functions
for the sum over copies of branes is the sum of the one-point functions
for the individual copies.
We thus sum over $l \in \zi_k$ the boundary states labeled
by $(\psi_0+ 2 \pi l / k, \hat{n}+2 l)$.
We obtain the one-point functions:
\begin{eqnarray}
\langle \Phi^j_{nw} V^{j'}_{-n-k\tilde{w}}
\rangle_{r_0\psi_0,\hat{\jmath\,} \hat{n}}
=  \delta_{w,0} \nu^{1/2-j} \frac{1}{ \sqrt{2}}\left[ e^{-r_0(1-2j)}+(-1)^n
e^{r_0(1-2j)} \right]
\nonumber \\
\frac{\Gamma(1-2j)\Gamma(1+\frac{1-2j}{k})}{\Gamma(1-j+\frac{n}{2})\Gamma(1-j-\frac{n}{2})}
\frac{
\sin(\frac{\pi(2\hat{\jmath}+1)(2j'+1)}{k+2})}{\sqrt{
\sin(\frac{\pi(2j'+1)}{k})}} 
e^{i n \psi_0}
\nonumber \\
 e^{- \pi i \frac{\hat{n} n}{k}}
e^{- \pi i \tilde{w} \hat{n}},
\end{eqnarray}
and we can absorb the second to last phase factor into a redefinition
of the label $\psi_0$.
Note that $\tilde{w}$ keeps track of
whether the $U(1)$ quantum numbers of the two cosets sum up to an even or
and odd multiple of $k$. In the GSO-projected closed string
spectrum~(\ref{contspecbulk}), there is an additional $\zi_2$ orbifolding
such that the charges of the cosets are identified modulo $2k$.
In the lightcone gauge,
the boundary conditions along the flat directions corresponding
to worldvolume of the five-brane of type A (label $\hat{s}_1$) for the
coordinates $(x^2 , x^3)$ and B (label $\hat{s}_2$) for $(x^4,x^5)$.
Then putting everything together, and using the labeling of the
quantum number of the closed string partition function~(\ref{contspecbulk}),
we find that the one-point function  for the D3-brane in the NS5-brane 
background is given by:
\begin{eqnarray}
\langle \ V_{j'm\bar{m}\, j w_L w_R,{\bf p}}^{(s_i)\
 (\bar s_i)}\
 \rangle^{D3}_{\hat{\jmath},r_0,\psi_0,\hat{s}_i,{\bf \hat{y}}}
=  \frac{ \nu_{k}^{\frac{1}{2}-j}}{2}   \
\delta_{m,-\bar{m}} \delta_{m+2kw_L,-\bar{m}+2kw_R} \
\delta_{s_1,\bar{s_1}} \delta_{s_2, -\bar{s_2}} \delta_{s_3,-\bar{s_3}}
\delta_{s_4,\bar{s_4}} \, \delta (p^5 )
\nonumber\\ e^{i \sum_{i=0}^{4} p^i \hat{y}^i}
e^{i\frac{\pi}{2} \sum_{i} s_i \hat{s}_i}\
e^{-i \psi_0 ( m+kw )}
\frac{\sin \pi \frac{(1+2j')(1+2\hat{\jmath})}{k}}{
\sqrt{\sin \pi \frac{1+2j'}{k}}}
\nonumber\\
\left\{ e^{-r_0 (1-2j)} + (-)^{m-s_4}
e^{r_0 (1-2j)} \right\} \frac{\Gamma (1-2j) \Gamma (1 + \frac{1-2j}{k} )}{
\Gamma(1-j - \frac{m-s_4}{2}-kw ) \Gamma (1-j + \frac{m-s_4}{2}+kw)}.\nonumber \\
\end{eqnarray}
Note that this one-point function contains only poles of the ``bulk''
type associated to infinite volume divergences, and consequently the D3-branes do not couple to the discrete
representations of \slc.
\subsection*{Open string partition function}
We wish to to compute the open string partition function coming from these boundary states.
The overlap between two of those boundary states give the following annulus amplitude in the
closed string channel:
\begin{eqnarray}
Z_{closed}^{D3} (\tilde{\tau} ) = \frac{1}{\sqrt{2k}}\ 
\int d^5 p\ 
\frac{\tilde{q}^{\frac{1}{2} {\bf p}^2}}{\eta (\tilde{\tau})^4}
\ e^{i {\bf p \cdot (\hat y - \hat{y}')}} \nonumber\\
\frac{1}{2} \sum_{a,b=0}^{1}
\sum_{ \{ \nu_i \}  \in (\zi_2 )^4} (-)^a (-)^{b (1+ \sum_{i} \nu_i)}
e^{i\frac{\pi}{2} \sum (a+2\nu_i) (\hat{s}_{i}' - \hat{s}_i )}
\chi^{(a+2\nu_1)} \chi^{(a+2\nu_2)} \nonumber\\
\sum_{2j'=0}^{k-2} \sum_{m \in \zi_{2k}}
\frac{S_{\hat{\jmath}}^{\ j'} S_{\hat{\jmath\,}'}^{\ j'}}{S_{0}^{\   j'}}
\mathcal{C}_{m}^{j\ (a+2\nu_3)} (\tilde{\tau}) \nonumber\\
\int_{0}^{\infty} dP
\sum_{w \in \zi}
\frac{\cosh 2\pi P + (-)^{m-a}}{\sinh 2\pi P \sinh 2\pi P/k}
e^{i  (\varphi_{0}' - \varphi_{0}) ( m+2kw )}\nonumber\\
\left[ \cos 2P (r_0-r_0 ') + (-)^{m-a} \cos  2P(r_0+r_0 ')  \right]
ch_{c}^{(a+2\nu_4)} (P,\frac{m}{2}+kw,\tilde{\tau} ).
\label{cardyD3}
\end{eqnarray}

To obtain a well-defined expression in the open string channel,
we consider the relative partition function with respect to a
reference one with parameters $(r_\ast , r_{\ast}')$, and, to
(possibly) preserve  supersymmetry, we choose the parameters
$(\psi_0 , \psi_{0}')$ of the reference branes to be equal to those
of the original branes. 
After a modular transformation
we get the following annulus amplitude in the open string channel:

\begin{eqnarray}
Z_{open} = \sqrt{-i\tau}\ 
\frac{q^{\frac{1}{2} \left( \frac{\bf \hat{y} - \hat{y}'}{2\pi}
\right)^2}}{ \eta (\tau)^4} \nonumber\\
\sum_{ \{ \upsilon_i \}  \in (\zi_2 )^4}
\frac{1}{2} \sum_{a,b=0}^{1} (-)^b (-)^{a (1+ \sum_{i} \upsilon_i)}
\chi^{(b+2\upsilon_1 + \hat{s}_{1}' - \hat{s}_{1})}
\chi^{(b+2\upsilon_2 + \hat{s}_{2}' - \hat{s}_{2}))}
\int_{0}^{\infty} d P \
\sum_{j=0}^{k-2} \sum_{n \in \zi_{2k}}
N^{j}_{\hat{\jmath} \ \hat{\jmath\,}'} \sum_{W \in \zi} \nonumber\\
\mathcal{C}^{j\ (b+2\upsilon_3  + \hat{s}_{3}' - \hat{s}_{3})}_{n}
\left\{
\frac{\partial
\log \frac{R (P|r_0 ,r_0 ')}{R (P|r_{\ast},r_{\ast}')} }{2i\pi \partial
P}\,
ch_{c}^{(b+2\upsilon_4 + \hat{s}_{4}' - \hat{s}_{4})} \left( P, \frac{n}{2} + kW
- \frac{k(\psi_{0}' - \psi_{0} )}{2\pi} \right)\right.
\qquad \qquad  \nonumber\\
+ \left.
\frac{\partial
\log \frac{R (P|r_0 ,-r_0 ')}{R (P|r_{\ast},-r_{\ast}')} }{2i\pi\partial
P}\,
ch_{c}^{(b+2\upsilon_4 + \hat{s}_{4}' - \hat{s}_{4}+2)} \left( P, \frac{n}{2} +
k(W+\nicefrac{1}{2}) - \frac{k(\psi_{0}' - \psi_{0} )}{2\pi} \right)
\right\}\nonumber \\ \label{openparttube}
\end{eqnarray}
using the modular transformations of the continuous characters,
and the reflection amplitude computed in~\cite{Ribault:2002ti},
both given in appendix~\ref{modtrans}.
Thus we have constructed the exact one-point function of the D-brane
whose profile in the T-dual NS5-brane picture we have discussed before.
We thought it useful to present the Cardy check for the regular brane
to illustrate the relevant techniques in detail, although it is known to
be satisfied by construction (i.e. by the fact that it is a sum over
branes for which the Cardy computation has been performed).

This channel duality also gives explicitly the open string partition function.
We observe that, to get a supersymmetric spectrum of open strings stretched between the 
two D-branes, one has to impose
the condition:
\begin{equation}
\frac{\psi_{0}' - \psi_{0}}{\pi} + \sum_i (\hat{s}_i ' - \hat{s}_i ) =
0 \mod 4
\end{equation}
such that the angular positions of the D3 branes at infinity in the
cigar directions and their orientations coincide. Note also that the parameter $r_0$, describing
the distance of nearest approach of the D3-branes to the plane of NS5-branes
has no role in the supersymmetry of the open string spectrum.
Indeed D3-branes with different parameters are separated but parallel in
the flat coordinates discussed in detail in section \ref{semiclass}.

The supersymmetric open string spectrum contains both states with integer windings (first term) and half-integer
windings (second term). The latter correspond to open strings stretched between the two
halves of the D1-brane of the cigar. By going back to the conformally flat Cartesian
coordinates~(\ref{cartcoords}) we observe that these open strings are
of arbitrary length. Indeed the
cigar is "unfolded" to the plane $(x^{8},x^{9})$. Because of the curved background generated by the
NS5-branes, they are of finite mass.

%%%%%%%%%%%%%%%%%%%%%%%%%%%%%%%%%%%%%%%%%%%%%%%%%%%%%%%%%%%%%%%%%%5
%%%%%%%%%%%%%%%%%%HW branes%%%%%%%%%%%%%%%%%%%%%%%%%%%%%%%%%%%%%%%

\subsection{Boundary states for the second class of D3-branes}
In section~\ref{semiclass}, we discussed the geometry of a second
D3-brane, see~(\ref{profilehwD3}). We will provide its exact description in this section.
It is constructed by tensoring
D1-branes in the trumpet with D1-branes of the bell.
We will need first to discuss the former.
\subsubsection*{The D1-branes in the trumpet}
The D1-branes of the trumpet --~the vector coset $SL(2,\mathbb{R})/U(1)$~--
are T-dual to D2-branes of the cigar (the axial coset).
The "cut" D2-branes of the supersymmetric
cigar (i.e. with $c<1$, see eq.~(\ref{magD2}))
were constructed in~\cite{Israel:2004jt}, following the
results of~\cite{Ribault:2003ss} for the bosonic case.
These D2-branes carry a magnetic charge localized near the tip
of the cigar, inducing a D0-brane charge. They descend
from the $H_2$ brane in Euclidean AdS$_3$ which is consistent
with a factorization constraint~\cite{Ponsot:2001gt}. However
the open string annulus amplitude for these D2-branes in the
cigar~\cite{Israel:2004jt} contains a D0-like contribution
--~induced by the magnetic flux~--  with  negative
multiplicities, so they are not consistent with the Cardy condition.
Therefore for generic $k$ only the D2-brane
without magnetic field is consistent.\footnote{
Another class of D2-branes --~coming presumably from dS$_2$ branes
of AdS$_3$~-- was proposed in~\cite{Fotopoulos:2004ut}
using the modular bootstrap method. It remains to be shown that
these branes are consistent with factorization.
In the semi-classical limit they correspond presumably to the "uncut"
D2-branes discussed already.} However, as we
show in appendix~\ref{D2integ}, when $k$ is {\it integer} --~and
this is the case in the present setup~-- all the unwanted features
disappear, and one obtains perfectly consistent boundary states,
containing only couplings to the continuous representations.

Let us now translate the result from the cigar CFT (the axial coset)
to the trumpet CFT (the vector coset).
The vector coset is characterized by left and right momenta
$( \nicefrac{\tilde{n}-k\tilde{w}}{2},
\nicefrac{\tilde{n}+k\tilde{w}}{2})$, which are related
to the $SL(2,\mathbb{R})$ quantum numbers as $m+\bar{m}=\tilde{n}$ and
$m-\bar{m}=-k\tilde{w}$. This theory can be obtained, for $k$ integer, by a
$\zi_k$ orbifold of the cigar, followed by a T-duality as has been
shown in~\cite{Israel:2003ry}.\footnote{More precisely this is for the
bosonic theory. In the supersymmetric case, we do
a $\zi_k \times \zi_2$ orbifold.}
In this case $\tilde{n}/k$ is identified
with the fractional winding of the orbifold of the cigar and
$k\tilde{w}$ with the momentum modes allowed by the $\zi_k$
projection. From these D2-branes covering
the cigar we will obtain by T-duality D1-branes of the trumpet of equation:
\begin{equation}
\cosh r \sin (\psi - \psi_0) = \sin \sigma
\end{equation}
thus reaching the singularity $r=0$. By this construction it is clear that the
zero-mode $\psi_0$ is quantized as $\psi_0 = \sigma +\nicefrac{2\pi \hat{p}}{k}$. From
the semi-classical point of view, it is related to the issue vortex singularities
for the D2-branes reaching the tip of the cigar discussed in sect.~\ref{cosetgeometries}.
Thus starting from the one-point function for the "cut" D2-brane in the cigar given in appendix~\ref{D2integ},
we obtain the following one-point function for the D1-branes of the trumpet reaching
the singularity:\footnote{We consider a type 0B-like theory with diagonal boundary conditions for the 
worldsheet fermions.} 
\begin{eqnarray}
\langle \Phi^{j,(s)}_{\tilde{n}\, \tilde{w}} \rangle_{\hat{s},\hat{n}} =
\frac{\nu_{k}^{1/2-j}}{4\pi k \sqrt{2}} \ 
\delta_{\tilde{w},0} \ \delta_{s,\hat{s}} \
e^{-i\pi \frac{s\hat{s}}{2}}  e^{\frac{i\pi \hat{n} \tilde{n}}{k}} \nonumber\\
 \Gamma \left( 2-2j  \right)
\Gamma \left( \frac{1-2j}{k}\right)
\frac{\Gamma \left( j+\frac{\tilde n -s}{2}\right)}{\Gamma \left( 1-j+\frac{\tilde n -s}{2}\right)}
   \left[ e^{i \sigma(1-2j)}  + (-)^{\tilde{n}-s}
      e^{-i \sigma(1-2j)}  \right]
      \label{boundtrump}
\end{eqnarray}
This one-point function contains only couplings to the
continuous representations.
Indeed, we find the following annulus amplitude in the closed string
channel:
\begin{align}
Z^{D2}_{\sigma \sigma'} (-1/\tau, \nu/\tau )
&=  \frac{1}{4k} \int dP \sum_{\tilde{n} \in \zi} \sum_{s \in \zi_4}\
e^{-i\pi \frac{s(\hat{s}' - \hat{s})}{2}} \ e^{\frac{i\pi (\hat{n}' - \hat{n}) \tilde{n}}{k}}\
\nonumber \\
\hskip-2.6cm &
 \frac{  \cosh 2P(\sigma+\sigma') +(-)^{\tilde{n}-s} \cosh 2P(\sigma-\sigma')}{
 \sinh 2\pi P \sin 2\pi P/k}\
 \ ch_{c}^{(s)} \left(j,\frac{\tilde{n}}{2};-1/\tau, \nu/\tau \right)
\label{cardD1trump}
\end{align}
Thus in the open string channel we find the following consistent
result:\footnote{As usual we consider a {\it relative} partition function
to get rid of the universal infrared divergence associated to the infinite volume.}
\begin{eqnarray}
Z^{D1_T}_{open} (\tau) =
\int_{0}^{\infty} dP' \sum_w  \left[
\frac{\partial
\log \frac{R (P|i\sigma ,i\sigma  ')}{R (P|i\sigma_{\ast},i\sigma_{\ast}')} }{2i\pi \partial
P}\ ch_{c}^{(\hat{s}' - \hat{s})} (P,\hat{n}'-\hat{n}+kw ; \tau ) \right. \\
+ \frac{\partial
\log \frac{R (P|i\sigma ,-i\sigma  ')}{R (P|i\sigma_{\ast},-i\sigma_{\ast}')} }{2i\pi \partial
P}\ ch_{c}^{(2+\hat{s}' - \hat{s})} (P,\hat{n}'-\hat{n}+k(w+1/2) ; \tau )
\label{opentrumpet}
\end{eqnarray}
checking the normalization of the one-point function.
Thus we have a family of D1-branes in the vector coset $SL(2,\mathbb{R})$ consistent
both with factorization and the modular bootstrap. Note that they give almost the
same open string partition function as the
D1-branes of the cigar, except that the parameter of the D-brane entering into the
reflection amplitudes is imaginary.

At least for parallel D1-branes ($\hat{n}' = \hat{n}$),
we can identify the terms with integer windings $w$ in eq.~(\ref{opentrumpet}) as
corresponding to open strings with both ends on the same half D1-brane, and the terms
with half-integer windings as corresponding to open strings with one end on each
half D1-brane. The two bracketed terms in~(\ref{boundtrump}) correspond
to the two different halves. In opposition to the D1-branes in the cigar, these two pieces should
be thought as independent D-branes since their worldvolume are not connected.

\subsubsection*{The one-point function of the D3-brane}
We would like now to construct the boundary state for the second class of D3-branes orthogonal to the
NS5-branes discussed in section~\ref{semiclassD3two}.
This D3-brane is obtained from the alternative
T-dual geometry~(\ref{tphidual}). It is made of a "cut" D1-brane of the
trumpet and a D1-brane of the bell. After T-duality we obtain the complicated
hypersurface given by eq.~(\ref{profilehwD3}), with $c=\sin \sigma$. The parameter
$\sigma$ is quantized as $\sigma \in \pi \mathbb{Z}/k$ as we shall discuss again
in the next section. The shape of the D3-brane is depicted in fig.~\ref{hwD3fig} for the
simplest case $\sigma=0$ (more precisely we have two disconnected
antipodal copies of this D3-brane).
This D3-brane connects a finite-size D1-brane inside the
ring of fivebranes with a "conical" D3-brane at infinity, see
eq.~(\ref{profilehwD3}). In particular the intersection with the plane $x^8=x^9=0$ is 
made of two straight half-lines. 
\FIGURE{
\epsfig{figure=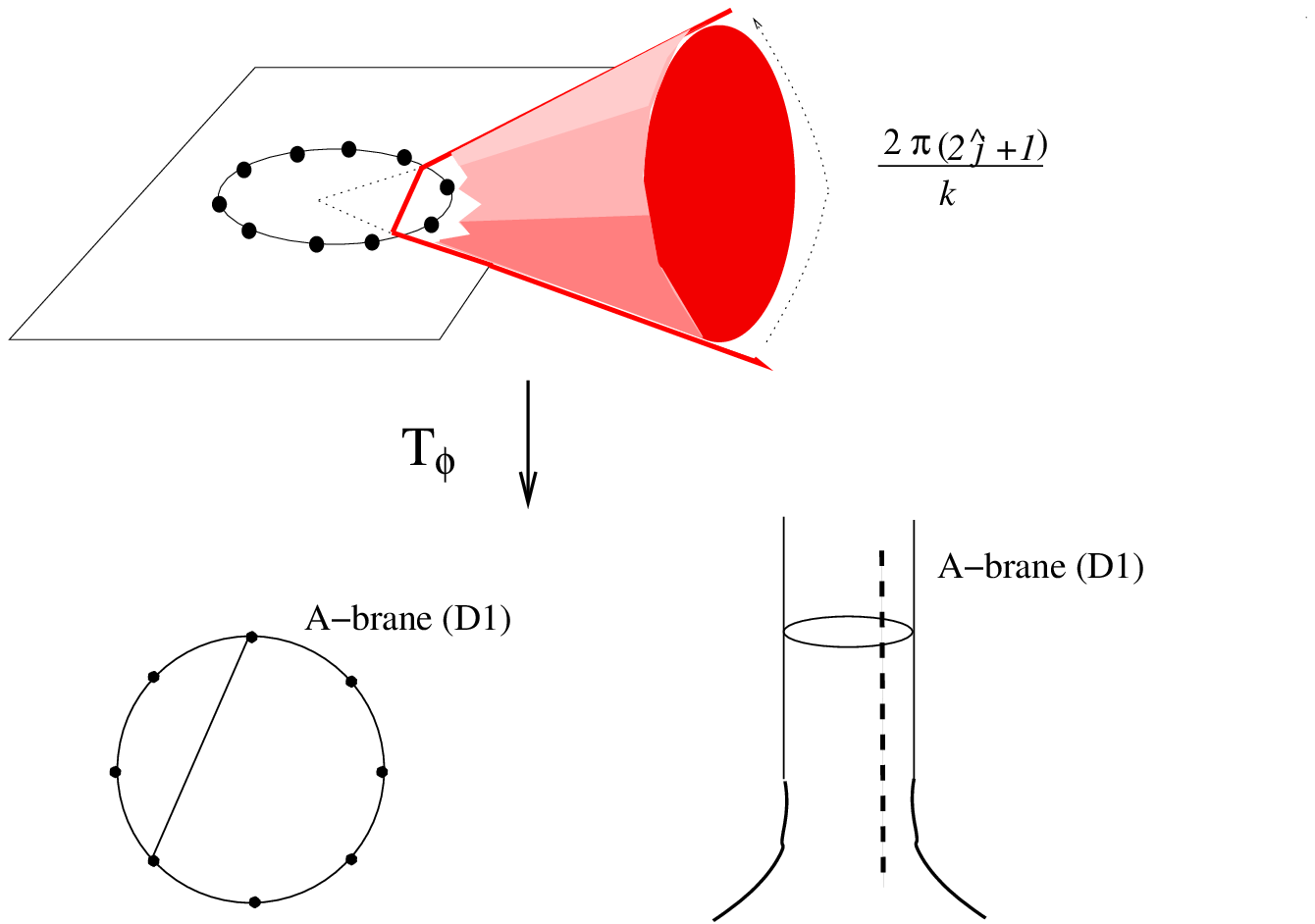, width=110mm}
\caption{Geometry of the D3-branes and
its T-dual, in the case $\sigma=0$. For large $r$ the
D3-brane asymptotes a  cone, corresponding to a symmetric D2-brane of $SU(2)$ and
a Neumann D-brane (FZZ) of the linear dilaton direction. The intersection
of the D3-brane with the plane $x^8=x^9=0$ is made of a D1-brane of the bell --~for the
interior of the ring of fivebranes~-- and two (half) D1-branes of the trumpet --~for the exterior of the ring.}
\label{hwD3fig}}

We are now ready to write the complete boundary states starting from the
boundary state for the D1-brane of the trumpet discussed
above. Following the same logic as the previous example, we gather the
contributions from the various factors and obtain after the $\zi_k$ orbifold the
following one-point function for the Hanany-Witten D3-brane:\footnote{A phase $(-)^{kw}$ has been
added for consistency with the Cardy condition for the overlaps with suspended D1-branes, see below.}
\begin{eqnarray}
\langle \ V_{j'm\bar{m}j w_L w_R,{\bf p}}^{(s_i)\
 (\bar s_i)}\
 \rangle^{}_{\hat{\jmath},\psi_0,\hat{s}_i,{\bf \hat{y}}}
= \frac{\nu_{k}^{ \frac{1}{2}-j}}{\pi k \sqrt{2} }\ (-)^{kw}
 \delta_{m,\bar{m}} \delta_{m+2kw_L,\bar{m}-2kw_R} \
\delta_{s_1,\bar{s_1}} \delta_{s_2, -\bar{s_2}} \delta_{s_3,\bar{s_3}}
\delta_{s_4,-\bar{s_4}} \, \delta (p^5 )
\nonumber\\
e^{i \sum_{i=0}^{4} p^i \hat{y}^i}
e^{i\frac{\pi}{2} \sum_{i} s_i \hat{s}_i}\
\frac{\sin \pi \frac{(1+2j')(1+2\hat{\jmath})}{k}}{
\sqrt{\sin \pi \frac{1+2j'}{k}}}\ e^{\frac{i\pi (\hat{n}' - \hat{n}) \tilde{n}}{k}}\nonumber\\
 \Gamma \left( 2-2j  \right)
\Gamma \left( \frac{1-2j}{k}\right) 
\frac{\Gamma \left( j+\frac{\tilde n -s_4}{2}+kw\right)}{\Gamma \left( 1-j+\frac{\tilde n -s_4}{2}+kw\right)}
   \left[ e^{i \sigma(1-2j)}  + (-)^{\tilde{n}-s_4}
      e^{-i \sigma(1-2j)}  \right] \nonumber\\
\end{eqnarray}
This one-point function contains only couplings to the continuous
representations of the $SL(2,\mathbb{R})/U(1)$ theory. The computation
of the annulus amplitude is quite similar to the previous D3-brane. Indeed
as we saw the only difference being the parameters of the reflection
amplitude, and the fact that the $U(1)$ label is quantized to
$\nicefrac{2\pi \hat{n}}{k}$. This last point is related to the fact
that, while the $SO(2)$ rotational isometry is preserved in the $(x^{8},x^{9})$
plane, it is broken to $\zi_k$ in the $(x^{6},x^{7})$ plane by the ring of five-branes.

So, we can skip the details of the computation of the overlap of boundary states,
and give the result for the open string partition function:
\begin{eqnarray}
Z_{open} = \sqrt{-i\tau} \  
\frac{q^{\frac{1}{2} \left( \frac{\bf \hat{y} - \hat{y}'}{2\pi}
\right)^2}}{ \eta (\tau)^4} \nonumber\\
\sum_{ \{ \upsilon_i \}  \in (\zi_2 )^4}
\frac{1}{2} \sum_{a,b=0}^{1} (-)^b (-)^{a (1+ \sum_{i} \upsilon_i)}
\chi^{(b+2\upsilon_1 + \hat{s}_{1}' - \hat{s}_{1})}
\chi^{(b+2\upsilon_2 + \hat{s}_{2}' - \hat{s}_{2}))}
\int_{0}^{\infty} d P \
\sum_{2j=0}^{k-2} \sum_{n \in \zi_{2k}}
N^{j}_{\hat{\jmath} \ \hat{\jmath\,}'} \sum_{W \in \zi} \nonumber\\
\mathcal{C}^{j\ (b+2\upsilon_3  + \hat{s}_{3}' - \hat{s}_{3})}_{n}
\left\{
\, \frac{\partial
\log \frac{R (P|i\sigma ,i\sigma  ')}{R (P|i\sigma_{\ast},i\sigma_{\ast}')} }{2i\pi \partial
P}\ 
ch_{c}^{(b+2\upsilon_4 + \hat{s}_{4}' - \hat{s}_{4})} \left( P, \frac{n}{2} + kW + \hat{n} - \hat{n}'
\right)\right.
 \nonumber\\
+ \left.
\frac{\partial
\log \frac{R (P|i\sigma ,-i\sigma  ')}{R (P|i\sigma_{\ast},-i\sigma_{\ast}')} }{2i\pi \partial
P}\ 
ch_{c}^{(b+2\upsilon_4 + 2+\hat{s}_{4}' - \hat{s}_{4})} \left( P, \frac{n}{2} +
k(W+\nicefrac{1}{2}) +\hat{n} - \hat{n}'  \right)
\right\}\nonumber \\
\label{specD3HW}
\end{eqnarray}
The supersymmetry conditions here reads
\begin{equation}
\sum_{i} (\hat{s}_{i}' - \hat{s}_i ) - \frac{2(\hat{n}'-\hat{n})}{k}
= 0 \ \text{mod} \ 4.
\end{equation}
Again the sector with half-integer windings correspond to open strings stretched
between the two halves of the D-branes, which have disconnected
worldvolumes.

The similarity of the result between both types of D3-branes
is expected since both branes asymptote to isomorphic
D3-branes of $SU(2) \times \mathbb{R}_Q$. Indeed in both cases we get a
D2-brane of $SU(2)$ times a Neumann brane of the linear dilaton, the
only difference being the position of the S$^2$-brane in the three-sphere,
so only the reflection amplitudes --~which depends on the physics near the
NS5-branes~-- are different.
Due to their different positions they cannot appear simultaneously in a
supersymmetric configuration. One can check indeed that the spectrum 
of open strings stretched between two D3-branes of different types is not 
supersymmetric.

\subsection{Boundary states for the D-rays}
Now we wish to consider the semi-infinite D1-branes ending on NS5-branes.
As we saw in the semi-classical discussion they are associated
to D1-branes in the trumpet, reaching the singularity $r=0$. In the
present case it means that they reach the ring of five-branes. As
for the D2-branes, the $U(1)$ label of the D1-brane
is quantized as $\psi_0 =\sigma+ \nicefrac{2\pi \hat{p}}{k}$. Moreover, as already stated, if we assume
also that the parameter $\sigma$ is quantized as $\sigma  \in \pi \zi/k$,
we find that the D1-branes have to end on the NS5-branes, see
fig.~\ref{figDrays}.
\FIGURE{
\epsfig{figure=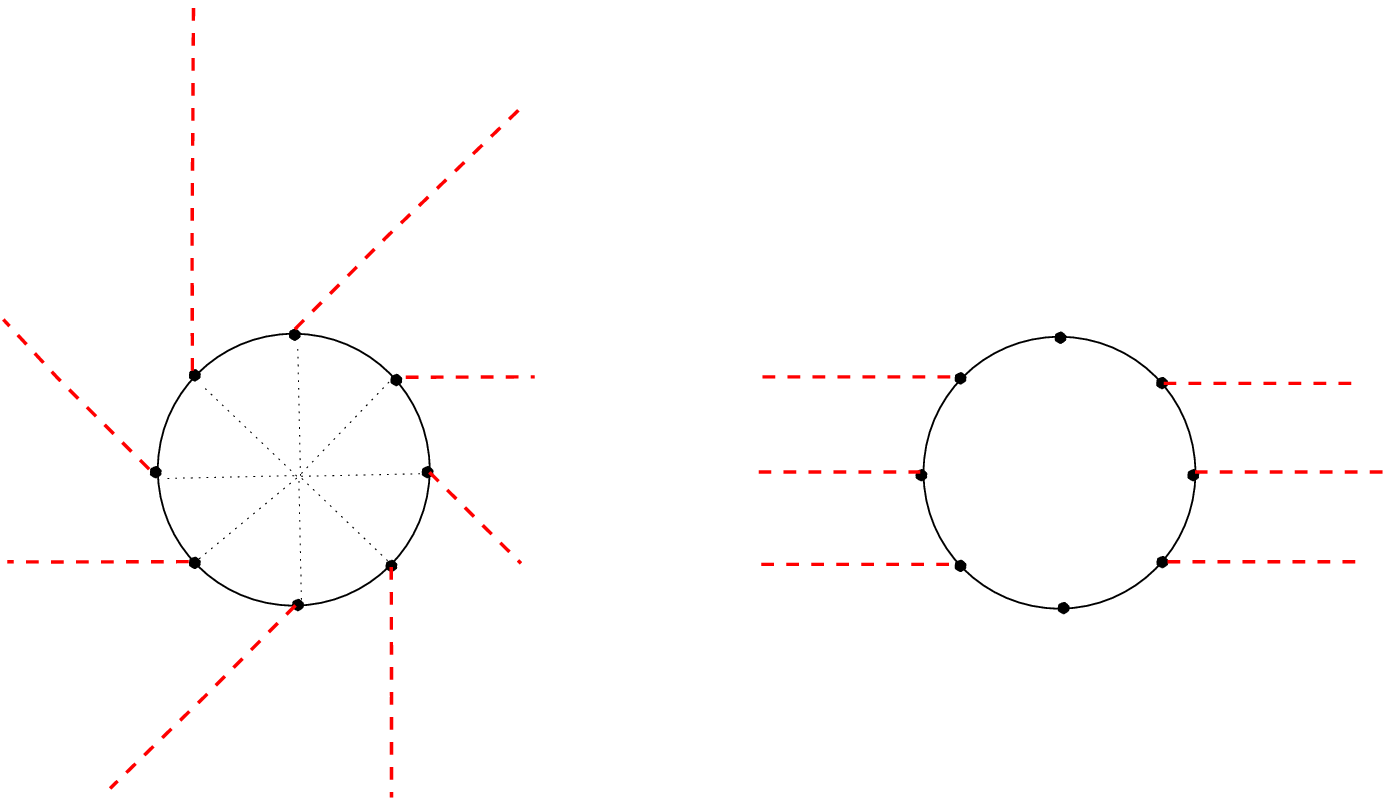, width=120mm}
\caption{Configurations of D-rays, for fixed $\sigma$ (left) and
supersymmetric (right).}
\label{figDrays}}
The D1-branes ending on the NS5-branes are just the special case $\hat{\jmath}=0$ of
the previous D3-branes. So they have the following one-point function:
\begin{eqnarray}
\langle \ V_{j'm\bar{m}j w_L w_R,{\bf p}}^{(s_i)\
 (\bar s_i)}\
 \rangle_{\hat{\jmath},\psi_0,\hat{s}_i,{\bf \hat{y}}}
= \frac{\nu_{k}^{ \frac{1}{2}-j}}{\pi k \sqrt{2}} \ (-)^{kw} \ 
 \delta_{m,\bar{m}} \delta_{m+2kw_L,\bar{m}-2kw_R} \
\delta_{s_1,\bar{s_1}} \delta_{s_2, -\bar{s_2}} \delta_{s_3,\bar{s_3}}
\delta_{s_4,-\bar{s_4}} \, \delta (p^5 )
\nonumber\\
e^{i \sum_{i=0}^{4} p^i \hat{y}^i}
e^{i\frac{\pi}{2} \sum_{i} s_i \hat{s}_i}\
\sqrt{\sin \frac{\pi(1+2j')}{k}}\ e^{\frac{i\pi (\hat{n}' - \hat{n}) \tilde{n}}{k}} \nonumber \\
\Gamma \left( 2-2j  \right)
\Gamma \left( \frac{1-2j}{k}\right) 
\frac{\Gamma \left( j+\frac{\tilde n -s_4}{2}+kw\right)}{\Gamma \left( 1-j+\frac{\tilde n -s_4}{2}+kw\right)}
   \left[ e^{i \sigma(1-2j)}  + (-)^{\tilde{n}-s_4}
      e^{-i \sigma(1-2j)}  \right] \nonumber\\
\end{eqnarray}
and the open string spectrum is the same as~(\ref{specD3HW}) by replacing
$N^{j}_{\hat{\jmath} \ \hat{\jmath\,}'} \to \delta_{j,0}$. Thus it consists only in
continuous representations of \slc, with a non-trivial density of states. Asymptotically
these D-branes are D0-branes of $SU(2)$ times Neumann D-branes of the linear dilaton.\footnote{As already mentioned 
we obtain two antipodal D0-branes.}
In the semi-classical analysis there was another class of these D1-branes, avoiding the ring of NS5-branes.
They are obtained from descent of dS$_2$ D-branes of Euclidean AdS$_3$ and a candidate 
boundary state was found using the
modular bootstrap method in~\cite{Fotopoulos:2004ut}. However up to now there
is no check of factorization constraints for those D-branes.

\subsubsection{The anomalous creation of branes}
These D3-branes of the second type are the natural D3-branes to consider the Hanany-Witten effect
of anomalous creation
of D1-branes. Indeed it is clear from the previous analysis --~see fig.~\ref{hwD3fig} and eq.~(\ref{profilehwD3})~-- 
that such D3-branes of label $\hat{\jmath}$ trap $(2\hat{\jmath}+1)$ NS5-branes inside their
worldvolume.\footnote{There is an issue of whether the D3-branes intersect or not NS5-branes
when they cross the ring of fivebranes; however we believe that it is not the case, since
the case $\hat{\jmath}=0$ corresponds to a D-ray brane discussed above and has to end on an NS5-brane.
Thus the D3-branes for $\hat{\jmath} \neq 0$ are presumably also centered around the position of an
NS5-brane and thus their worldvolume do not intersect fivebranes.} It has been argued in~\cite{Hanany:1996ie}
from the study of the theory living on the worldvolume of the D3-branes that in such a configuration
D1-branes are created whenever the D3-brane crosses a stack of $n$ NS5-branes and traps them. Because the
D3-branes are repelled by the NS5-branes a long tube is formed when the D3-branes are pushed far away,
and corresponds effectively to a stack of $n$ D1-branes with one end on the NS5-stack and the
other end on the D3-brane. 
\FIGURE{\centering
\epsfig{figure=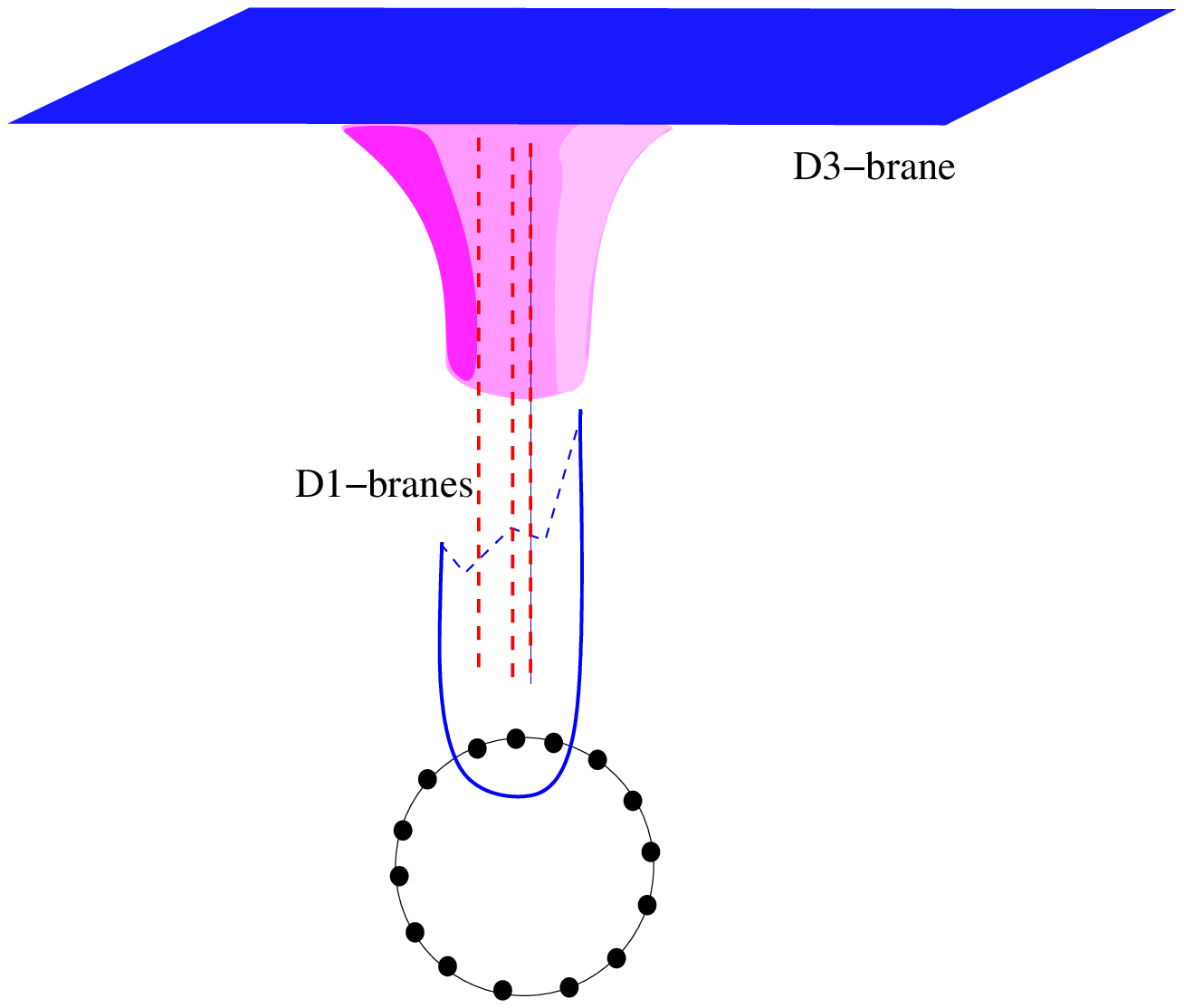, width=120mm}
\caption{Hanany-Witten configuration with creation of D1-branes.}
\label{hwtube}}
As was shown in~\cite{Pelc:2000kb}, in the supergravity approximation, for coincident NS5-branes it gives
simply a cone in the near-horizon limit. In the setup of the ring of NS5-branes that we consider
we can have a handle on the worldsheet theory which is weakly coupled, and the
distribution of fivebranes is "resolved". It allows to show explicitly the link between the
label $\hat{\jmath}$ of the D2-brane of $SU(2)$ and the number of NS5-branes trapped by
the D3-brane (because we know
exactly the shape of the D3-branes around the NS5-branes),
and to construct the exact boundary state for this configuration.
As it is well known~\cite{Bachas:2000ik,Alekseev:2000fd} a D2-brane of SU(2) of label
$\hat{\jmath}$ can be constructed as a bound state of $(2\hat{\jmath}+1)$ D0-branes. Translated to
our setup it means that we can construct our D3-brane trapping $(2\hat{\jmath}+1)$
N55-branes as a bound state of $(2\hat{\jmath}+1)$ D-rays.
This fits perfectly with the
Hanany-Witten picture.
Note that in the near-horizon limit we see only the tip of the
tube --~made with the D1-branes that have been created~-- and so it is natural that the D3-brane we consider
in the CFT analysis
is made only of a bound state of D1-branes. The D3-brane itself lives in the asymptotically flat region
that is absent from the double scaling limit considered in this paper, see fig.~\ref{hwtube}. It would
be interesting to find explicitly the supersymmetric solution of the DBI action, in the
full asymptotically flat spacetime, corresponding to this D3-brane geometry.

For the first class of D3-branes (the "cylindrical" D-branes of section~\ref{boundcyld3})
the story is actually very different. This configuration should be
thought --~after adding the asymptotic flat region ~-- as a configuration of two parallel D3-branes
widely separated, which are on both sides of the NS5-branes ring. They are connected by
a tube of D1-strings going through (for $r_0 = 0$) the ring of fivebranes, see fig.~\ref{ringtube}.
\FIGURE{\centering
\epsfig{figure=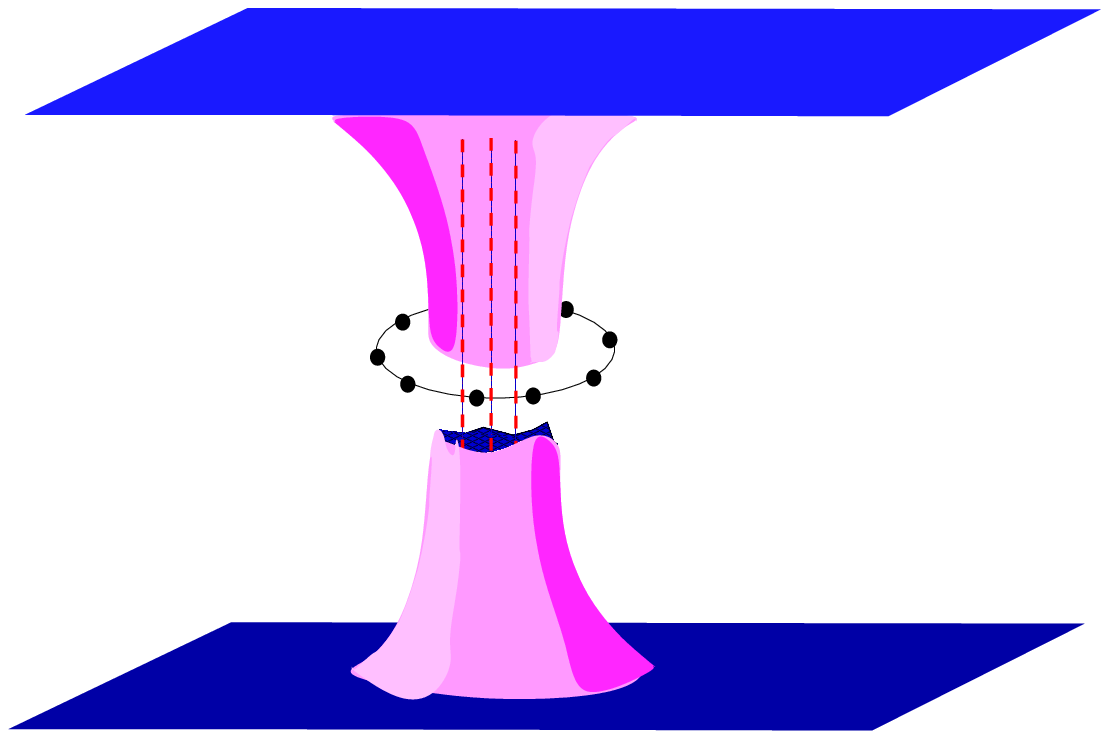, width=120mm}
\caption{Configuration of two D3-branes linked by a tube of D1-branes going through the ring of fivebranes.}
\label{ringtube}}
In the double scaling limit considered here we focus on the portion of the tube near the ring of fivebranes.
However in this case we can move the D3-brane continuously out of the ring of five-brane in the plane $(x^{8},x^{9})$
by changing the parameter $r_0$ of the D3-brane. Indeed as we saw in the section of the D3-brane in this plane is
simply a straight line, whose minimal distance to the stack of fivebranes is $\rho_0 \sinh r_0$. At the level
of the exact CFT nothing dramatic happens for $r_0 = 0$, i.e. when the tube is indeed lassoed by the fivebranes ring.
It corresponds simply to a particular value of the density of states in the open string partition
function of eq.~(\ref{openparttube}). It would be interesting to understand
such a (Euclidean) wormhole-like configuration
from the point of view of the worldvolume gauge theory on the D3-branes.

%%%%%%%%%%%%%%%%%%%%%%%%%%%%%%%%%%%%%%%%%%%%%%%%%%%%%%%%%%%%%%%%%%%%%%%%%%%%%%%%%%%%%%%%%
%%%%%%%%%%%%%%%%%%%%%%%%%%%%%%%%%%%%HYPERS%%%%%%%%%%%%%%%%%%%%%%%%%%%%%%%%%%%%%%%%%%%%%%%
%%%%%%%%%%%%%%%%%%%%%%%%%%%%%%%%%%%%%%%%%%%%%%%%%%%%%%%%%%%%%%%%%%%%%%%%%%%%%%%%%%%%%%%%%

\subsubsection{Hypermultiplets on D3-branes associated to NS5-branes}
There is an interesting issue that has been addressed in~\cite{Hanany:1996ie}
in the low-energy limit.
Configurations of D3-branes ending on NS5-branes were studied (it is simply a
T-dual of our setup of D1-branes along the directions $x^{1,2}$)
and it was argued that when two stacks of D3-branes ending on two sides
of a NS5-brane become "aligned", a new massless hypermultiplet appears in the low energy effective
action on the D-branes. We would like to study such a phenomenon in
 our setup, see
fig.~\ref{hyperD3}.\FIGURE{\centering \phantom{aaaaaaaaaaaaaaaaaaaaaaaa}
\epsfig{figure=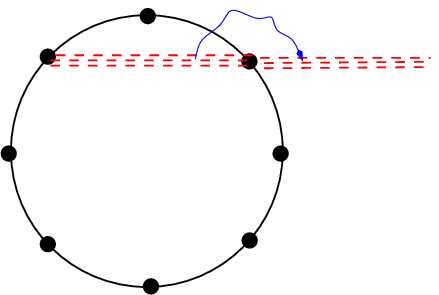, width=100mm}
\caption{Configuration of two D3-brane stacks ending on both sides of an NS5-brane, with an open string
stretched between them.}
\label{hyperD3}}
The stacks of D3-branes are parallel in the $(x^{6},x^7)$ plane, 
coincident in the plane $(x^8,x^9)$ and 
at a distance $\delta$ in the directions $x^{3,4,5}$ longitudinal to the NS5-branes.
We are interested in computing the
spectrum of open strings stretched between a suspended D3-brane of parameters $(\hat{\jmath},\hat{n}=0,\hat{s}=0)$
and a semi-infinite D3-brane of parameters
$(\sigma,\hat{n}=0,\hat{s}=0)$. This D-brane setup
is supersymmetric. The value of $\sigma$ that we will choose is expected to correspond to aligned
D3-branes, thus giving a massless hypermultiplet.

Let us consider for convenience the configuration of D1-branes obtained by T-duality
in the directions $x^{1,2}$. The overlap of boundary states for a suspended D1-brane of parameter $\hat{\jmath}$ and
a semi-infinite D1-brane of parameter $\sigma$ give, in the closed string channel:
\begin{eqnarray}
Z_{closed}^{\textsc{hyper}} (\tilde{\tau} ) =  \sqrt{\frac{2}{k}}
\int d^5 p \ 
\frac{\tilde{q}^{\frac{1}{2} {\bf p}^2}}{\eta (\tilde{\tau})^4}
\ e^{i {\bf p \cdot (\hat y - \hat{y}')}} \nonumber\\
\frac{1}{2} \sum_{a,b=0}^{1}
\sum_{ \{ \nu_i \}  \in (\zi_2 )^4} (-)^a (-)^{b (1+ \sum_{i} \nu_i)}
\chi^{(a+2\nu_1)} \chi^{(a+2\nu_2)}
\sum_{2j'=0}^{k-2} \sum_{m \in \zi_{2k}}
S_{\hat{\jmath}}^{\ j'} \mathcal{C}_{m}^{j'\ (a+2\nu_3)} (\tilde{\tau}) \nonumber\\
\int dj \ \frac{1}{\sin \pi (j-\frac{m-a}{2})}
\sum_{w \in \zi}
\left[  e^{i \sigma (1-2j)}+(-)^{m-a} e^{-i \sigma (1-2j)}
\right]
\
ch^{(a+2\nu_4)} (j,\frac{m}{2}+kw,\tilde{\tau} ).\nonumber \\
\label{hyperclosed}
\end{eqnarray}
and this expression contains indeed poles for the discrete representations, in 
contrast to
the D2-D2 overlap for integer $k$. The
 overlap does not contain "`bulk" poles at
zero radial momentum $j=1/2$ corresponding to infinite volume divergences, which is natural since
at the end we will find localized open string states.
The integral $\int dj$ contains both and
integral over the continuous representations
$j=\nicefrac{1}{2}+iP$ and contributions from the discrete representation whenever there is
a pole with non-zero residue. The poles in this expression arises whenever $j-(m-a)/2 \in \mathbb{Z}$,
i.e. we have a pole for each discrete representation appearing in the closed string spectrum.
In the following we will assume that the parameter $\sigma$ is quantized as:
\begin{equation}
\sigma = \frac{\pi (2J-1)}{k} \ , \ \ 2J \in \mathbb{N} \ , \ \ \nicefrac{1}{2} < J < \nicefrac{k+1}{2}
\end{equation}
As we argued earlier for the D1-branes of the trumpet, each of  the bracketed terms in eq.~(\ref{hyperclosed})
correspond to one semi-infinite D1-brane. So let us consider only the first term.
Then, using the modular transformation of the extended discrete characters given
in appendix~\ref{modtrans}, we find that the contribution from the poles of this expression 
(together with the continuous couplings) comes from the
following discrete spectrum in the open string channel:
\begin{eqnarray*}
Z^\textsc{hyper}_{\text{open}} = \sqrt{-i\tau} \  \frac{q^{\frac{1}{2} \left( \frac{\delta}{2\pi}
\right)^2}}{\eta (\tau)^4}
\sum_{ \{ \upsilon_i \}  \in (\zi_2 )^4}
\frac{1}{2} \sum_{a,b=0}^{1} (-)^b (-)^{a (1+ \sum_{i} \upsilon_i)}
\chi^{(b+2\upsilon_1) } \chi^{(b+2\upsilon_2 ) } \nonumber\\
\sum_{n \in \zi_k+\frac{b}{2}}
\mathcal{C}^{\hat{\jmath} \ (b+2\upsilon_3)}_{2n}  (\tau)\
 Ch_{d}^{(b+2\upsilon_4)} (J,n-J;\tau )
\end{eqnarray*}
Let us assume that the D1-branes are coincident in the directions $x^{1,2,3,4,5}$.
Then --~using a similar analysis as for the suspended D1-branes~--
this spectrum contains one massless hypermultiplet
if $J=\hat{\jmath}+1$, which is perfectly consistent with our
geometrical picture. 
Indeed the massless hypermultiplet corresponds to the open string sectors
stretched between the aligned
 D1-branes of the bell and the trumpet. In the present case
we have chosen $\sigma=\nicefrac{\pi}{2}-\theta_0$ and the two D1-branes indeed end 
on the same NS5-brane according to the geometrical picture given in the previous sections 
for the various D-branes.\footnote{ 
The second term in~(\ref{hyperclosed}) gives a similar partition function  
with the replacement: 
$Ch_{d}^{(b+2\upsilon_4)} (J,n-J;\tau ) \to Ch_{d}^{(b+2+2\upsilon_4)} (1-J,n+k-1+J;\tau )$. 
It corresponds to the hypermultiplet attached to the other half-infinite D1-brane.}

We have then proven that when the D3-branes ending on both sides of the NS5-brane are coincident, a new
massless hypermultiplet in the bifundamental appears. This constitutes
 a worldsheet CFT demonstration
 of 
the property understood in~\cite{Hanany:1996ie}.

%%%%%%%%%%%%%%%%%%%%%%%%%%%%%%%%%%%%%%%%%%%%%%%%%%%%%%%%%%%%%%%%%%
%%%%%%%%%%%%%%%%%%NON-BPS%%%%%%%%%%%%%%%%%%%%%%%%%%%%%%%%%%%%%%%%%
%%%%%%%%%%%%%%%%%%%%%%%%%%%%%%%%%%%%%%%%%%%%%%%%%%%%%%%%%%%%%%%%%%%%
\section{Non-BPS D-branes}
In this section we comment briefly on the non-BPS D-branes in this
background. Some of the relevant boundary states have already be given
in~\cite{Eguchi:2004ik} but we would like to focus on geometrical
aspects, and discuss in some more detail the non-compact branes. As we
have seen in the previous sections, the BPS D-branes in the NS5-brane
background are, for the non-trivial \slc and $SU(2)/U(1)$ factors,
either of A-A type or of B-B type in the T-dual backgrounds. In this case the projection
induced by the $\zi_k$ orbifold will translate into a spectrum of
odd-integral N=2 charges in the open string channel, thus
into a space-time supersymmetric spectrum.
Apart from the obvious D-branes of wrong dimensionality
(i.e. odd-dimensional in type IIA and even-dimensional in type IIB)
the non-BPS D-branes are of type A-B or B-A for the CFT
of the transverse directions to the five-branes.

\subsection{Compact D2-branes}
Using the procedure that is by now familiar, we start from D-branes
in the T$_{\psi}$-dual geometry~(\ref{cosetprod}), i.e. the orbifold
of the cigar times the T-bell. We can construct a first non-BPS D-brane
by starting with a D1-brane of the T-bell and a D0-brane of the
cigar CFT (in type IIB). After T-duality, we obtain in type IIA
superstrings a non-BPS D2-brane in the NS5-brane
background of embedding equations and gauge field:
\begin{eqnarray}
r&=&0\ , \ \ \theta < \theta_0 \nonumber \\
F &=& \frac{k}{2\pi} \ \frac{\tan \theta \cos \theta_0}{\sqrt{\cos^2 \theta -
\cos^2 \theta_0}} d\theta \wedge d \psi
\end{eqnarray}
They correspond to D2-branes of finite size, living in the plane
$x^8=x^9 = 0$ where the fivebrane sit. Their worldvolume are
discs defined by $\theta <\theta_0$, contained inside the ring of fivebranes.

Their one-point function is:
\begin{eqnarray}
\langle \ V_{j'm\bar{m}j w_L w_R,{\bf p}}^{(s_i)\
 (\bar s_i)}\
 \rangle_{\hat{\jmath},\hat{n},\hat{s}_i,{\bf \hat{y}}}
=  \sqrt{\frac{2}{k}} \ \nu_{k}^{\frac{1}{2}-j} \
\delta_{m,-\bar{m}}
\delta_{m+ 2kw_L,\bar{m}-2k w_R} \
\delta_{s_1,\bar{s_1}} \delta_{s_2, -\bar{s_2}} \delta_{s_3,-\bar{s_3}}
\delta_{s_4,-\bar{s_4}} \, \delta (p^5 )
\nonumber\\ e^{i \sum_{i=0}^{4} p^i \hat{y}^i}
e^{i\frac{\pi}{2} \sum_{i} s_i \hat{s}_i }\
\ \frac{\sin \pi \frac{(1+2j')(1+2\hat{\jmath})}{k}}{
\sqrt{\sin \pi \frac{1+2j'}{k}}} \ \frac{\Gamma \left( j -\frac{s_4}{2}+ kw
\right) \Gamma \left( j +\frac{s_4}{2}- kw \right)}{\Gamma (2j-1)
\Gamma ( 1 - \frac{1-2j}{k} )}.\nonumber\\
\end{eqnarray}
and are normalized with a factor of $\sqrt{2k}$ w.r.t. the BPS
suspended D1-branes to satisfy the Cardy condition.
Their overlap give the following non-supersymmetric amplitude in the
open string channel:
\begin{eqnarray*}
Z^\textsc{w}_{\text{open}} = \sqrt{-i\tau}\ \frac{q^{\frac{1}{2} \left( \frac{\bf \hat{y} -
\hat{y}'}{2\pi}
\right)^2}}{\eta (\tau)^4}
\sum_{ \{ \upsilon_i \}  \in (\zi_2 )^4}
\frac{1}{2} \sum_{a,b=0}^{1} (-)^b (-)^{a (1+ \sum_{i} \upsilon_i)}
\chi^{(b+2\upsilon_1 + \hat{s}_{1}' - \hat{s}_{1}
)} \chi^{(b+2\upsilon_2 + \hat{s}_{2}' - \hat{s}_{2})} \nonumber\\
\sum_{2j=0}^{k-2} \sum_{n \in \zi_{2k}}
N^{j}_{\hat{\jmath} \ \hat{\jmath\,}'} \
\mathcal{C}^{j \ (b+2\upsilon_3+
\hat{s}_{3}' - \hat{s}_{3})}_{n}  (\tau)\
\sum_{r \in \zi_{k}+\frac{b}{2}}
Ch_\mathbb{I}^{(b+2\upsilon_4 + \hat{s}_{3}' - \hat{s}_{3})} (r;\tau ),
\end{eqnarray*}
In this open string spectrum the non-integral part of the N=2 charges
of the two coset models are uncorrelated, thus the spectrum is
not supersymmetric and contains tachyons (see also~\cite{Eguchi:2004ik}). This is related to the
instability of the D2-brane of $SU(2)/U(1)$ towards moving to
the boundary of the disc.

\subsection{Non-BPS non-compact D4-branes}
This non-compact D4-brane of type IIA is obtained from the T$_{\psi}$-dual
background (\ref{cosetprod}), out of a D1-brane of the cigar and a D2-brane of the T-bell.
Using the same T-duality techniques as for the BPS D-branes, it gives a non-BPS 
D4-brane with a non-trivial gauge field:
\begin{equation}
A = \frac{k}{2\pi} \left[ \arcsin \left( \frac{\sin \theta_0}{\sin \theta} \right) - \theta_0 \right] 
\left[ d\phi - d \left(  \arcsin   \frac{\sinh r_0}{\sinh r} \right)  \right]
+ \frac{k}{2\pi} \left[ \arcsin  \left( \frac{\sinh r_0}{\sinh r} \right) + \psi_0 \right] d\psi
\end{equation}
whose worldvolume is restricted to the region $r>r_0$ and $\theta > \theta_0$. 
For large $r$, all these D4-branes asymptote symmetry-breaking D3-branes 
of $SU(2)$ (i.e. a B-brane, see~\cite{Maldacena:2001ky})
of gauge field 
\begin{equation}
A = \frac{k}{2\pi} \left[ \arcsin \left( \frac{\sin \theta_0}{\sin \theta} \right) -\theta_0  \right] d\phi
\end{equation}
covering the region $\theta > \theta_0$, times a Neumann D-brane of the linear dilaton.

The one-point function for these D4-branes is quite similar to the
one-point function for cylindrical D3-branes, and is given by:
\begin{eqnarray}
\langle \ V_{j'm\bar{m}\, j w_L w_R,{\bf p}}^{(s_i)\
 (\bar s_i)}\
 \rangle^{D3}_{\hat{\jmath},\theta_0,r_0,\hat{s}_i,{\bf \hat{y}}}
=   \sqrt{\frac{k}{2}}\  \nu_{k}^{\frac{1}{2}-j}   \
\delta_{m,\bar{m}} \delta_{m+2kw_L,-\bar{m}+2kw_R} \
\delta_{s_1,\bar{s_1}} \delta_{s_2, -\bar{s_2}} \delta_{s_3,-\bar{s_3}}
\delta_{s_4,\bar{s_4}} \, \delta (p^5 )
\nonumber\\ e^{i \bf{ p \cdot y}}
e^{i\frac{\pi}{2} \sum_{i} s_i \hat{s}_i}\
e^{-2i k \psi_0 w}
\frac{\sin \pi \frac{(1+2j')(1+2\hat{\jmath})}{k}}{
\sqrt{\sin \pi \frac{1+2j'}{k}}}
\left[e^{-r_0 (1-2j)}+(-)^{s_4} e^{r_0 (1-2j)} \right]   \nonumber\\
 \frac{\Gamma (1-2j) \Gamma (1 + \frac{1-2j}{k} )}{
\Gamma(1-j -kw+\frac{s_4}{2} ) \Gamma (1-j +kw-\frac{s_4}{2})}.
\nonumber \\
\end{eqnarray}
Thus giving in the open string channel the following non-supersymmetric
partition function:
\begin{eqnarray}
Z_{open} = \sqrt{-i\tau} \ 
\frac{q^{\frac{1}{2} \left( \frac{\bf \hat{y} - \hat{y}'}{2\pi}
\right)^2}}{\eta (\tau)^4} \nonumber\\
\sum_{ \{ \upsilon_i \} }
\frac{1}{2} \sum_{a,b=0}^{1} (-)^b (-)^{a (1+ \sum_{i} \upsilon_i)}
\chi^{(b+2\upsilon_1 + \hat{s}_{1}' - \hat{s}_{1})}
\chi^{(b+2\upsilon_2 + \hat{s}_{2}' - \hat{s}_{2}))}
\sum_{j=0}^{k-2} \sum_{n \in \zi_{2k}}
N^{j}_{\hat{\jmath} \ \hat{\jmath\,}'}
\mathcal{C}^{j\ (b+2\upsilon_3  + \hat{s}_{3}' - \hat{s}_{3})}_{n}\nonumber\\
\int_{0}^{\infty} d P  \
\sum_{N \in \zi}
\left\{
\frac{\partial
\log \frac{R (P|r_0 ,r_0 ')}{R (P|r_{\ast},r_{\ast}')} }{2i\pi \partial
P}\,
ch_{c}^{(b+2\upsilon_4 + \hat{s}_{4}' - \hat{s}_{4})} \left( P,  \frac{N}{2}
- \frac{k(\psi_{0}' - \psi_{0} )}{2\pi} \right)\right.
\qquad \nonumber\\
+ \left.
\frac{\partial
\log \frac{R (P|r_0 ,-r_0 ')}{R (P|r_{\ast},-r_{\ast}')} }{2i\pi\partial
P}\,
ch_{c}^{(b+2\upsilon_4 + 2+\hat{s}_{4}' - \hat{s}_{4})} \left( P,
 \frac{N}{2} - \frac{k(\psi_{0}' - \psi_{0} )}{2\pi} \right)
\right\}\nonumber \\ \label{openpartD4nonbps}
\end{eqnarray}

\subsubsection*{Second type of non-BPS D4-branes}
These D4-branes are constructed  from the alternative
T$_{\phi}$-dual background (\ref{tphidual}). Then our D4-brane
is constructed out of a D2-brane of the bell and a D1-brane of the trumpet.
It gives a D4-brane with the following gauge field:
\begin{eqnarray}
A = \frac{k}{2\pi} \left[ \arcsin \left( \frac{\cos \theta_0}{\cos \theta} \right) -\frac{\pi}{2}+\theta_0 \right] 
\left \{ d \left(\arcsin \frac{c}{\cosh r} \right)  - d\psi \right\}\nonumber \\
+ \frac{k}{2\pi} \left[  \arcsin \left( \frac{c}{\cosh r} \right) + \phi_0 - \psi \right] 
d\phi
\end{eqnarray}
which also asymptotes a D3 symmetry-breaking brane of SU(2) covering the region $\theta < \theta_0$.
For the case $c = \sin \sigma$
(the ``cut'' branes), the open string partition function is similar
to the eq.~(\ref{openpartD4nonbps}) with the replacement
$r_o \to i\sigma$.

Besides all these static non-BPS D-branes,  interest was
 raised recently for the time-dependent D-branes in the
NS5-background following~\cite{Kutasov:2004dj} (see also~\cite{Ghodsi:2004wn}).
The works~\cite{Nakayama:2004yx,Chen:2004vw,Nakayama:2004ge} have already
investigated the exact CFT description. It would be interesting to study
these setups further.

%%%%%%%%%%%%%%%%%%%%%%%%%%%%%%%%%%%%%%%%%%%%%%%%%%%%%%%%%%%%%%%%%
%%%%%%%%%%%%%CONCLUSIONS%%%%%%%%%%%%%%%%%%%%%%%%%%%%%%%%%%%%%%%%%
%%%%%%%%%%%%%%%%%%%%%%%%%%%%%%%%%%%%%%%%%%%%%%%%%%%%%%%%%%%%%%%%
\section{Conclusions}
\label{conclusions}
The T$_{\psi,\phi}$-duals to the doubly scaled Neveu-Schwarz five-brane background are
 orbifolds of products of two cosets, see eqs~(\ref{cosetprod},\ref{tphidual}). One coset is equivalent to an
$N=2$ minimal model, while the other corresponds to a non-rational $N=2$
$SL(2,\mathbb{R})/U(1)$ conformal field theory. We used our extended knowledge
of both coset backgrounds to construct branes in the Neveu-Schwarz five-brane
geometry. In particular, since we have a good understanding of both
the semi-classical geometry of branes in the cosets and their exact conformal
field theory description, we were able to construct examples of highly non-trivial
configuration of D3-branes in NS5-brane backgrounds, and their exact boundary
state description. The techniques we used can be extended to other
examples of branes in the doubly scaled NS5-brane background. In this work
we have constructed the D-branes that are expressed simply in
the T-dual geometry as product of D-branes in the cosets \slc and $SU(2)/U(1)$.

We have given a precise analysis of both the full exact boundary states, and
the semi-classical geometry, and we made the links between them manifest.
The subtle issues concerning the quantization of the D-branes
parameters in the \slc and $SU(2)/U(1)$ cosets have a 
nice geometrical interpretation.
We were able to identify a type of D3-brane orthogonal to the NS5-branes giving a
realization of the effect of anomalous creation of D1-branes. 
We showed how this effect translates into the properties of the boundary
conformal field theory. We also find that with another kind of D3-branes --~that can
be also described in the boundary CFT~-- we can construct wormhole-like configuration of
two D3-branes linked by a tube of D1-branes going through the ring of NS5-branes. It would
be very interesting to understand this configuration from the point of view of the field
theory living on the D3-branes.

We also analyzed in detail the D-branes stretching between
NS5-branes. We constructed the D1-branes that are the W-bosons of the type IIB
little string theory (which are at fixed mass in the double scaling limit)
These can be viewed as fractional branes in the T-dual orbifold singularity.
Our boundary state construction, and the analysis of the open string annulus
amplitude allows for a precise verification of the open string spectrum on
these branes, from first principles.
We verified the massless spectrum on branes between NS5-branes \cite{Hanany:1996ie},
as well as the appearance of new massless hypermultiplets when D-branes end on the
same point on the worldvolume of NS5-branes from different sides. In our exact
boundary CFT analysis we constructed the full massive tower of open strings
that completes the spectrum at high energies.

Making use of the fact that the NS5-brane backreacts when we attach
D4-branes and that the backreaction
encodes the beta-function of $N=2$ super Yang-Mills theory, we
demonstrated
 that the D4-brane one-point function (which can be argued to induce the
closed string backreaction) also encodes the beta-function. We did a
qualitative computation to confirm this claim.

To obtain further non-trivial information on Little String Theory, it would be useful
to obtain the open string field theory on the D1-branes stretching between
NS5-branes. The spectrum is known, but we would need to determine the
interactions between these open strings, which requires more work on boundary
non-rational conformal field theory (see e.g.~\cite{Ponsot:2001ng}).
We expect that open string field theory to give a good handle on the physics of
Higgsed little string theory.
We proposed that
the low-energy effective field theory on these D1-branes in 
the appropriate regime
describes  dynamics of higgsed $\mathcal{N}=(1,1)$ Little
String Theory.

\section*{Acknowledgements}
We would like to thank Costas Bachas, Shmuel Elitzur, Amit Giveon,
Ami Hanany, Emiliano Imeroni, David Kutasov,
Sameer Murthy, Marios Petropoulos, Boris Pioline, Sylvain Ribault
and Amit Sever for interesting comments, discussions and correspondence.

\appendix

\section{Modular data}
\label{modtrans}
We gather in this appendix some conventions and modular properties
of characters that we use abundantly in the bulk of the paper,
and in particular in the transformations of the annulus amplitude from
open to closed string channels.
\subsection*{Free fermions}
We define the theta-functions as:
$$\vartheta \oao{a}{b} (\tau,\nu )
= \sum_{n \in \zi} q^{\frac{1}{2} (n+\frac{a}{2})^2}
e^{2 i\pi (n+\frac{a}{2})(\nu+\frac{b}{2})},$$
where $q=e^{ 2 \pi i \tau}$.
In the following, we will assume that $a,b \in \{ 0,1 \}$
for convenience. The theta-functions modular transform as follows:
\begin{equation}
\vartheta \oao{a}{b} (-1/\tau,\nu/\tau ) = e^{i\pi (-\frac{ab}{2} +
  \frac{\nu^2}{\tau} )} \vartheta \oao{b}{a} (\tau , \nu ) \ , \ \
\vartheta \oao{a}{b} (\tau+1,\nu ) = e^{\frac{i\pi}{4} a }
\vartheta \oao{a}{a+b+1} (\tau , \nu ).
\end{equation}
Recall that the chiral partition functions for fermions with Neveu-Schwarz
or Ramond boundary conditions, and possibly weighted by the fermion number
$\zi_2$ operator $(-1)^F$ where $F$ denotes worldsheet fermion number are
given by:
\begin{eqnarray}
Z^{a}_{b} (\tau) &=& \frac{1}{\eta(\tau)} \vartheta \oao{a}{b} (\tau,\nu )
\end{eqnarray}
where $a=0,1$ indicates the NS or R-sector respectively, and $b=0,1$ denotes
whether we did not or did insert the operator $(-1)^F$ into the chiral
partition sum.
It will be convenient to split the states inside the R and NS sectors
according to their fermion number, and to define characters as follows:
\begin{equation}
\begin{array}{ccllll}
\chi^{(0)} &=& \frac{1}{2\eta} \left\{ \theta \oao{0}{0} - \theta \oao{0}{1} \right\} & = &
\frac{\Theta_{0,2}}{\eta}
\\
\chi^{(2)} &=& \frac{1}{2\eta} \left\{ \theta \oao{0}{0} + \theta \oao{0}{1} \right\} & = &
\frac{\Theta_{2,2}}{\eta}
\\
\chi^{(1)} &=& \frac{1}{2\eta} \left\{ \theta \oao{1}{0} - i\theta \oao{1}{1} \right\} & = &
\frac{\Theta_{1,2}}{\eta}
\\
\chi^{(3)} &=& \frac{1}{2\eta} \left\{ \theta \oao{1}{0} + i\theta \oao{1}{1} \right\} & = &
\frac{\Theta_{3,2}}{\eta}
\\
\end{array}
\end{equation}
in terms of the theta functions of $\hat{\mathfrak{su}} (2)$
at level $2$:
$$\Theta_{m,k} (\tau,\nu) = \sum_{n \in \zi}
q^{k\left(n+\frac{m}{2k}\right)^2}
e^{2i\pi \nu k \left(n+\frac{m}{2k}\right)}.$$
It is natural that we can rewrite the characters of two fermions in this
way, since we can bosonize two fermions and obtain a compact boson at
radius $R= \sqrt{\alpha'/2}$, to which is associated an extended chiral
$U(1)$ algebra at level $2$.
The modular transformation property is then:
\begin{equation}
\chi^{(s)} (-1/\tau , \nu/\tau ) = \frac{1}{2} e^{i\pi \nu^2 /\tau}
\sum_{s' \in \zi_4}
e^{-i\pi ss'/2} \chi^{(s')} (\tau , \nu).
\end{equation}
We will often work in terms of these characters for fermions in NS or
R sector, projected onto even or odd fermion number states.
\subsection*{N=2 minimal models}
The N=2 minimal models correspond to the supersymmetric gauged WZW model
$SU(2)_k / U(1)$, and are characterized by the level $k$ of the supersymmetric
WZW model.
The $N=2$ minimal models characters are determined implicitly through the
identity:
\begin{equation}
\sum_{m \in \zi_{2k}} \mathcal{C}^{j}_{m} \oao{a}{b} \Theta_{m,k} =
\chi^{j} \vartheta
\oao{a}{b},
\end{equation}
where $\chi^j$ denote the characters for $SU(2)_{k-2}$ at level $k-2$.
Since we have the modular transformation properties:
\begin{eqnarray}
\Theta_{m,k} (-1/\tau,\nu/\tau) &=& \frac{\sqrt{- i \tau}}{\sqrt{2k}}
\sum_{m' \in \zi_{2k}} e^{- \pi i m' m/k} e^{\pi i k \nu^2/2 \tau}
\Theta_{m',k}(\tau,z) \nonumber \\
\chi^{j} (-1 / \tau) &=& \sum_{2j'=0}^{k-2}
\sqrt{\frac{2}{k}} \sin \pi \frac{(1+2j)(1+2j')}{k}
\chi^{j'} (\tau),
\end{eqnarray}
we can derive:
\begin{equation*}
\mathcal{C}^{j}_{m} \oao{a}{b} (-1/\tau,0 )
= \frac{1}{\sqrt{2k}}\  e^{-i\pi \frac{ab}{2}}
\sum_{2j'=0}^{k-2}
S^{j}_{\ j'} \sum_{m' \in \zi_{2k}}  e^{i\pi \frac{mm'}{k}}
\mathcal{C}^{j'}_{m'} \oao{b}{a} (\tau,0 ),
\end{equation*}
in terms of the modular S-matrix of SU(2):
\begin{equation}
 S^{j}_{\ j'} = \sqrt{\frac{2}{k}} \sin \pi \frac{(1+2j)(1+2j')}{k}.
\end{equation}
Note also that the fusion rules of $SU(2)$ are given by:
\begin{equation}
N_{\hat{\jmath}\hat{\jmath\,}'}^j = 1 \ \text{for}\ |\hat{\jmath}-\hat{\jmath\,}'|\leqslant 
j \leqslant \text{min}\{\hat{\jmath}+\hat{\jmath\,}',k-\hat{\jmath}-\hat{\jmath\,}'\}\ \text{and}\ 
j+\hat{\jmath}+\hat{\jmath\,}' \in \zi \ , \quad 0 \ \text{otherwise.}
\end{equation}
Similarly we can define characters through the decomposition:
\begin{equation}
\sum_{m \in \zi_{2k}} \mathcal{C}^{j\ (s)}_{m}  \Theta_{m,k} = \chi^{j}
\Theta_{s,2}\, ,
\end{equation}
In this case the characters are labeled by the triplet $(j,m,s)$.
The following identifications apply:
\begin{eqnarray*}
(j,m,s) &\sim &(j,m+2k,s)\\
(j,m,s) &\sim &(j,m,s+4)\\
(j,m,s) &\sim &(k/2-j-1,m+k,s+2)\\
\end{eqnarray*}
as well as the selection rule
\begin{equation}
2j+m+s =  0  \mod 2
\label{selruleMM}
\end{equation}
The weights of the primaries states are as follows:
\begin{equation}
\begin{array}{cccccc}
h &=& \frac{j(j+1)}{k} - \frac{m^2}{4k} + \frac{s^2}{8} \ & \text{for} & \ -2j \leqslant m-s \leqslant 2j \\
h &=& \frac{j(j+1)}{k} - \frac{m^2}{4k} + \frac{s^2}{8} + \frac{m-s-2j}{2}
\ & \text{for} & \ 2j \leqslant m-s \leqslant 2k-2j-2 \\
\end{array}
\end{equation}
We have the following modular S-matrix for these characters:\footnote{The reader may notice that the 
S-matrix of the N=2 minimal model given here may differ by a factor of two with the literature. Indeed in 
our conventions, the S-matrix is defined as acting on all triplets 
$(j,n,s)$, while it is often defined as acting on the fundamental domain w.r.t. the identification
$(j,n,s) \sim (k/2-j-1,n+k,s+2)$.
}
\begin{equation}
S^{j\, m\, s}_{\quad j' \, m' \, s'} = \frac{1}{2k} \sin \pi
\frac{(1+2j)(1+2j')}{k} \ e^{i\pi \frac{mm'}{k}}\ e^{-i\pi ss'/2}.
\end{equation}
That concludes our review of the minimal model characters.
\subsection*{Supersymmetric $SL(2,\mathbb{R})/U(1)$}
The characters of the $SL(2,\mathbb{R})/U(1)$ super-coset
at level $k$ come in
different categories corresponding to the classes of
irreducible representations
of the $SL(2,\mathbb{R})$ algebra in the parent theory. In all cases
the quadratic Casimir of the representations is $c_2=-j(j-1)$.

Firstly we consider \emph{continuous representations}, with $j = 1/2 + ip$,
$p \in \mathbb{R}^+$. The characters are denoted by
 $ch_c (p,m) \oao{a}{b}$,
where the $N=2$ superconformal
 $U(1)_R$ charge of the primary is $Q=2m/k$, $m \in \zi/2$. Explicitely they are given by:
\begin{equation}
ch_c (p,m;\tau,\nu) \oao{a}{b} = q^{\frac{p^2+m^2}{k}}e^{4i\pi\nu \frac{m}{k}} \frac{\vartheta \oao{a}{b} (\tau, \nu)}{\eta^3 (\tau)}
\end{equation}
We can define characters labeled by a $\zi_4$
valued quantum number for
$SL(2,\mathbb{R})/U(1)$, following the method we used to define
these characters for the free fermions. In other words, we define
these characters by summing over untwisted and twisted NS or R
sectors with the appropriate signs.

Then we have \emph{discrete representations} with $\nicefrac{1}{2} \leqslant j \leqslant \nicefrac{k+1}{2}$,
of characters $ch_d (j,r) \oao{a}{b}$, where the $N=2$ $U(1)_R$
charge is $Q= (2j+2r+a)/k$, $r\in \zi$. The characters read: 
\begin{equation}
ch_d (j,r;\tau,\nu) \oao{a}{b} = q^{\frac{-(j-1/2)^2+(j+r+a/2)^2}{k}} e^{2i\pi\nu \frac{2j+2r+a}{k}} \frac{1}{1+(-)^b \, 
e^{2i\pi \nu} q^{1/2+r+a/2}} \frac{\vartheta \oao{a}{b} (\tau, \nu)}{\eta^3 (\tau)}
\end{equation}
In the text we will take the convenient convention that this character is identically zero if $r$ is non-integer. 
The primaries in the NS sector are:
\begin{eqnarray*}
|j,m=j+r\rangle&=&  |0\rangle_\textsc{ns} \otimes |j,m=j+r\rangle_\textsc{bos} \quad r \geqslant 0 \\
|j,m=j+r\rangle &=& \psi^{-}_{-\frac{1}{2}}|0\rangle_\textsc{ns} \otimes (J^{-}_{-1} )^{-r-1}
|j,j\rangle_\textsc{bos} \quad r<0 \\
\end{eqnarray*}
Thus in the $\zi_4$ formalism the former are in the $s=0$ sector and the latter in the $s=2$ sector.
These primary states have weights
\begin{eqnarray*}
h&=& \frac{j(2r+1)+r^2}{k}   \quad \ \ \qquad \qquad r \geqslant 0\\
h&=& \frac{j(2r+1)+r^2}{k} -r - \frac{1}{2} \qquad r<0\\
\end{eqnarray*}
The Ramond sector is obtained by one-half spectral flow.
 
The third category corresponds to the \emph{finite
representations}, with $j=(u-1)/2$ and where $u = 1,2,3, \dots$
denotes the dimension of the finite representation.
These representations are not unitary except for the trivial representation with $u=1$.
The character for this identity representation
we denote by $ch_\mathbb{I} (r) \oao{a}{b}$. It is given by:
\begin{equation}
ch_\mathbb{I} (r;\tau,\nu) \oao{a}{b} =  \frac{(1-q)\  q^{\frac{-1/4+(r+a/2)^2}{k}} e^{2i\pi\nu \frac{2r+a}{k}}}{\left( 1+(-)^b \, 
e^{2i\pi \nu} q^{1/2+r+a/2} \right)\left( 1+(-)^b \, e^{-2i\pi \nu} q^{1/2-r-a/2}\right)} \frac{\vartheta \oao{a}{b} (\tau, \nu)}{\eta^3 (\tau)}
\end{equation}
The primaries in the NS sector for this identity representation are as follows.
First we have the identity operator $|j=0,r=0\rangle \otimes
| 0\rangle_\textsc{ns}$ belonging to the sector $s=0$.
The other primary states are:
\begin{eqnarray*}
|r\rangle&=& \psi^{+}_{-\frac{1}{2}} |0\rangle_\textsc{ns} \otimes (J^{+}_{-1} )^{r-1} |0,0\rangle_\textsc{bos} \quad r>0 \\
|r\rangle &=& \psi^{-}_{-\frac{1}{2}}|0\rangle_\textsc{ns} \otimes (J^{+}_{-1} )^{-r-1} |0,0\rangle_\textsc{bos} \quad r<0 \\
\end{eqnarray*}
They belong to the sector $s=2$ and have weights
\begin{eqnarray*}
h&=& \frac{r^2}{k} +r - \frac{1}{2} \quad r>0\\
h&=& \frac{r^2}{k} -r - \frac{1}{2} \quad r<0\\
\end{eqnarray*}

It is often convenient to define \emph{extended characters} by summing
over the spectral flow by $k$ units (for $k$ integer). These characters
correspond to an extended chiral algebra which can be constructed along
the line of the extended chiral algebra for a $U(1)$ boson at rational
radius squared. For example, for
the continuous characters we define the corresponding extended characters
(denoted by capital letters) by:
\begin{equation}
  Ch_c (p,n)\oao{a}{b} = \sum_{w \in \zi} ch_c (p,n+kw) \oao{a}{b}.
\end{equation}
They carry a $\zi_{2k}$ charge given by $2n$. In contrast with standard
characters, their modular transformations involve only a discrete set
of $N=2$ charges. We omit the details but record their modular 
properties.
For instance, the extended character associated to the trivial
representation (in the $\zi_4$ formalism for fermions)
transform as~\cite{Eguchi:2003ik}
\begin{eqnarray}
Ch_{\mathbb{I}}^{(s)} (r ;-1/\tau,0 ) \  = \sum_w ch_{\mathbb{I}}^{(s)} (r+kw ;-1/\tau,0 )
  \hskip5cm
\nonumber\\
= \frac{1}{k} \sum_{s' \in \zi_4} e^{-i \pi \frac{ss'}{2}}
\left\{
\int_{0}^{\infty} dp' \, \sum_{m' \in \zi_{2k}} e^{-2i\pi \frac{rm'}{k}}
\frac{ \sinh 2\pi p' \, \sinh 2\pi p' /k}{\cosh 2\pi p' + \cos \pi (m'+s')}
\, Ch_{c}^{(s')} (p',\frac{m'}{2};\tau,0)  \right. \qquad \nonumber\\
\left.  +  \sum_{2j'=2}^{k} \sum_{r' \in \zi_k}
\sin \left(\pi \frac{1-2j}{k}\right)\,  e^{-i 2\pi \frac{r(2j'+2r')}{k}}
\, Ch_{d}^{(s')} (j',r' , \tau ,0) \right\}.\nonumber\\
\end{eqnarray}
The extended characters of the discrete representations have a quite similar modular transformation law.
It reads~\cite{Eguchi:2003ik,Israel:2004xj}:
\begin{eqnarray}
Ch_{d}^{(s)} (j,r ;-1/\tau,0 ) \  = \sum_w ch_{d}^{(s)} (j,r+kw ;-1/\tau,0 )
  \hskip5cm
\nonumber\\
= \frac{1}{2k} \sum_{s' \in \zi_4} e^{-i \pi \frac{ss'}{2}}
\left\{
\int_{0}^{\infty} dp' \, \sum_{m' \in \zi_{2k}} e^{2i\pi \frac{(j+r)m'}{k}}
\frac{\cosh \pi \left(p' \frac{k+2(1-2j)}{k}+i\frac{m'+s'}{2}\right) }{\cosh \pi (p' +i\frac{m'+s'}{2} )}
\, Ch_{c}^{(s')} (p',\frac{m'}{2};\tau,0)  \right.  \nonumber\\
\left.  + \ i  \sum_{2j'=2}^{k} \sum_{r' \in \zi_k}
e^{- i\pi \frac{(2j+2r)(2j'+2r')}{k}} e^{-i\pi \frac{(2j-1)(2j'-1)}{k}}
\, Ch_{d}^{(s')} (j',r' , \tau ,0) \right.\nonumber\\
\left. + \frac{1}{2} \sum_{r' \in \zi_k} e^{- i\pi \frac{(2j+2r)(2r'+1)}{k}}
\left[ Ch_{d} (\nicefrac{1}{2},r';\tau,0)-Ch_{d} (\nicefrac{k+1}{2},r';\tau,0)\right] \right\}\nonumber\\
\end{eqnarray}
Finally we will also use the modular transform of the unextended continuous
characters:
\begin{eqnarray}
ch_{c}^{(s)} (p,m ;-1/\tau,0 )
=\nonumber \\ \frac{2}{k}  \sum_{s' \in \zi_4} e^{-i \pi \frac{ss'}{2}} \
\int_{-\infty}^{+\infty} d m' \ e^{4 i\pi \frac{m' m}{k}}\
\int_{0}^{\infty} dp' \ \cos \left( \frac{4\pi pp'}{k} \right)
ch_{c}^{(s')} (p',m' ;\tau,0).
\end{eqnarray}

\subsection*{Reflection amplitude}
In the course of the transformation of the annulus amplitude
from closed to open string channel, we need the boundary reflection
amplitude for two (Euclidean) $AdS_2$ branes in (Euclidean) $AdS_3$.
It corresponds to the non-trivial part of the boundary two-point
function in the presence of two $AdS_2$ branes of parameters $(r,r')$
It is given, for a boundary field of momentum $P$,
by~\cite{Ribault:2002ti}:
\begin{equation}
R (P|r,r') = \frac{S_{k}^{(0)} \left(\frac{k}{2\pi} (r+r') +P \right)}{
S_{k}^{(0)} \left(\frac{k}{2\pi} (r+r') -P \right)}
\frac{S_{k}^{(1)} \left(\frac{k}{2\pi} (r-r') +P \right)}{
S_{k}^{(1)} \left(\frac{k}{2\pi} (r-r') -P \right)}
\frac{S_k^{(0)}(-P)}{S_k^{(0)}(P)}
\, ,
\end{equation}
in terms of the special functions:
\begin{eqnarray}
\log S^{(0)}_k (x) &=& i \int_{0}^{\infty} \frac{dt}{t}
\left( \frac{\sin 2tx/k}{2\sinh t\,  \sinh t/k} - \frac{x}{t} \right)
\\
\log S^{(1)}_k (x) &=& i \int_{0}^{\infty} \frac{dt}{t}
\left( \frac{\cosh t\, \sin 2tx/k}{2\sinh t\,  \sinh t/k} - \frac{x}{t}
\right)
\end{eqnarray}
The reflection amplitude has extra $(r,r')$-independent terms but they cancel from the relative
open string partition function.

\section{Free fermion boundary states}
\label{boundfree}
In this appendix, we show in some familiar cases how the formalism
in which the fermion characters are labeled by a $\zi_4$ quantum number
operates in practice. These are warm-up exercises to acquaint 
ourselves and the reader
with the formalism that is extensively used in the bulk of the text.

For a free complex fermion, with characters $\chi^{s} = \eta^{-1}
\Theta_{s,2}$,
the A-type one-point function, which has left-moving
momentum equal to right-moving momentum in the boundary state, is:
\begin{equation}
\langle V^{(s,\bar{s})} \rangle_{A,\hat s} = \frac{1}{\sqrt{2}}
\delta_{s,\bar{s}}
 e^{-i\pi \frac{s\hat s}{2}}.
\end{equation}
The phase factor follows from a standard construction of Cardy
states from modular transformation matrices in rational conformal
field theories.
Let us consider a theory with diagonal bulk partition function, i.e.
$Z= \sum_{s} \chi^{(s)} \bar{\chi}^{(s)}$.
Then the overlap between two A-type boundary states labeled by
$\hat{s}$ and $\hat{s}'$ is:
\begin{equation}
\frac{1}{2} \sum_{s}  e^{i\pi \frac{s(\hat{s}' - \hat{s})}{2}} \chi^{(s)}
(\tau )
= \chi^{(\hat{s}' - \hat{s})} (-1/\tau).
\end{equation}
The B-type one-point function, where
the left-moving momentum is
equal to minus the right-moving momentum in the Ishibashi state is:
\begin{equation}
\langle V^{(s,\bar{s})} \rangle_{B,\hat s} = \delta_{s,-\bar{s}}
\  e^{-i\pi \frac{s\hat s}{2}}.
\end{equation}
Therefore the overlap between two boundary states $\hat{s}$ and
$\hat{s}'$ is:
\begin{equation}
\sum_{s=0,2}  e^{i\pi \frac{s(\hat{s}' - \hat{s})}{2}} \chi^{(s)} (\tau )
= \sum_{s' \in \zi_4} \frac{1 + (-)^{\hat{s}' - \hat{s} - s'}}{2}
\chi^{(s')} (\tilde{\tau}) = \sum_{\epsilon = 0,1} \chi^{(\hat{s}' -
\hat{s} +
2 \epsilon)}.
\end{equation}
We have used the fact that only the states with $s=0=-\bar{s}$ or
$s=2=-\bar{s}$ satisfy both the constraint of having opposite momentum
while simultaneously belonging to the particular bulk partition function
that we chose.

\subsection*{Branes in flat space}
In this subsection, we show how the formalism applies to branes in flat
space. Note first of all that
in the $\zi_4$ formalism, the GSO-projected partition function for the
fermions (only) of the type IIB superstring is:
\begin{eqnarray}
\frac{1}{2}\sum_{a,b \in \zi_2} \sum_{\{\eta_i\} \in (\zi_{2})^4}
(-)^a (-)^{b(1+\sum_i \eta_i)} \prod_i \chi^{(a+2\eta_i)} \nonumber\\
\times
\frac{1}{2}\sum_{\bar{a},\bar{b} \in \zi_2} \sum_{\{\bar{\eta}_i\} \in
(\zi_{2})^4}
(-)^{\bar{a}} (-)^{\bar{b}(1+\sum_i \bar{\eta}_i)}
\prod_i \bar{\chi}^{(\bar{a}+2\bar{\eta}_i)}.
\end{eqnarray}
The factor $(-1)^a$ is due to spin-statistics while the factor
$(-1)^b$ is due to a ghost contribution to the fermion number. The
labels $\eta_i$ label terms in the projector onto even worldsheet
fermion number.
The type IIA partition function is similar, but with an extra phase
$(-)^{\bar{a}\bar{b}}$ inverting the right GSO projection for the
spacetime spinors.

We concentrate on the fermions only in the following since it is the
fermionic part of the boundary state that incorporates most of the
properties we wish to highlight, and which are typical of the formalism.
We remind the reader that we work in light-cone gauge, where the light-cone
directions are implicit and have Dirichlet boundary conditions.
Let us define the AAAA-type boundary states through
the one-point function:
\begin{equation}
\langle V^{s_i,\bar{s}_i} \rangle^{A}_{\hat{s}_i} =
\prod_{i=1}^4
e^{-i \pi \frac{\pi s_i \hat{s}_i}{2}} \ \delta_{s_i ,\bar{s}_i} .
\end{equation}
Then the overlap between two of them in the GSO-projected theory is,
in the diagonal type IIB theory:
\begin{equation}
\frac{1}{2^5} \sum_{a,b \in \zi_2} (-)^a \sum_{\{\eta_i\} \in (\zi_{2})^4}
 (-)^{b(1+\sum_i \eta_i)} \
\prod_i\ e^{i\frac{\pi}{2} (a+2\eta_i) (\hat{s}_{i}' - \hat{s}_i )} \
\chi^{(a+2\eta_i)} (\tau).
\end{equation}
In the open string channel we find:
\begin{equation}
\frac{1}{2} \sum_{a,b \in \zi_2} (-)^b \sum_{\{\nu_i\} \in (\zi_{2})^4}
 (-)^{a(1+\sum_i \nu_i)} e^{- i \pi a b /2} \
\prod_i \
\chi^{(b+2\nu_i+ \hat{s}_{i}' - \hat{s}_i)} (-1/ \tau).
\end{equation}
We need to impose the constraint that all four pairs of
fermions in the open string
channel are either NS or R: this
amounts to setting $\hat{s}_{i}' - \hat{s}_i = 0 \mod 2$, $\forall i$.
Then, the open string spectrum is supersymmetric and consistent with
spin statistics provided that
$\sum_{i} (\hat{s}_{i}' - \hat{s}_i) =0 \mod 4$ as can be seen by shifting
the $\nu_i$ summation variables appropriately.
We can interpret this result as follows. Four A-type boundary conditions are
for example appropriate for a $D(-1)$-brane or a $D7$-brane (as one can
see by relating the boundary conditions on the fermions via worldsheet
supersymmetry to the boundary conditions on the bosons on the worldsheet).
We thus find that
odd dimensional branes are consistent with supersymmetry in type IIB string
theories.

Let us now consider a D-brane of the type AAAB still in type IIB
superstrings.
The relevant one-point function is:
\begin{equation*}
\langle V^{s_i,\bar{s}_i} \rangle^{A}_{\hat{s}_i} = 2
\prod_{i=1}^{3} e^{-i \pi \frac{\pi s_i \hat{s}_i}{2}} \ \delta_{s_i
,\bar{s}_i} .
 e^{-i \pi \frac{\pi s_4 \hat{s}_4}{2}} \ \delta_{s_4 ,-\bar{s}_4}
\end{equation*}
The B-type Ishibashi states will impose the constraint
$a + 2\eta_4 = -\bar{a} - 2\bar{\eta}_4 \mod 4$ on the left- and rightmoving
momenta, with solutions:
$a=0,\ \eta_4 = \bar{\eta}_4$ or $a=1,\ 1+\eta_4 =  \bar{\eta}_4 \mod 2$,
where we assumed that only NSNS and RR sectors are present in the boundary
states.
In the case $a=1$, the condition is never compatible with the chiral
GSO projections (because if $\sum \eta_i$ is odd, it implies
that $\sum \bar{\eta}_i$ is even which is inconsistent with the chiral
type IIB GSO projection in which these have the same parity),
so the R sector is not present in the closed string channel
annulus amplitude.
Therefore the cylinder amplitude will be non-zero and the spectrum in the
open string channel will be non-supersymmetric. This can be checked explicitly
using the techniques described above.
By contrast, in type IIA superstrings, while $\sum \eta_i$ is odd, $\sum \bar{\eta}_i$
is even, so the AAAB boundary conditions are compatible with the GSO
projections, leading to supersymmetric branes of even dimension.
Thus we have shown in a few basic examples how the formalism operates.

\section{On NS5-branes localization from instanton corrections}
\label{localization-appendix}
We mentioned in Section 2 that the factor $\Lambda_k$ that takes into account the localization
of the NS5 branes along the circle must come from instanton corrections when the background is
obtained by T-duality.

For the case of an infinite array of NS5-branes along a line, (or equivalently, a single
NS5-brane in a transverse space compactified to $S^1 \times \mathbb{R}^3$), the T dual space is
the Taub-NUT background, and the localization phenomenon has been studied in \cite{Gregory:1997te},
and  rigorously proved in \cite{Tong:2002rq}.
Although we do not study in detail here the way these instantons corrections arise in our geometry, we
will support our claim of an instantonic origin for $\Lambda_k$,
by showing that in a certain limit our background becomes that studied in \cite{Gregory:1997te,Tong:2002rq},
and the instanton corrections take the same form.

Let us define $\psi = \frac{\xi}{k}$
and consider the background of the $k$ NS5-branes in a circle:
\be
ds^2 &=& \eta_{\mu\nu}dx^{\mu}dx^{\nu} + H \left(d\rho^2 + \frac{\rho^2}{k^2} d \xi^2 + dR^2 + R^2 d\phi^2 \right)
\label{ns5circle}
\ee
where the harmonic function is either
\be
H_{h}&=& 1+  \frac{\alpha' k }{2 \rho \rho_0 \sinh y }
\ee
for an homogeneous distribution, with $y$ defined in (\ref{defy}), or
\be
H_{l}&=& 1+ \sum_{a=0}^{k-1} \frac{\alpha'}{R^2 +\rho^2+\rho_0^2-2 \rho_0 \rho \cos(\frac{2 \pi a}{k}-\frac{\xi}{k})}
\label{loca} \\
&=& \frac{\alpha' k}{2 \rho \rho_0 \sinh y} \left(1 + \sum_{\pm} \sum_{m =1}^{\infty} e^{-m ky \pm imk \psi} \right)
\label{locainstanton}
\ee
for the $k$ NS5 branes localized at equally spaced points along the circle.

We consider an observer situated between two consecutive localized NS5-branes.
Since he only sees the periodicity $\psi \sim \psi + \frac{2 \pi}{k}$, we
let $\xi = k\psi \in [0,2\pi)$. We want to take a limit where the NS5-branes in the circle become an infinite array of NS5-branes
along a line. This corresponds to the background of a single NS5, where one of the transverse directions, $\xi$,
has been compactified. This limit is obtained by defining $\rho= \rho_0 + t$, and taking
\be
k,\rho,\rho_0 & \rightarrow & \infty \nn \\
\frac{\rho}{k}  = \frac{\rho_0}{k} &=& 1
\label{limit}
\ee
where we keep the coordinate distance $t$ to the NS5-branes
fixed. Note that we had  $\rho >0$ but $t \in \mathbb{R}$.
The space transverse to the NS5's    becomes conformal to flat $S^1 \times \mathbb{R}^3$,
\be
ds^2 &=& \eta_{\mu\nu}dx^{\mu}dx^{\nu} + H \left(dt^2 +  d \xi^2 + dR^2 + R^2 d\phi^2 \right)
\ee
Let $u$ be the radial direction in the three transverse non-compact directions, i.e., $u^2 = t^2 + R^2$.
Then the harmonic functions become, in the limit (\ref{limit}),
\be
H_{h}^{li}&=& 1+  \frac{\alpha' }{2 u }\,,  \\
H_{l}^{li}&=&1 + \sum_{n \in \mathbb{Z}} \frac{\alpha'}{u^2 + (\xi -2\pi n)^2} \nn \\
&=& 1 + \frac{\alpha'}{2u} \left( 1 + \sum_{m=1}^{\infty} \sum_{\pm} e^{ - mu \pm i m \xi } \right)
\label{instantcor}
\ee

Let us see now what happens in the T-dual side.
We can always choose a gauge where the B field has only components of $B_{\xi,*}$ type.
Before taking the limit (\ref{limit}), a T duality of (\ref{ns5circle}) along
$\xi$ yields in the homogeneous case:
\be
ds^2 &=& H_{h}\left(  d\rho^2 + dR^2 + R^2 d \phi^2  \right) + \frac{k^2}{\rho^2 H_{h}}
\left( d \tilde{\xi} +  B_{\tilde{\xi}\rho} d \rho + B_{\tilde{\xi}\phi}d \phi + B_{\tilde{\xi}R}d R  \right)^2 \nn \\
\tilde{\Phi} &=& \Phi_0 - \log(\rho/k)
\label{td-metric}
\ee
and no $B$ field.
Taking now the limit (\ref{limit}), we get the Taub-NUT ALF metric
\be
ds^2 &=& H_{h}^{li} d \mathbf{r} \cdot d \mathbf{r}  +
\frac{1}{H_{h}^{li}} \left( d \tilde{\xi} +  \mathbf{B} \cdot d \mathbf{r}  \right)^2
\ee
When making a T-duality back along $\tilde{\xi}$ in this background,
it was shown in \cite{Tong:2002rq}
how instanton corrections generate the infinite sum
in the second line of (\ref{instantcor}), thus correcting $H_{h}^{li}$ to $H_{l}^{li}$.
Our claim is that a similar phenomenon occurs before taking the limit (\ref{limit}), so that
the T-duality of (\ref{td-metric}) leads to (\ref{ns5circle}), but with $H_{h}$ to corrected to $H_{l}$.
Just notice that the background in (\ref{td-metric}) is dilatonic, so the techniques of \cite{Tong:2002rq}
have to be adapted along the lines of \cite{Hori:2001ax}.

\subsection*{Remark on instantons corrections and D-branes}
The function $\Lambda_k$, whose geometric meaning is to keep track of the
discrete positions of the five-branes are reinterpreted as the
instantons corrections to the effective action of the
null coset $[SU(2)\times SL(2,\mathbb{R})]/ U(1)^2$. If that is
accurate we might expect
that we can extract from it information
 about the exact shape of D-branes.

Let us first consider background in the plane $r=0$. Then from
solution~(\ref{NS5geom}) we obtain the background of
the vector coset $SU(2)/U(1)$, i.e. the bell. Thus we propose that
the exact effective action of the D1-brane should be written in terms
of the following closed string background:
\begin{eqnarray}
\nicefrac{ds^2}{2k} &=& \left[ 1 + 2 \sum_{m=1}^{\infty}
\sin^{km} \theta \cos (mk \psi) \right] \left\{ d\theta^{2} + \tan^2 \theta d\psi^2 \right\}
\nonumber\\
e^{2\phi} &=& \frac{k}{\cos \theta} \left[ 1 + 2 \sum_{m=1}^{\infty}
\sin^{km} \theta \cos (mk \psi) \right]
\end{eqnarray}
In particular we observe that when we approach the singularity of the
classical action of the bell ($\theta \to \pi / 2$) the instanton corrections
becomes $\frac{2\pi}{k} \sum_n \delta ( \psi - \frac{2\pi n}{k} )$ such that
the singularity is replaced by $k$ punctures on the boundary of the disc as we expect from
the CFT analysis. However, these instantons corrections have no effect on the
Dirac-Born-Infeld action for the suspended D1-branes living in this plane. Indeed
in appropriate coordinates, we obtain:
\begin{equation}
S_{D1} =  \tau_1 \int d\tau d\sigma e^{-\Phi} \sqrt{\det g} =
\tau_1 \ e^{-\Phi_0} \int d\tau d\sigma \ \sqrt{\frac{u^2-1}{\Lambda_k (u,\psi)}}
\ \sqrt{\Lambda_k (u,\psi) \frac{u_{\sigma}^2 + u^2 \psi_{\sigma}^2}{u^2-1} }
\end{equation}
And we obtain also straight lines as solutions. However since the metric and
the dilaton blows up at the boundary of the disc, the DBI effective action
is no longer trustworthy. 

We can further discuss the background geometry,
by focusing on the limit $\theta \to 0$ of the background geometry in
(\ref{NS5geom}). We find
the cigar solution and at the same time $\Lambda_k \to 1$. This is consistent
with the fact that the $U(1)$
isometry of the cigar is unbroken and that the supersymmetric
cigar effective action is just the classical action.

Another region
 of interest in the closed string background
is the plane  $\theta =\pi/2$, where semi-infinite D1-branes
have been constructed. It gives a corrected
solution of the trumpet, i.e. the vector gauging \slc.  In that case we now that instantons corrections appear,
first because the non-conservation of windings in the cigar maps to
a breaking of the $U(1)$ isometry in the trumpet in the strong coupling region.
We know also that the exact effective action for the supersymmetric trumpet should be
the N=2 Liouville.
We find the following form for the closed string
background seen by the D1-branes (after a change
of coordinates):
\begin{eqnarray}
ds^2 &=&
\frac{\left( 1-e^{-\sqrt{\frac{k}{2}}\, Z + \bar{Z}} \right)
e^{\sqrt{\frac{2}{k}}(Z +\bar{Z})} \ dZ d\bar{Z}
}{\left( 1-e^{-\sqrt{\frac{k}{2}}\, Z}\right)
\left( 1-e^{-\sqrt{\frac{k}{2}}\, \bar{Z}} \right) \left( e^{\sqrt{\frac{2}{k}}(Z +\bar{Z})}-1 \right)}
\nonumber \\
e^{2\Phi} &=&   \frac{ e^{2\Phi_0} \left(1-e^{-\sqrt{\frac{k}{2}}\, Z + \bar{Z}} \right)
}{\left( 1-e^{-\sqrt{\frac{k}{2}}\, Z}\right)
\left( 1-e^{-\sqrt{\frac{k}{2}}\, \bar{Z}} \right) \left( e^{\sqrt{\frac{2}{k}}(Z +\bar{Z})}-1 \right)}
\end{eqnarray}
and all the corrections are given in terms of the N=2 Liouville potential.

\section{D2-branes of the cigar for integer level}
\label{D2integ}
The exact D2-branes of the supersymmetric coset \slc were found in~\cite{Israel:2004jt} in
a similar way as for the bosonic coset in~\cite{Ribault:2003ss}. 
In a careful analysis
it was found that
the boundary state coupling leads to an open string partition function containing
a contribution from the trivial representation,
with {\it negative} multiplicities.
However whenever the level $k$ is integer --~this is the case in the present context~-- these
unwanted features disappear as we shall see below.

The one-point function for the D2-branes covering all the cigar are obtained from the known
H$_2$-branes of Euclidean AdS$_3$ (and are 
therefore consistent with part of the conformal bootstrap):
\begin{multline}
\label{oneptD2}
\langle \Phi^{j\ (s)}_{nw} (z,\bar{z} )\rangle^{D2}_{\sigma, \hat{s}}  = \delta_{n,0}\ \delta_{s,0 \, \text{mod}\, 2}
\frac{\Psi_{\sigma} (j,w)}{|z-\bar{z}|^{\Delta_{j,w}}} \ , \quad
\text{with}\ :\\
\Psi_{\sigma} (j,w)
= \frac{1}{\sqrt{k}} \ \nu^{\frac{1}{2}-j} \ e^{i\pi \frac{s\hat{s}}{2}}
\frac{\Gamma(j+\frac{kw}{2}-\frac{s}{2})\Gamma(j-\frac{kw}{2}+\frac{s}{2})
}{\Gamma(2j-1) \Gamma(1-\frac{1-2j}{k})}\\
\frac{e^{i \sigma(1-2j)} \sin \pi (j-\frac{kw}{2}+\frac{s}{2}) +
      e^{-i \sigma(1-2j)} \sin \pi (j+\frac{kw}{2}-\frac{s}{2})
}{\sin \pi(1-2j) \sin \pi \frac{1-2j}{k} }
\end{multline}
For generic $k$ this one point function has poles corresponding to the discrete
representations, and therefore will couple both to localized and asymptotic
states. This is expected on general grounds since these
 D2-branes carry D0-brane charge.

The annulus partition function in the closed string channel, for general
Casimir labeled by $j$, becomes:
\begin{align}
Z^{D2}_{\sigma \sigma'} (-1/\tau, \nu/\tau )
&= - \ \int dj \sum_{w \in \zi}\sum_{\nu=0}^{1}
(-)^{\nu (\hat{s}'-\hat{s})}\
\frac{ch^{(2\nu)} \left(j,\frac{kw}{2};-1/\tau, \nu/\tau \right)}{\sin \pi (1-2j) \sin \pi
  \frac{(1-2j)}{k}}
\nonumber \\
\hskip-2.6cm &
 \left\{ 2 \cos(\sigma+\sigma')(1-2j) -2 \cos (\sigma-\sigma') (1-2j) \cos 2 \pi j
\phantom{\frac{1}{2}}  \right.
\nonumber \\
\hskip1.4cm
&\left.
 + \ \frac{2 \cos (\sigma -\sigma') (1-2j) \sin^2 2 \pi j}{ \cos \pi kw-\cos 2 \pi
j}
- \frac{2 i \sin(\sigma-\sigma')(1-2j) \sin 2 \pi j \sin \pi kw}{
\cos \pi kw - \cos 2 \pi j} \label{cardD2}
\right\}
\end{align}
However for k integer  it can be recast in a simpler form:
\begin{align}
Z^{D2}_{\sigma \sigma'} (-1/\tau, \nu/\tau )
&= - 2\ \int dj \sum_{w \in \zi} \sum_{\nu=0}^{1} (-)^{\nu (\hat{s}'-\hat{s})}\
\frac{ch^{(2\nu)} \left(j,\frac{kw}{2};-1/\tau, \nu/\tau \right)}{\sin \pi (1-2j) \sin \pi
  \frac{(1-2j)}{k}}
\nonumber \\
\hskip-2.6cm &
 \left[  \cos(\sigma+\sigma')(1-2j) +(-)^{kw} \cos (\sigma-\sigma') (1-2j)
\phantom{\frac{1}{2}}  \right]
\label{cardD2integer}
\end{align}
and thus there are no couplings at all to the discrete representations.
The generic problem of negative multiplicities
disappears for this particular case. 
Operationally, this is because the residues at the poles vanish.
After a modular transformation we obtain the following relative amplitude in the
open string channel:
\begin{eqnarray}
Z^{D2}_{\sigma \sigma'} (\tau, \nu) = \sum_{t \in \zi_4}
\frac{1+(-)^{\hat{s}'-\hat{s}+t}}{2} \int dP \
\left\{ \frac{\partial}{2i\pi \partial P} \log \frac{R (P|i \frac{\sigma +\sigma'}{2})}{
R (P|i\frac{\sigma_0 +\sigma_0 '}{2})}
\sum_{n \in \zi} ch^{(t)}_c ( P, n;\tau, \nu)\right. \phantom{aaaaa}
\nonumber \\ +
\left.
\frac{\partial}{2i\pi \partial P} \log \frac{R (P|i \frac{\sigma -\sigma'}{2})}{R (P|i\frac{\sigma_0-\sigma_0 '}{2})}
\sum_{n \in \zi} ch_{c}^{(t)} ( P, n+\frac{k}{2};\tau, \nu) \right\}\nonumber \\
\end{eqnarray}
which is perfectly consistent. So we conclude that the D2-brane
is consistent for
$k$ integer.

\bibliography{LSTbranes}

\end{document}